\documentclass[a4paper,11pt]{article}
\pdfoutput=1 
\usepackage{afterpage}

\usepackage{jheppub} 
\usepackage{url}
\usepackage[T1]{fontenc} 
\graphicspath{{img/}} 

\usepackage[utf8x]{inputenc}
\usepackage[T1]{fontenc}
\usepackage{jheppub}
\usepackage{amsthm}
\usepackage{slashed}
\usepackage{mathtools}
\usepackage{float}
\usepackage{empheq}
\usepackage{bm}
\usepackage{framed}
\usepackage{comment}
\usepackage{xcolor}

\definecolor{Mblue}{rgb}{0.37,0.51,0.71}
\definecolor{Morange}{rgb}{0.88,0.61,0.14}
\definecolor{Mgreen}{rgb}{0.56,0.69,0.19}

\def\ss{\subsection}
\def\sss{\subsubsection}
\def\pg{\paragraph}

\def\ie{\emph{i.e.} }
\def\eg{\emph{e.g.} }

\def\tz{{\tilde{z}}}

\def\nt{\notag}
\def\ol{\overline}
\def\wt{\widetilde}
\def\wh{\widehat}

\newcommand{\norm}[1]{\left\lVert#1\right\rVert}

\def\R{\mathbb{R}}
\def\Z{\mathbb{Z}}
\def\bA{\mathbb{A}}

\def\cE{\mathcal{E}}
\def\cF{\mathcal{F}}

\def\cH{\mathcal{H}}

\def\cN{\mathcal{N}}

\def\cP {\mathcal{P}}
\def\cQ {\mathcal{Q}}

\def\cS{\mathcal{S}}
\def\cT{\mathcal{T}}

\def\bA{\mathbb{A}}

\def\bH{\mathbb{H}}

\def\bN{\mathbb{N}}

\def\bQ {\mathbb{Q}}

\def\fh {\mathfrak{h}}

\def\dg {\dagger}
\def\p{\partial}

\def\/{\over}
\def\ov{\over}

\def\rn{\rangle}
\def\ln{\langle}

\def\t{\theta}
\def\s{\sigma}
\def\e{\epsilon}

\def\vphi{\varphi}
\def\a{\alpha}

\def\d{\delta}

\def\g {\gamma}
\def\la {\lambda}

\def\z{\zeta}

\def\l{\ell}

\def\n {\nabla}
\def\L{\Lambda}
\def\D{\Delta}
\def\G {\Gamma}
\def\Om {\Omega}

\def\ra{\rightarrow}

\def\r{\mathrm}
\def\hc{\text{h.c.}}

\def\_{\hspace{2cm}}
\def\-{\\\notag}
\def\={&=&}

\newcommand\be{\begin{equation}}
\newcommand\ee{\end{equation}}

\newcommand{\bea}{\begin{eqnarray}}
\newcommand{\eea}{\end{eqnarray}}

\newcommand{\bpm}{\begin{pmatrix}}
\newcommand{\epm}{\end{pmatrix}}

\newcommand{\bit}{\begin{itemize}}
\newcommand{\eit}{\end{itemize}}

\newcommand{\ben}{\begin{enumerate}}
\newcommand{\een}{\end{enumerate}}

\newcommand\bsp{\begin{split}}
\newcommand\esp{\end{split}}

\def\le{\left}
\def\ri{\right}

\def\l{\ell}

\def\qq{\qquad}

\newcommand{\subf}[2]{%
  {\small\begin{tabular}[t]{@{}c@{}}
      #1\\#2
    \end{tabular}}%
  }

\title{Möbius randomness in the Hartle-Hawking state}

\author[a]{Victor Godet}

\affiliation[a]{Laboratoire de Physique Théorique et Hautes Energies (LPTHE, UMR 7589),
Sorbonne Université and CNRS, Campus Pierre et Marie Curie,
4 place Jussieu 75005, Paris, France}

\emailAdd{vgodet@lpthe.jussieu.fr}

\abstract{We consider quantum cosmology for toroidal universes in $d+1$  dimensions. The Hilbert space is the space of square-integrable automorphic forms for GL($d$). The Hartle-Hawking state is defined as a Poincar\'e sum over the no-boundary geometries. We obtain its representation in the  Langlands spectral decomposition. This leads to an expression as a sum over the Riemann zeta zeros and implies that its near singularity dynamics is governed by the Hilbert-Pólya Hamiltonian. It also takes the form of  a Möbius average of CFT$_d$ partition functions which suggests a similar interpretation for the de Sitter entropy. We briefly discuss the  relationship between quantum cosmology and the Langlands program.

}

\def\SLZ{\r{SL}(3,\Z)}

\def\bs{\backslash}
\def\tb{{\ol{b}}}

\def\tQ{{\wt{Q}}}

\newcommand{\eqd}[1]{\underset{(d=#1)}{=}}

\newcommand{\bsm}{\Big(\begin{smallmatrix}}

\newcommand{\esm}{\end{smallmatrix}\Big)}

\def\fm{{\mathfrak{m}}}
\def\vn{{\vec{n}}}
\def\vm{{\vec{m}}}
\def\vu{{\vec{u}}}
\def\vnu{{\vn\cdot\vu}}

\begin{document} 
\maketitle
\flushbottom

\section{Introduction}

 The quantum theory of closed universes remains elusive, despite tremendous progress for quantum gravity in spacetimes with boundaries \cite{Maldacena:1997re,Witten:1998qj}. Recently, this has led to a renewed understanding of the semi-classical gravity path integral, suggesting that it computes a disorder average \cite{Saad:2019lba,Almheiri:2020cfm} whose  true nature is still unclear.  The application of the gravitational path integral to cosmology leads to the definition of a special state: the Hartle-Hawking wavefunction \cite{Hartle:1983ai}. This gives a notion of gravitational vacuum and a natural candidate for the wavefunction describing our universe, although in tension with slow-roll inflation  \cite{Maldacena:2024uhs}. The Hartle-Hawking state can be defined for any choice of spatial topology, and various topologies beyond the sphere have been considered
\cite{Laflamme:1986bc,Anninos:2012ft,Castro:2012gc,Banerjee:2013mca,Conti:2014uda,Turiaci:2025xwi}. In this paper we will focus on the Hartle-Hawking state for the $T^d$ topology, a potentially realistic model of our universe for $d=3$, although only pure gravity will be considered.

Whether or not holographic ideas can be generalized to cosmology has been a subject of  discussion and  debate \cite{Witten:2001kn,Strominger:2001pn,Strominger:2001gp,Maldacena:2002vr,Goheer:2002vf,Anninos:2011ui,Hertog:2011ky,Harlow:2011ke,Anninos:2012qw,Anninos:2014lwa,Maldacena:2019cbz}. Part of the confusion comes from the lack of a consistent framework for semi-classical quantum cosmology. The canonical quantization of gravity does offer such a framework based on first principles \cite{Arnowitt:1962hi}. The Hilbert space is the space of solutions of the Wheeler-DeWitt equation \cite{DeWitt:1967yk}, a second order functional equation which has proven difficult to handle in a systematic manner.

Recently, it was shown that this equation can be solved systematically in the large volume limit $T=\int d^du\,\sqrt{g}\to+\infty$. The solutions take the form  \cite{Chakraborty:2023yed}
\be
\Psi[g]  = e^{iS_\text{div}}Z_\r{CFT}[g]
\ee
where $S_\r{div}$ is a divergent universal phase and $Z_\r{CFT}[g]$ can be thought of as a CFT partition function. This is because in the large volume limit,  the Wheeler-DeWitt equation  becomes the Ward identity for the Weyl symmetry of a CFT$_d$.  Time evolution corresponds to the RG flow and in $d=2$ is equivalent to the $T\bar{T}$ deformation   \cite{Godet:2024ich}.

Note that the CFTs that are relevant in cosmology are rather exotic. To be viewed as a wavefunction, the partition function must be in $L^2$, which is generally not the case for unitary CFTs. Nonetheless we may expect to obtain interesting wavefunctions from unitary CFTs at analytically continued values of the parameters. For example, the Liouville CFT$_2$ at $c\in 13+i\R$ gives an interesting state for pure three-dimensional cosmology which is normalizable and in which some cosmological observables are computed by a matrix model \cite{Collier:2024kmo,Collier:2025lux}. We will see in this work that  the Liouville CFT$_2$ and its higher-dimensional generalizations \cite{Levy:2018bdc,Kislev:2022emm} are related to the Hartle-Hawking state.

In the mini-superspace approximation, the Wheeler-DeWitt equation becomes a Klein-Gordon equation on the space $\r{GL}(d,\R)/\r{O}(d)$. This fact has been known for a long time \cite{DeWitt:1967yk,Hawking:1983hj,Martinec:1984fs,Halliwell:1984eu,PhysRevD.38.2468,Pioline:2002qz, Witten:2022xxp} although it doesn't seem to have been exploited systematically, with the global $\r{SL}(d,\R)$ symmetry made manifest. We will take the canonical time to be the spatial volume $T$ and the unit spatial metric to be parametrized by $z\in \fh^d  = \r{SL}(d,\R)/\r{SO}(d)$, the generalized upper half-plane. We view the mini-superspace approximation as a WKB-like approximation of the wavefunctional, which could in principle be corrected systematically.

The canonical formalism  defines a separate Hilbert space for each choice of spatial topology. The focus of this paper will be the torus topology $T^d$ in which the $d$ spatial coordinates are taken periodic. This case is particularly nice because it has a moduli space with finite volume. 
This is because the $\r{SL}(d,\Z)$ symmetry acting on $\fh^d$ is gauged, as it is part of the spatial diffeomorphism group. The  Hilbert space  is then
\be
\cH=L^2(\r{SL}(d,\Z)\bs \fh^d)~.
\ee
This is the Hilbert space of automorphic forms for $\r{GL}(d,\R)$. The spectral theory for this Hilbert space is due to Langlands \cite{langlandsFunctionalEquationsSatisfied1976} as a generalization of Selberg's theory in the $d=2$ case \cite{Selberg}. In physics, we always  classify states according to representations of the global symmetry group, which is here $\r{SL}(d,\R)$, so we want to view $\cH$ as a representation space and are interested in the irreducible representations. This is precisely Langlands' original perspective on $\cH$, that of \emph{automorphic representations}  \cite{Langlands1971Euler, langlandsProblemTheoryAutomorphics,muellerGenesisLanglandsProgram2021,borel1979automorphic,Fleig:2015vky}. His spectral decomposition   then provides a basis for the Hilbert space that  diagonalizes the global symmetries. Thus automorphic representations are the physical states here and Langlands' spectral theory, which we review in  Appendix~\ref{app:aut}, provides a framework for  quantum cosmology which can be applied whenever the moduli space has finite volume. Further comments about this relationship are included in Section \ref{sec:Langlands}.

The no-boundary geometries are filled tori, obtained by choosing a circle and making it contractible. They are quotients of (minus) the hyperbolic space $H_{d+1}$. The Hartle-Hawking state takes the form of the Poincaré sum
\be
\psi_\r{HH}(z) = \tfrac12\sum_{\g\in P\bs \r{SL}(d,\Z)} \r{det}(\g z)^{1/2}e^{-C \,\r{det}(\g z)}~,
\ee
where $C\sim 1/G$ comes from the classical action. This is a sum over $P\bs\r{SL}(d,\Z)$ where $P$ is the maximal parabolic subgroup. This can be written as a sum over $(n_1,\dots,n_d)\in \Z^d$ with a condition $\r{gcd}(n_1,\dots,n_d)=1$.  This corresponds to the linear combinations of the elementary circles. The prime condition is required by the no-boundary condition, as non-primitive circles would lead to conical singularities.

We will see that the spectral tools developed by Selberg and Langlands will be invaluable to study the Hartle-Hawking state. In fact the spectral representation of the Hartle-Hawking state
\be
\psi_\r{HH}(z)=-{2\sqrt{\pi C}\/d\L(\tfrac{d}{2})}+ {1\/2\pi i}\int_{({1\/2})} ds\,\G(\tfrac{1}{2}-s) C^{s-{1\/2}} E_s(z)~,
\ee
directly follows from the Cahen-Mellin integral, given that the maximal parabolic Eisenstein series for $\r{GL}(d)$ is defined as
\be
E_s(z) = \tfrac12 \sum_{\g\in P\bs \r{SL}(d,\Z)} \r{det}(\g z)^s~.
\ee
 Higher order corrections can be  discussed in terms of the $g$-function, a $\r{GL}(d)$ generalization of the Dedekind eta function \cite{efratGL3AnalogIz1992,liuKroneckerLimitFormula2015}, reviewed in \ref{app:gfunction}.

We  can now examine the pole structure of the integrand. There are integer poles in the region $\r{Re}(s)>\tfrac12$ and we get the expansion
\be
\psi_\r{HH}(z) = \sum_{n\geq 1} {(-C)^n\/n! } E_{n+{1\/2}}(z)~.
\ee
Note that this expansion is actually what follows  from the power series expansion of the exponential.  On the other half-plane, the Eisenstein series has poles at the location of the non-trivial zeros of the Riemann zeta function. This follows by rewriting it in terms of the completed Eisenstein $\cE_s(z)=\L(\tfrac{d}{2}s)E_s(z)$ whose poles are only at $s=0,1$ and the completed Riemann zeta function  \eqref{defLambda}. Summing the residues we get 
\be
\psi_\r{HH}(z) = -{2\sqrt{\pi C}\/d\L({d\/2})} -\sum_\rho  {C^{ {\rho/d}-{1\/2}}\G({1\/2}-{\rho\/d})\/d\,\pi^{-\rho/2}\G({\rho\/2})\z'(\rho)} \cE_{\rho\/d}(z)
\ee
where $\rho$ runs over the non-trivial zeros of the Riemann zeta function. This sum is convergent in $d=3$ so this expresses the Hartle-Hawking state as a constant contribution with additional small fluctuations. The first expansion does reproduce this constant prediction in a region of convergence as shown in Figure~\ref{Fig:HH3intplot}. The small fluctuations are governed by the non-trivial zeta zeros and the Riemann hypothesis is the statement that they are of order $G^{1/3}$ in $d=3$ and $G^{1/4}$ in $d=2$.

The near-singularity limit is particularly interesting. It can be obtained by evolving back the Hartle-Hawking state in time as $T\to0$, by simply solving the Klein-Gordon equation. In this limit, we get 
\be
\psi_\r{HH}^T(\tau) \underset{T\to0}{=} -{3\, e^{-{i\pi/4}}\/\sqrt{G}}+ e^{-t/2} \sum_{\rho} e^{-i\, \r{Im}(\rho)\, t}\,\psi_{\rho}(\tau) + O(e^{-3t/2}),\qq (d=2)
\ee
where the sum is over the non-trivial zeros of the zeta function. This is written in terms of the time variable $t= -\tfrac12 \log T$ in which the singularity corresponds to $t\to+\infty$.  Thus  the Hilbert-Pólya Hamiltonian, defined to have the imaginary parts as eigenvalues, is the time evolution operator for the Hartle-Hawking state near the singularity.  
The contribution of $\psi_\rho$ decays with the imaginary part so   the sum is actually well approximated by the first term. Thus the near-singularity wavefunction oscillates in an effective two-dimensional Hilbert space with frequency $\r{Im}(\rho_1)\approx 14.13$ as we display in Figure \ref{Fig:HHplotsing}.

The origin of this expansion on the zeta zeros can be traced back to the relatively prime condition in the Poincaré sum. In the spectral representation with all poles manifest, it is responsible for a factor of the inverse zeta function. This prime condition is ultimately due to the no-boundary condition which requires the contracting circle to be primitive to avoid a conical singularity. So using the Dirichlet expansion
\be
{1\/\z(s)} = \sum_{m\geq 1}{\mu(m)\/m^s}~,
\ee
we obtain a different interpretation of this inverse zeta function. This gives a representation of the Hartle-Hawking as a Möbius average. The Möbius function $\mu(m)$, despite having a simple deterministic definition, is known to be pseudo-random. It behaves to a remarkable extent like  a random sequence, a fact that has important consequences in number theory \cite{sarnakThreeLecturesMobius,TaoBlogPost}. In section \ref{sec:Mob}, we show that the Hartle-Hawking state is the Möbius average
\be
\psi_\r{HH}=\sum_{m\geq 1}\mu(m)  m^{{d\/2}-1} Z_\r{CFT}[Cm^d]
\ee
where $Z_\r{CFT}$ is the partition function of a CFT on $T^d$, explicitly   given in \eqref{ZCFTgen}. At leading order, we can identify this CFT as the  Liouville CFT in $d$ dimensions \cite{Levy:2018bdc,Kislev:2022emm}, a generalization of the well-studied Liouville CFT$_2$ \cite{Polyakov:1981rd,Nakayama:2004vk}.   
A similar interpretation of the de Sitter entropy is also given as a Möbius average of the Liouville CFT partition function on $S^d$.
The fact that the Möbius function arises naturally in the sum over geometries suggests that it could explain the randomness we expect in the gravity path integral.  The Möbius averaging may be viewed as a form of gauging and we might expect it to be quenched, which would predict wormhole corrections. Finally, as we cannot have true randomness in an exact theory, pseudo-randomness may be ultimately required, and number theory is the most natural source.

The norm of the Hartle-Hawking state is computed in section \ref{Sec:norm}. For $d=3$, we only have the leading order answer which takes the form
\be
\norm{\psi_\r{HH}}^2= {(\tfrac23 \pi)^5\/\z(3)G} + O(1)~,\qq (d=3)~.
\ee
For $d=2$, we can compute the exact norm including higher order corrections, and the third-quantized norm takes the form of a suggestive product formula \eqref{finalHH3norm}.

  Our results should also apply in the Anti-de Sitter context. They provide new expressions for the Maloney-Witten partition function \cite{Maloney:2007ud}  and its generalization to higher dimensions, \ie the partition function of pure AdS$_{d+1}$ gravity on $T^d$. The relation between Poincaré sums and the gravity path integral has a long history  \cite{Dijkgraaf:2000fq,Cheng:2011ay,Cheng:2012qc,Castro:2011zq, Castro:2011xb,Keller:2014xba,
  Benjamin:2021wzr,Meruliya:2021utr} and the spectral representation of non-holomorphic Poincaré sums may be applied to related problems. The spectral decomposition is generally useful to deal with automorphic objects, and has also been applied  to study properties of the AdS$_3$ average \cite{Benjamin:2021ygh,Collier:2022emf,Collier:2021rsn,DiUbaldo:2023qli,Haehl:2023mhf,Haehl:2023xys,Boruch:2025ilr}.

\section{Quantization of toroidal universes}

We consider pure Einstein gravity in $d+1$ dimensions as described by the action
\be
S_\text{grav}= {1\/16\pi G}\int d^{d+1}x\,\sqrt{-\hat{g}} (\hat{R}-2\L) -{1\/8\pi G} \int d^{d}u \sqrt{g} \,K
\ee
with a positive cosmological constant $\L ={d(d-1)\/2}$ in units where $\l_\r{dS}=1$.

\ss{Wheeler-DeWitt equation}

The quantum states in canonical gravity are functionals $\Psi[g]$ of the spatial metric $g$ that satisfy the gravitational constraints. The momentum constraint imposes that $\Psi[g]$ is invariant under spatial diffeomorphisms. We will always impose it from the start, by considering only objects invariant under spatial diffeomorphisms.

The Hamiltonian constraint is the Wheeler-DeWitt equation \cite{DeWitt:1967yk}
\be
\le[{16\pi G\/\sqrt{g}}\le(  \pi_{ij} \pi^{ij} -{1\/d-1}\pi^2\ri)  - {1\ov 16\pi G}\sqrt{g}(R-2\L)\ri]\Psi[g] = 0
\ee
and defines the physical Hilbert space.

We take the spatial topology to be $T^d$ so that the spacetime has the metric 
\be
ds^2 = -N^2 dt^2 + g_{ij}du^i du^j,\qq u^i\sim u^i+2\pi~.
\ee
where $u_i$ are periodic coordinates on $T^d$.  The $\r{SL}(d,\Z)$ action on the coordinates preserves the periodicity conditions so they constitute global diffeomorphisms of the torus. Thus $\r{SL}(d,\Z)$ is a gauge symmetry and we should only consider $\r{SL}(d,\Z)$-invariant objects.

In the small $G$ limit, we use an approximation of the wavefunction
\be\label{PsiWKBmini}
\Psi[g] = \r{exp}\Big({i\/G}F(g^{(0)})+O(1)\Big)
\ee
where the leading order term only depends on the constant modes of the metric in the Fourier decomposition on $T^d$. In this approximation, the Wheeler-DeWitt equation reduces to a Klein-Gordon equation: this is the mini-superspace approximation \cite{DeWitt:1967yk,Hawking:1983hj,Martinec:1984fs,Halliwell:1984eu,PhysRevD.38.2468}. Quantum cosmology then becomes the theory of a particle, which we will now make more explicit. Our viewpoint is that the mini-superspace approximation  computes the leading term of the wavefunction and can be systematically corrected in $G$, in a way similar to the large volume expansion described in \cite{Chakraborty:2023yed}.

\ss{Cosmology as a particle}

In the mini-superspace approximation, the wavefunction becomes a function on the generalized upper half-plane
\be
\fh^d =\r{GL}(d,\R)/(\r{O}(d,\R)\cdot\R^\times)\simeq\r{SL}(d,\R)/\r{SO}(d),\qq \r{dim}\,\fh^d=\tfrac12(d-1)(d+2)
\ee
known in this context as  DeWitt's mini-superspace \cite{DeWitt:1967yk}. This is the space that parametrizes symmetric $d\times d$ real matrices, the spatial metrics in canonical gravity. 

The standard coordinates are obtained from the Iwasawa decomposition of an element of $\r{GL}(d,\R)$. For $d=3$, it is
\be\arraycolsep=3pt\def\arraystretch{1.1}
z=  \bpm 1& x_{2} & x_{3} \\ 0 & 1 &  x_{1} \\ 0 & 0 & 1 \epm\cdot   \bpm y_1 y_2 & 0 & 0\\ 0 & y_1 & 0 \\ 0 &0 & 1\epm =\bpm y_1 y_2 & y_1 x_2 & x_{3} \\ 0 & y_1 & x_{1} \\ 0 & 0 & 1 \epm \in \fh^3
\ee
and we can restrict to representatives in the upper triangular form. The five coordinates $(x_1,x_2,x_3,y_1,y_2)$ are three-dimensional generalizations of $\tau=x+iy\in \fh^2 \simeq \bH$ which is the upper half-plane.

 The unit spatial metric is then defined as
\be\label{defunitmetric}
g_z ={1\/\r{det}(z)^{2/d}}\: z\cdot {}^tz,\qq \r{det}\,g_z=1~,
\ee
which is indeed invariant under the right action of $\r{O}(d)$. This allows us to parametrize the space of spatial metrics by $z\in \fh^d$ and  the spatial volume
\be
T = {1\/(2\pi)^d}\int_{T^d} d^du\sqrt{g}~.
\ee
We will take $T$  to be our definition of canonical time which  is appropriate for an expanding spacetime \cite{Chakraborty:2023yed,Godet:2024ich}.

For the $T^d$ topology of interest here, the mapping class group is $\r{SL}(d,\Z)$. Indeed  the action of $\g\in \r{SL}(d,\Z)$ on $\vec{u}$ is a global spatial diffeomorphism of $T^d$ so it is gauged in gravity. This means that we should quotient by the left   action of $\r{SL}(d,\Z)$ on $\fh^d$ so the relevant moduli space is
\be
\r{SL}(d,\Z)\bs\fh^d~.
\ee
This is the $\r{SL}(d,\Z)$ generalization of the modular surface in $d=2$.

The simplest way to obtain the particle theory is from a path integral perspective, as Kaluza-Klein reduction on the $T^d$ \cite{Martinec:1984fs, Pioline:2002qz}. The spacetime metric ansatz is
\be
ds^2 = -N^2 dt^2 +T^{2/d} (g_z)_{ij}du^i du^j~.
\ee
We can compute the Einstein-Hilbert action of this spacetime. The Gibbons-Hawking-York term gives a contribution that cancels terms of the form $\p_t N$. We can integrate out $N$ and we obtain
\be
S_\text{grav}=M \int dt\,\sqrt{G_{ab}\dot{X}^a\dot{X}^b}~,
\ee
which is the action of  a particle of mass $M$ on an  auxiliary spacetime with metric
\be\label{auxmetric}
ds^2_\r{aux} =G_{ab}dX^a dX^b= dT^2- {d\/2(d-1)}T^2 ds^2(\fh^d)~.
\ee
The auxiliary spacetime is $\r{GL}(d,\R)/\r{O}(d)$ which is the same as $\fh^d$ together with the  spatial volume as   the time variable $T$. The mass of the particle is found to be
\be\label{defMgen}
M={(d-1)(2\pi)^{d-1}\/4G}~,
\ee
which gives for the first dimensions
\be
M\eqd{2}{\pi\/2 G},\qq M \eqd{3} {2\pi^2\/ G} ,\qq M \eqd{4} {6\pi^3\/ G} ~.
\ee
The metric on $\fh^d$ is the invariant metric defined in \eqref{defsymmetrichd} and for $\fh^3$ it takes the form
\be\label{h3metric}
ds^2(\fh^3) = {dx_1^2\/y_1^2} + {dx_2^2\/y_2^2}+ {(dx_3-x_2 dx_1)^2\/y_1^2 y_2^2}+ {4\/3}\Big({dy_1^2\/y_1^2}+{dy_1 dy_2\/y_1 y_2}+{dy_2^2\/y_2^2}\Big)~.
\ee
The Wheeler-DeWitt equation then reduces to the   Klein-Gordon equation on \eqref{auxmetric}:
\be\label{KGauxWDW}
(G^{ab}\n_a\n_b +M^2)\psi(T,z)=0
\ee
This is also the BRST constraint of the particle theory. In $d=3$, this is explicitly
\be
 (T^2 \p_T^2 + 5 T \p_T + M^2 T^2-\tfrac43 \D_1)\psi =0
\ee
in terms of the Laplacian in the metric \eqref{h3metric}
\be\label{Delta1def}
\D_1 = y_1^2 \p_{y_1}^2 + y_2^2\p_{y_2}^2 - y_1 y_2 \p_{y_1}\p_{y_2}+y_1^2(x_2^2 +y_2^2)\p_{x_3}^2 + y_1^2 \p_{x_1}^2+ y_2^2 \p_{x_2}^2+ 2 y_1^2 x_2 \p_{x_1}\p_{x_3}~,
\ee
which is the first Casimir element of the $\r{SL}(3,\R)$ algebra, see \eqref{GL3Delta2} for the second. 
At higher orders, the particle gets corrected by the propagating gravitons and the matter. Since the particle remains the leading effect, it should be possible to formulate corrections as a  non-trivial background for the particle.

\ss{Harmonic analysis}

Let $\Psi[g]$ be a late-time solution of the Wheeler-DeWitt equation. We can think of the metric as 
\be
g=\Om^2 g_z+ h
\ee
which is the constant metric $g_z$ defined in \eqref{defunitmetric}, up to a Weyl factor, perturbed by a gravitational correction $h$. We can then write the  wavefunctional as
\be
\Psi[g] = \Psi(z) \cF_z[h]
\ee
where $\Psi(z)$ contains the leading mini-superspace term in \eqref{PsiWKBmini} and $\cF_z[h]$ accounts for the gravitational corrections.  At late time, the canonical inner product then gives \cite{Chakraborty:2023los}
\be\label{caninnerGF}
\int {Dg\/\r{vol}(\r{diff}\times \r{Weyl}) }|\Psi[g]|^2 = \int d^\ast z\, Z_{bc}(z)|\Psi(z)|^2 \int Dh_\r{TT}\,|\cF_z[h_\r{TT}]|^2~.
\ee
where the gauge symmetry has been fixed in the RHS. The gauge-fixing has introduced the Faddeev-Popov determinant, interpreted as the partition function $Z_{bc}$ of the $b c$ ghosts  on $T^d$. The last factor  comes from the propagating gravitons $h_\r{TT}$, the transverse-traceless piece of $h$, and is non-trivial in $d\geq 3$. The mini-superspace approximation corresponds to discarding it.

The $d^\ast z $ is the invariant measure associated with the metric \eqref{defsymmetrichd}. We have for $d=2$ and $d=3$
\be
d^\ast z \eqd{2} {dxdy\/y^2},\qq d^\ast z  \eqd{3} {dy_1 dy_2 dx_1 dx_2dx_3\/(y_1y_2)^3}~.
\ee
As $\r{SL}(d,\Z)$ is a gauge symmetry here,  we should write everything in terms of $\r{SL}(d,\Z)$ invariant objects, a rather constraining requirement. In particular this fixes the measure $d^\ast z$. The $bc$ ghost partition function is computed in \ref{app:Zbc} for a suitable gauge-fixing  as
\be
\sqrt{Z_{bc}(z)} = \r{det}(z)^{1/2}|g(\tz)|^d
\ee
in terms of the $g$-function, a higher-rank generalization of the Dedekind eta function defined in \ref{app:gfunction}. This combination is $\r{SL}(d,\Z)$ invariant.

We see that we should define the wavefunction
\be
\psi(z)=\sqrt{Z_\r{bc}(z)}\,\Psi(z)
\ee
so that its inner product becomes the Petersson inner product on $\fh^d$:
\be\label{PetInner}
(\psi_1,\psi_2) = \int_{\r{SL}(d,\Z)\bs\fh^d} d^\ast z \,\psi_1(z)\psi_2^\ast(z)~.
\ee
Thus the wavefunction $\psi$ lives in the Hilbert space 
\be
\cH= L^2(\r{SL}(d,\Z)\bs \fh^d)~.
\ee
This is the space of square-integrable automorphic forms for $\r{GL}(d,\R)$, whose theory we review in \ref{app:aut}.
In particular the volume $V_d =\r{vol}(\r{SL}(d,\Z)\bs \fh^d)$ is finite and the formula for $V_d$ is given in \eqref{volhdSLd}.

The spectral decomposition for $d=2$ was obtained by Selberg \cite{Selberg,iwaniecSpectralMethodsAutomorphic2002}. This is the spectral decomposition of the modular surface $\r{SL}(2,\Z)\bs\fh^2$ in terms of Eisenstein series and Maass cusp forms \cite{terrasHarmonicAnalysisSymmetric1985,iwaniecSpectralMethodsAutomorphic2002}, as reviewed in \ref{app:gl2}.

For $d=3$, there are two Casimir elements in $\r{SL}(3,\R)$ corresponding to two invariant differential operators:  $\D_1$ which is the Laplacian \eqref{Delta1def} on $\fh^3$  and $\D_2$ is given in  \eqref{GL3Delta2}. The spectral decomposition gives a basis of $\cH$ in which $\D_1$ and $\D_2$ are diagonal. It was obtained by  Langlands   as \cite{langlandsFunctionalEquationsSatisfied1976,Goldfeld_2006}
\begin{eqnarray}\label{LanglandsDec3}
\psi(z) &=& \text{const}+{1\/(4\pi i)^2}\int_{({1\/3})}ds_1\int_{({1\/3})}ds_2 \,(\psi, E_{s_1,s_2}) E_{s_1,s_2}(z) \-
&& \hspace{2cm}+ {1\/2\pi i }\sum_{j\geq 0} \int_{({1\/2})} ds \,( \psi, E_s^{(j)}) E_s^{(j)}(z) +\sum_{k\geq 1} \,(\psi,v_k)v_k(z)~.
\end{eqnarray}
The continuous spectrum consists of the Eisenstein series:  the minimal parabolic Eisenstein series $E_{s_1,s_2}$ and the maximal parabolic Eisenstein series $E_s^{(j)}$ twisted by the $\r{SL}(2,\Z)$ Maass forms $u_j$. The discrete spectrum is the set $\{v_k\}$ of $\r{SL}(3,\Z)$ Maass forms. This is reviewed in Appendix \ref{app:aut}.

The  spectral decomposition is really a decomposition in automorphic representations. Since $\r{SL}(d,\R)$ acts as a global symmetry on the particle, the Langlands spectral decomposition furnishes the elementary states, as it gives a basis of the Hilbert space in which the symmetries  are diagonalized.

The Wheeler-DeWitt equation   relates the time evolution to the Laplace eigenvalues so we see that the spectral decomposition also diagonalizes the time evolution. Thus it is  the right basis for third quantization.

\ss{Third quantization}\label{sec:third}

Third quantization involves viewing the wavefunction $\psi\in \cH$ as a field operator $\phi$. The modes of $\phi$ are then creation/annihilation operators for universes (single-particle states). The third-quantized Hilbert space is a Fock space corresponding to states with an arbitrary number of universes, see for example \cite{Giddings:1988cx,Giddings:1988wv,Marolf:2020xie}.

Since mini-superspace quantum cosmology is the worldline theory of a particle, the third-quantized version is the second quantization of the particle. It is just a free scalar field on the auxiliary spacetime \eqref{auxmetric} with mass $M$ given in \eqref{defMgen}.

The third-quantized action  is
\be
\cS = {1\/2}\int d^D X  \sqrt{G} (G^{ab}\p_a \phi \p_b \phi - M^2\phi^2)
\ee
and its equation of motion is the Wheeler-DeWitt equation \eqref{KGauxWDW}. Despite being free, this theory is rather non-trivial due to the $\r{SL}(d,\Z)$ quotient on $\fh^d$. This means that instead of considering the standard eigenmodes on $\fh^d$ we have to use the Langlands spectral decomposition on $\r{SL}(d,\Z)\bs\fh^d$.

Thus the field operator should be written as
\begin{eqnarray}
\phi(T,z) &=& {1\/(4\pi i)^2}\int_{({1\/3})}ds_1\int_{({1\/3})}ds_2 \, \,a_{s_1,s_2}\chi^+_{s_1,s_2}(T) E_{s_1,s_2}(z) \-
&& \hspace{1cm}+ {1\/2\pi i }\sum_{j\geq 0} \int_{({1\/2})} ds  \,a_{s}^{(j)} \chi_s^{(j)}(T)E_s^{(j)}(z) +\sum_{k\geq 0} a_k \chi_k(T)v_k(z)+\hc ,
\end{eqnarray}
where we have dressed each term in the spectral decomposition with a time-dependent mode $\chi(T)$, such that $\phi(T,z)$ solves the Klein-Gordon equation. As is standard, we have included the constant mode as the constant $\r{SL}(3,\Z)$ Maass form $v_0$. The operators $a_{s_1,s_2},a_s^{(j)}$ and $a_k$ are annihilation operators and their Hermitian conjugates are the creation operators. These operators annihilate/create universes with the corresponding wavefunction.  In particular the third-quantized vacuum $|0\rn$ is defined to be annihilated by all the annihilation operators. The third-quantized Hilbert space is then the Fock space generated by acting with creation operators on $|0\rn$.

Thus we see that the Langlands spectral decomposition gives the right framework for third quantization, as it provides a basis of the single-particle Hilbert space which diagonalizes the global symmetries and the time evolution.

 One can study a third-quantized Hamiltonian. This leads to a time-dependent Hamiltonian with dissipative/diffusive properties \cite{Godet:2024ich}. As the theory has a global $\r{SL}(d,\R)$ symmetry, one can also compute third-quantized charges, obtained from the third-quantized stress tensor
\be\label{thirdStressTensor}
\cT_{ab}  =\p_a \phi\p_b\phi - {1\/2} G_{ab}( G^{cd}\p_c\phi\p_d\phi - M^2\phi^2)~.
\ee
Let us now discuss the notion of coherent states. For a given (single particle) wavefunction $\psi\in\cH$, whose Langlands spectral decomposition takes the form
\be
\psi = {1\/(4\pi i)^2} \int_{({1\/3})}ds_1\int_{({1\/3})}ds_2\,\rho(s_1,s_2)E_{s_1,s_2}+ {1\/2\pi i}\sum_{j\geq 0}\int_{({1\/2})} \rho^{(j)}(s) E_s^{(j)}+ \sum_{k\geq 0} \rho(k) v_k~,
\ee
we can define an associated coherent Fock state as
\be
\r{exp}\le(\int_{({1\/3})}ds_1\int_{({1\/3})}ds_2\,\rho(s_1,s_2)\,a_{s_1,s_2}^\dg +\sum_{j\geq 0}\int_{({1\/2})} \rho^{(j)}(s) (a_s^{(j)})^\dg + \sum_{k\geq 0} \rho(k) a_k^\dg\ri) |0\rn~.
\ee
This is an eigenstate of the annihilation operators and prepares a   coherent state of universes.  If we don't restrict to connected spacetimes, the Hartle-Hawking state should also involve disconnected contributions, as discussed for example in  \cite{Marolf:2020xie}. So the third-quantized Hartle-Hawking state $|\r{HH}\rn$ should be the state obtained from $\psi_\r{HH}$ using the above procedure.   Its third-quantized norm takes the form
\be
\ln \r{HH}|\r{HH}\rn = e^{(\psi_\r{HH},\psi_\r{HH})}
\ee
which follows from standard identities about coherent states in free field theory. It seems that $|\r{HH}\rn$ may be more natural to consider than $\psi_\r{HH}$ and we will see in section \ref{Sec:norm} that the norm $\ln \r{HH}|\r{HH}\rn$ in $d=2$ takes a particularly suggestive form.

The straightforward interpretation of the Hartle-Hawking wavefunction as a probability distribution appears inadequate to describe slow-roll inflation \cite{Maldacena:2024uhs}. We may hope that the third-quantized perspective is relevant here as it may change how we interpret probabilities in quantum cosmology. 

In the third-quantized theory, most observables will be computed by trace formulas. In the physics description, trace formulas arise from the trivial fact that the first and  second quantization of the particle should give identical answers. It is well known  \cite{Grosche} that the Selberg trace formula \cite{Selberg} can be viewed as an exact version of Gutzwiller's trace formula \cite{Gutzwiller:1971fy} for a quantum system consisting of a particle on $\r{SL}(2,\Z)\bs\fh^2$, and this perspective should also apply for the particle describing quantum cosmology. The Green function, for example, is simple to define in field theory as the  two-point function $\ln 0 |\phi(T,z)\phi(T',z')|0\rn$ but the worldline description involves summing over all particle paths, an integral that localizes to geodesics and provides the geometric side of the trace formula.


Trace formulas, pioneered by Selberg, are one of the major tools in the Langlands program and they were used to prove some cases of functoriality \cite{jacquet1970automorphic,langlands1980basechange,arthur2005introduction}. The basic third-quantized operators, such as the Hamiltonian or the global symmetries operators constructed from \eqref{thirdStressTensor}, always involve creation/annihilation operators for all the automorphic representations appearing in the spectral decomposition. Observables involving their correlations will then naturally involve trace formulas. The norm is one of the simplest observables and we will show in section \ref{Sec:norm} that the Kuznetsov  trace formula \cite{kuznecovPeterssonsConjectureCusp1981} gives a way to compute the finite loop-corrected Hartle-Hawking norm in $d=2$. 

\section{The Hartle-Hawking state}

We will now discuss a particular state in the Hilbert space that can be defined from the gravitational path integral \cite{Hartle:1983ai}. The Hartle-Hawking wavefunctional $\Psi[g]$ is defined by performing the path integral on all the  geometries with the spatial metric $g$ as a boundary condition. The requirement that the geometries are all regular, \ie have no additional boundaries, is the no-boundary condition. For the $S^d$ topology, the Hartle-Hawking state is the Bunch-Davies or Euclidean vacuum of dS$_{d+1}$. In this paper we consider it for the $T^d$ topology where it will involve a Poincaré sum over $\r{SL}(d,\Z)$.
\ss{No-boundary geometries}

In the mini-superspace approximation, quantum cosmology on $T^d$ is a particle and the Einstein equation becomes the geodesic equation on the auxiliary spacetime
\be
\r{GL}(d,\R)/\r{O}(d)
\ee
with the metric \eqref{auxmetric}. We will use this to obtain the no-boundary solution. The Einstein equation with $T^3$ spatial topology was discussed in \cite{Gowdy:1973mu,Moncrief:1980pq,Hervik:2000ed}. 

\pg{Geodesic solutions.}We consider only vertical geodesics satisfying $\dot{x}_a=0$. These are the relevant solutions for the Hartle-Hawking state. In any case, general solutions can be obtained from them by acting with the  $\r{SL}(3,\R)$ global symmetry. The general vertical solution can be parametrized as
\be
T(t) = c_0\sqrt{f(t)\/f'(t)},\qq y_1(t) = c_1 f(t)^{\la_1},\qq y_2(t) = c_2 f(t)^{\la_2}
\ee
and $f(t)$ must satisfy the Schwarzian equation
\be
\{f(t),t\}=0
\ee
so that we can take $f(t)= {at+b\/ct+d}$  and the exponents must satisfy
\be
\la_1^2+\la_1\la_2+\la_2^2={1\/4}~.
\ee
We will make the choice $f(t)= {t\/t+1}$ so that we can fix the constant $c_1,c_2$ from the asymptotic condition
\be
\lim_{t\to+\infty} (T(t),y_1(t),y_2(t)) = (+\infty,y_1,y_2)~,
\ee
leading to $c_1=y_1$ and $c_2=y_2$. The last constant $c_0$ is fixed by the no-boundary condition.

\pg{No-boundary condition.} 
The singularity $T\to 0$ corresponds to $t\to 0$ and we can study the near-singularity behavior. We can redefine the coordinates
\be
u_3 \ra u_3 - x_4 u_1 -x_1 u_2,\qq u_2 \ra u_2 -x_2 u_1
\ee
so that the metric becomes diagonal
\be
ds^2 = -{dt^2\/9t^2} + g_{11} t^{\nu_{1}} du_1^2 + g_{22} t^{\nu_{2}} du_2^2 + g_{33} t^{\nu_{3}} du_3^2~ ,\qq t\to 0~,
\ee
where the Kasner exponents are
\be
\nu_1 =\tfrac13(1+2\la_1+4\la_2),\qq 
\nu_2 =\tfrac13(1+2\la_1-2\la_2),\qq 
\nu_3 =\tfrac13(1-4\la_1+2\la_2)~.
\ee
To see that the geometry closes we write $t= {9\/4}\rho^2$ and we expand around $\rho=0$. We get
\be
ds^2=  -  d\rho^2 + \sum_{i=1}^3 g_{ii} (\tfrac94)^{\nu_i} ( \rho^2)^{\nu_i} du_i^2~ ,\qq \rho\to 0~.
\ee
We get a regular solution only if $\nu_i=1$ for one of the $i$. We have to choose a circle to close so we pick the last one $u_3$ as this will make the rest $\tau_2$ invariant, and be compatible with the maximal parabolic subgroup $P_{2,1}$ of $\r{SL}(3,\Z)$, as will be discussed.  This corresponds to the choice $\la_1= -\tfrac12,\la_2=0$. Other choices are related by an $\r{SL}(3,\Z)$ transformation.

The no-boundary condition requires the metric to be of the form $ds^2 = -d\rho^2 - \rho^2 du_3^2+\dots$ so that it is regular at the origin. This fixes
\bea
c_0 \= {8\/27}i y_1^2 y_2~.
\eea
We can then compute the on-shell action and the leading order contribution to the path integral is
\be
\Psi_0 =\r{exp}\Big({-} {8\pi^2\/27G}\,\r{det}(z)\Big)
\ee
where we rewrote $\r{det}(z) = y_1^2 y_2$.

\pg{General dimensions.} To generalize this to $d$ dimensions, the simplest is to consider an ansatz for the metric  of the form
\be
ds^2 = - d\rho^2 + f_1(\rho)( y_1^2 \dots y_{d-1}^2 du_1^2+\dots + y_1^2 du_{d-1}^2)+ f_2(\rho) du_d^2
\ee
corresponding to a geometry where $u_d$ is the contracting circle. The Einstein equation gives simple ODEs for $f_1$ and $f_2$. With the asymptotic conditions and the no-boundary condition $ds^2 =- d\rho^2-\rho^2 du_d^2 +\dots $ near $\rho=0$, we obtain a unique solution which is (minus) the $H_{d+1}$ metric (Euclidean AdS$_{d+1}$) and we find the classical contribution to be
\be
\Psi_0=e^{- C \,\r{det}(z)},\qq \r{det}(z)=y_1^{d-1} y_2^{d-2}\dots y_{d-1}
\ee
where the constant $C$ is found to be
\be\label{Cdefinition}
C=  -{(d-1)(4 i\pi)^{d-1}\/4 d^d G}~,
\ee
which gives for the first dimensions
\be
C \eqd{2} -{i\pi\/4G},\qq C \eqd{3} {8\pi^2\/27G},\qq C \eqd{4} {3i\pi^3\/16G}~.
\ee Note that the phase of $C$ has a periodicity in $d \text{ mod } 4$ which comes from the fact that the metric is minus the $H_{d+1}$ metric. In $d= 3$, since $C>0$, the Poincaré sum is actually convergent, unlike the $d=2$ case where it must be defined by analytic continuation in $C$.

Alternatively, this is identical to computing the contribution to the gravity path integral for $\L<0$. The action of the AdS$_{d+1}$ spacetime  with $T^d$ asymptotics has a fixed sign and is multiplied by $C\propto {\l_\r{AdS}^{d-1}/G}$ by dimensional analysis. The positive cosmological constant answer is then obtained using $\l_\r{AdS} = i\l_\r{dS}$ which produces the factor $i^{d-1}$.

Note that the metrics we consider don't satisfy the criterion for allowable metrics conjectured in \cite{Kontsevich:2021dmb, Witten:2021nzp}. This criterion is based on having a positive-definite action for $p$-forms. The gravitational action is already not positive-definite,  so it appears that the Wick rotation should also act on the space of fields so as to rotate the conformal mode, and this would extend the range of allowable metrics. In fact, we will see that the dS computation for the $T^d$ path integral leads to an interesting wavefunction which is square-integrable, while the physical interpretation of the AdS result (the Maloney-Witten partition function) is still unclear.  


\pg{Sum over geometries.}

Recall that the wavefunction we should consider is
\be
\psi_\r{HH}=\sqrt{Z_\r{bc}}\,\Psi_\r{HH}
\ee
to have the normalized inner product \eqref{PetInner}. Here $\sqrt{Z_\r{bc}} = \r{det}(z)^{1/2}|g(\tz)|^d$ is computed  in \ref{app:Zbc}. The $g$-function defined  below contains higher-order terms so we ignore it for now. We should still include the $\r{det}(z)^{1/2}$ factor.

As a result the contribution from this saddle-point is
\be\label{defpsiHHgcdsum}
\psi_0(z) = \r{det}(z)^{1/2}e^{-C \,\r{det}(z)}~.
\ee
This corresponds to filling in the $T^d$ along the circle in the direction of $u^d$. To obtain the full Hartle-Hawking state we should sum over all the possible smooth fillings of the $T^d$. This can be written as 
\be\label{HHsumgcd}
\psi_\r{HH}(z)=\tfrac12  \sum_{\substack{\vn\in \Z^d\\ \r{gcd}(n_1,\dots,n_d)=1}} \psi_0(\g_{\vec{n}} z)
\ee
where $\g_{\vec{n}}$ is the $\r{SL}(d,\Z)$ element whose bottom row is $(n_1,\dots,n_d)$. This corresponds to the sum over the geometries where we fill the circle which is the linear combination $(n_1,\dots,n_d)$ of the elementary circles. The primitive condition comes from the fact that using a non-primitive circle would lead to a conical singularity.  The factor $\tfrac12$ comes because we should identify $\vec{n}$ and $-\vec{n}$.

The elements $\g_{\vec{n}}\in \r{SL}(d,\Z)$ are precisely the representatives of $P\bs \r{SL}(d,\Z)$ where $P$ is the maximal parabolic subgroup, as reviewed in \ref{app:Eis}.  Thus we obtain the expression of the Hartle-Hawking state as a Poincaré sum
\be\label{defHHpoinc}
\psi_\r{HH}(z)= \tfrac12 \sum_{\g\in P\backslash \r{SL}(d,\Z)} \r{det}(\g z)^{1/2} \,e^{-C\, \r{det}(\g z)}~.
\ee
We take this to be our definition for the Hartle-Hawking state at leading order. This fixes the normalization. We are really interested in the dependence on $G$ of this quantity where $C\propto G^{-1}$ is given in \eqref{Cdefinition}.

 Note that similar sums where the one-loop factor is $\r{det}(\g z)^{\a}$ for different exponents $\a$ can also be considered, and our results do translate with small modifications.


\ss{Spectral decomposition}

The Hartle-Hawking state on $T^d$ defines an automorphic form for $\r{GL}(d)$, a function on $\fh^d$ invariant under $\r{SL}(d,\Z)$. Interpreted as a cosmological wavefunction, it should be in $L^2$ so we should be able to obtain its representation in Langlands' spectral decomposition \eqref{LanglandsDec3}. 

\pg{Direct derivation.} We first explain a simple way to obtain the Hartle-Hawking spectral decomposition. This is based on the fact that the Hartle-Hawking state is manifestly closely related to the maximal parabolic Eisenstein series
\be
E_s(z) = \tfrac12 \sum_{\g\in P\bs \r{SL}(d,\Z)} \r{det}(\g z)^s~.
\ee
The relation is just the Cahen-Mellin integral 
\be
{1\ov 2\pi i} \int_{(\a)} ds\,\G(-s) y^{s}=e^{-y} ~.
\ee 
This integral is valid if the contour $s\in \a+i\R$ is with $\a<0$. It can be derived for example by summing over the poles $s\in \bN$ of the gamma function. Performing the Poincaré sum on both sides  leads to a direct relationship between the Hartle-Hawking state and the maximal parabolic Eisenstein series. However we must first move the contour to $\a=1+\e$ as the  Poincaré sum for the Eisenstein series is only convergent for $\r{Re}(s)>1$. This picks up the first pole of the gamma function  so we should write
\be
\r{det}(z)^{1/2} e^{-C\,\r{det}(z)} = \r{det}(z)^{1/2}+ {1\/2\pi i}\int_{(1+\e)} ds\,\G(\tfrac12-s) C^{s-{1\/2}}\,\r{det}(z)^s~.
\ee
Then we can perform the Poincaré sum on both sides and we get
\be
\psi_\r{HH}(z) = {1\/2\pi i }\int_{(1+\e)}ds\,\G(\tfrac{1}{2}-s)C^{s-{1\/2}}\,E_s(z)~.
\ee
 We have removed the contribution of the first term $\r{det}(z)^{1/2}$ which leads to a divergent sum formally equal to $E_{1/2}(z)$. This  term is independent of $G$ and we don't expect to be able to fix such terms at leading order.

To obtain the spectral representation, we should further move the contour to the line $\r{Re}(s)={1\/2}$. So this picks up the pole of $E_s$ at $s=1$. The  completed Eisenstein series is defined as 
\be
\cE_s(z) = \L(\tfrac{d}{2}s)E_s(z)
\ee
where $\L$ is the completed Riemann zeta function
\be\label{defLambda}
\L(s) = \pi^{-s}\G(s)\z(2s)
\ee
which satisfies Riemann's functional equation $\L(s)  = \L(\tfrac12-s)$. We have that $\cE_s$ is meromorphic with only poles at $s=0,1$ and with residues \eqref{compEispoles}, so we get
\be
\r{Res}_{s=1}E_s = {1\/d \L(\tfrac{d}{2})} ~.
\ee
This produces the constant contribution and our final expression is
\be\label{HHspec1}
\psi_\r{HH}(z)=-{2\sqrt{\pi C}\/d\L(\tfrac{d}{2})}+ {1\/2\pi i}\int_{({1\/2})} ds\,\G(\tfrac{1}{2}-s) C^{s-{1\/2}} E_s(z)~.
\ee
This is the Langlands spectral representation of the Hartle-Hawking state. At leading order, we get only contributions from the constant and the maximal parabolic Eisenstein series, although this will be modified by higher order corrections. In this expression $E_s(z)$ is always the maximal parabolic Eisenstein series for $\r{GL}(d,\R)$ and  the coupling $C \propto 1/G$ is given in \eqref{Cdefinition}. For $d=2$ we have
\be
\psi_\r{HH}(\tau)\eqd{2} -{3\, e^{-i\pi/4}\/\sqrt{G}}+ {1\/2\pi }\int_\R d\mu\,\G(-i\mu) \Big({\pi\/4iG}\Big)^{i\mu} E_{{1\/2}+i\mu}(\tau)~,
\ee 
and for $d=3$ this takes the form
\be\label{HHspectraldec3}
\psi_\r{HH}(z) \eqd{3} -{2 (\tfrac23\pi)^{5/2}\/ \z(3) \sqrt{G}} +{1\/2\pi }\int_\R d\mu\,\G(-i\mu) \Big({8\pi^2\/27 G}\Big)^{i\mu} E_{{1\/2}+i\mu}(z)~.
\ee
In the following we will explain how to rederive this spectral decomposition  using the important technique of unfolding.

\pg{Unfolding.} The unfolding trick is a standard way to deal with Poincaré sums in the context of harmonic analysis \cite{Selberg}. In $d=2$, it is the fact that since $\psi_\r{HH}=\sum_{ P\bs\r{SL}(2,\Z)} \psi_0$, we have 
\be
(\psi_\r{HH}, \chi) = \int_{\r{SL}(2,\Z)\bs\fh^2} d^\ast\tau\, \psi_\r{HH} \chi^\ast = \int_{P\bs\fh^2} d^\ast\tau \,\psi_0 \chi^\ast = \int_{0}^1dx \int_0^{+\infty} {dy\/y^2} \psi_0\chi^\ast
\ee
which becomes a simple integral on the strip after unfolding the action of $\r{SL}(2,\Z)$. This was used  in \cite{Godet:2024ich} to derive the spectral representation for $d=2$, including the higher order terms.

In $d=3$, we are considering the maximal parabolic subgroup $P=P_{2,1}$, defined in \ref{app:Eis}. The unfolding technique gives
\be
(\psi_\r{HH},\chi) = \int_{\r{SL}(3,\Z)\backslash\fh^3} d^\ast z\,\psi_\r{HH} \chi^\ast = \int_{P\backslash\fh^3} d^\ast z\,\psi_0 \chi^\ast~.
\ee
 Unlike for the minimal parabolic subgroup, this doesn't unfold to a strip, and we should use the coordinates \cite{Goldfeld_2006}
\be
P_{2,1}\backslash\fh^3 = \r{SL}(2,\Z)\backslash\fh^2 \times (\R/\Z)^2 \times [0<\l<\infty]
\ee
in which the measure becomes
\be
d^\ast z = {3\/2}\, \l^{-3} d^\ast \tau_2 dx_3 dx_1 {d\l\/\l},
\ee
and we do not unfold the $\r{SL}(2,\Z)$ acting on $\tau_2=x_2+iy_2$. This then leads to the expression
\be
(\psi_\r{HH},\chi)= {3\/2} \int_0^{+\infty}{d\l\/\l^4}  \int_0^1 dx_1\int_0^1dx_3  \int_{\r{SL}(2,\Z)\backslash \fh^2} d^\ast \tau_2\,\psi_0 \, \chi
\ee
where the variable $y_1$ has been replaced by $\l$ where we have
\be
 \l^3 = y_1^2 y_2 =\r{det}(z)
\ee
Note that $\psi_0$ is independent of $x_1,x_3$. Thus we obtain
\be
(\psi_\r{HH},\chi)= {3\/2} \int_0^{+\infty}{d\l\/\l^4} \,\ln \psi_0, \chi_{P}\rn_{\tau_2},\qq \chi_{P} = \int_0^1 dx_1\int_0^1dx_3 \,\chi
\ee
where $\ln\cdot,\cdot\rn_{\tau_2}$ is the Petersson inner product in $\tau_2 \in \r{SL}(2,\Z)\bs\fh^2$ and $\chi_P$ is the projection of $\chi$ along $P$. In the variable $\l$ we have
\be
\psi_0 = \l^{3/2}\, e^{-C\l^3},\qq
\ee
which is independent of $\tau_2$, so our final formula is
\be
(\psi_\r{HH},\chi)= {3\/2} \int_0^{+\infty}{d\l\/\l} \l^{-3/2}  \,e^{-C\l^3}\ln 1, \chi_{P}\rn_{\tau_2}~.
\ee
This expresses the inner product of $\psi_\r{HH}$ with any $\r{GL}(3)$ automorphic form $\chi$ as the Mellin transform of the $\r{GL}(2)$ inner product $\ln 1,\chi_P\rn_{\tau_2}$.

 Let us use this to compute the constant contribution. By taking $\chi=1$, we get
\be
(\psi_\r{HH},1)= {\pi\/2} \int_0^{+\infty}{d\l\/\l}\l^{-3/2}   \,e^{-C\l^3}= -{\pi^{3/2}\sqrt{C}\/3}
\ee
using that $\ln 1,1\rn_{\tau_2}= V_2={\pi\/3}$. Here we have regularized this divergent integral by the analytic continuation of the gamma function. Since $V_3= {\z(3)/4}$, the normalized constant mode in $d=3$ is
\be
v_0= {2\/\sqrt{\z(3)}},\qq (v_0,v_0)=1
\ee
so that the constant term is
\be
\psi_\r{HH}^\r{const}(z)=(\psi_\r{HH},v_0) v_0= -{4\pi^{3/2} \sqrt{C}\/3\z(3)} = -{2 (\tfrac23\pi)^{5/2}\/ \z(3) \sqrt{G}} 
\ee
which matches the above answer \eqref{HHspectraldec3}.

To obtain the Eisenstein spectrum, we compute
\be
(\psi_\r{HH},E_s)= {3\/2} \int_{\R_+}{d\l\/\l} \l^{-3/2} \,e^{-C \l^3}\ln 1, E_s^P\rn_{\tau_2}~.
\ee
From the Fourier decomposition of the maximal parabolic Eisenstein series, we have \cite{friedbergGlobalApproachRankinSelberg1987}
\be\label{EismaxP}
E_s^P = \int_0^1 dx_1\int_0^1 dx_3 \,E_s = \l^{3s} +  \l^{3(1-s)\/2}{\L(\tfrac12(3s-1))\/\L(\tfrac32 s)} E^{(2)}_{3s-1\/2}(\tau_2)
\ee
where what appears is the $\r{GL}(2)$ Eisenstein series $E_s^{(2)}(\tau_2)$ which is orthogonal to the constant $\ln 1, E_s^{(2)}\rn_{\tau_2} = 0$. Thus we only get a contribution from the first term
\be
(\psi_\r{HH},E_s)= {\pi \/2} \int_0^{+\infty }{d\l\/\l}  \,\l^{3s^\ast-{3\/2}}  e^{-C\l^3}  = {\pi\/6}\,C^{{1\/2}-s^\ast} \G(s^\ast-\tfrac12)~,
\ee
and evaluated at $s={1\/2}+i\mu$  we get
\be
(\psi_\r{HH},E_{{1\/2}+i\mu}) = {\pi\/6} \,C^{i\mu}\G(-i\mu)~.
\ee
As reviewed in \ref{app:Eis}, the maximal parabolic Eisenstein series appears in the spectrum as the Eisenstein series twisted by the constant $\r{GL}(2)$ Maass form $u_0 = \sqrt{3\/\pi}$. Thus we have
\be
E_s^{(u_0)} = 2\sqrt{3\/\pi} \,E_s~.
\ee
As a result the spectral density is
\be
(\psi_\r{HH},E^{(u_0)}_{{1\/2}+i\mu}) =\sqrt{\pi\/3}  \,C^{i\mu}\G(-i\mu)~,
\ee
and the maximal parabolic Eisenstein piece of the Hartle-Hawking state is
\be
\psi_\r{HH}^\r{Eis}(z) = {1\/2\pi i}\int_{({1\/2})} ds\,(\psi_\r{HH},E^{(u_0)}_s) E^{(u_0)}_s(z)= {1\/2\pi }\int_\R d\mu\, \,C^{i\mu}\G(-i\mu) E_{{1\/2}+i\mu}(z)
\ee
which matches the answer from the direct derivation. The unfolding technique gives a way to also compute contributions from higher order corrections as  discussed in section~\ref{sec:higher}.

\ss{Zeta zeros expansion}

The integral \eqref{HHspec1} defining the Hartle-Hawking state has a complicated pole structure due to the Eisenstein series. We should rewrite it in terms of the completed Eisenstein series 
\be
\cE_s(z)=\L(\tfrac{d}{2}s)E_s(z)
\ee
where we recall that the completed Riemann zeta function is
\be
\L(s) = \pi^{-s}\G(s)\z(2s)
\ee
which satisfies the Riemann functional equation $\L(s) = \L(\tfrac12-s)$. This leads to  
\be\label{HHoverzeta}
\psi_\r{HH}(z)=-{2\sqrt{\pi C}\/d\L(\tfrac{d}{2})}+{1\/2\pi i}\int_{({1\/2})} ds\,{\G(\tfrac12-s) C^{s-{1\/2}} \/\pi^{-ds/2}\G(\tfrac{d}{2}{s})\z(ds)} \cE_s(z)
\ee
which is the representation  in which all the poles manifest. The gamma function gives poles at $s=\tfrac12+n,n\in \bN$. The completed Eisenstein series $\cE_s$ has only simple poles at $s=0,1$.  The pole at $s=0$ and the trivial zeta zeros $ds = -2n,n\geq 1$, are cancelled by the gamma function in the denominator. As a result the only poles of the integrand are at the locations
\be
s={\rho\/d},\qq s=1,\qq s=\tfrac12+n,\qq n\in \bN~.
\ee
where $\rho$ runs over the non-trivial zeros of the Riemann zeta function, the critical line being at $\r{Re}(s) = {1\/2d}$. The contour is at $\r{Re}(s)={1\/2}$ so it separates the two types of poles. Thus we obtain two different expressions for $\psi_\r{HH}$ by summing over the poles on each side of the contour.

The fact that the Hartle-Hawking state has such a representation with a zeta function in the denominator is responsible for many of the effects discussed in this paper. Its origin   lies in the relatively prime condition in the sum \eqref{defpsiHHgcdsum}, a geometric condition coming from the no-boundary condition.

\sss{$d=2$}

    \begin{figure}
\centering\includegraphics[width=14cm]{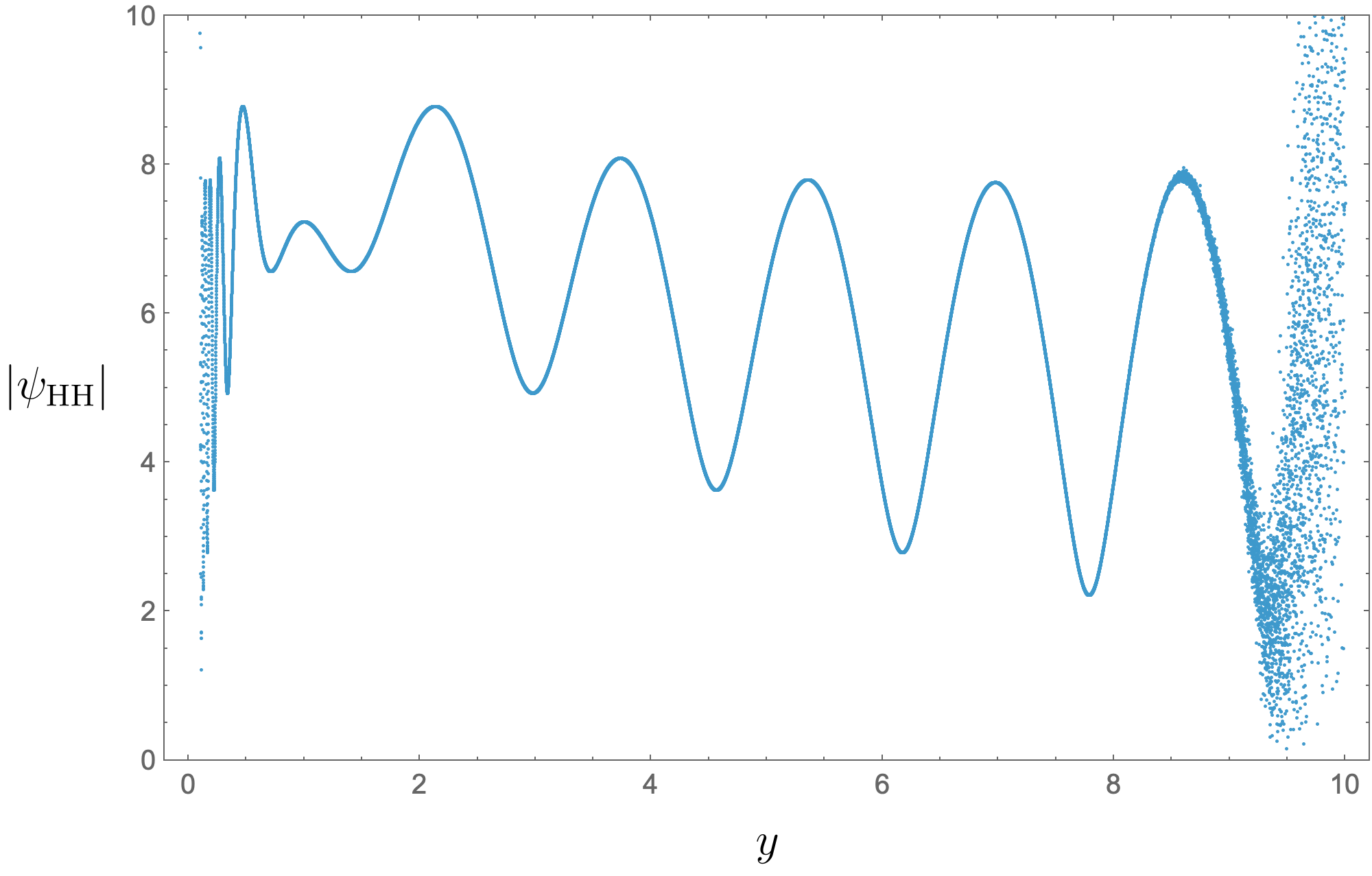}
\caption{We plot the magnitude $|\psi_\r{HH}(x=0,y)|$ as a function of $y$ on the line $x=0$ after summing the first 200 terms in  \eqref{HHintegerexp2} with $G=0.2$. The convergence is slow and worse for smaller $G$. }\label{Fig:plotHHintline}
    \end{figure}

\begin{figure}
  \centering
  \includegraphics[width=12cm]{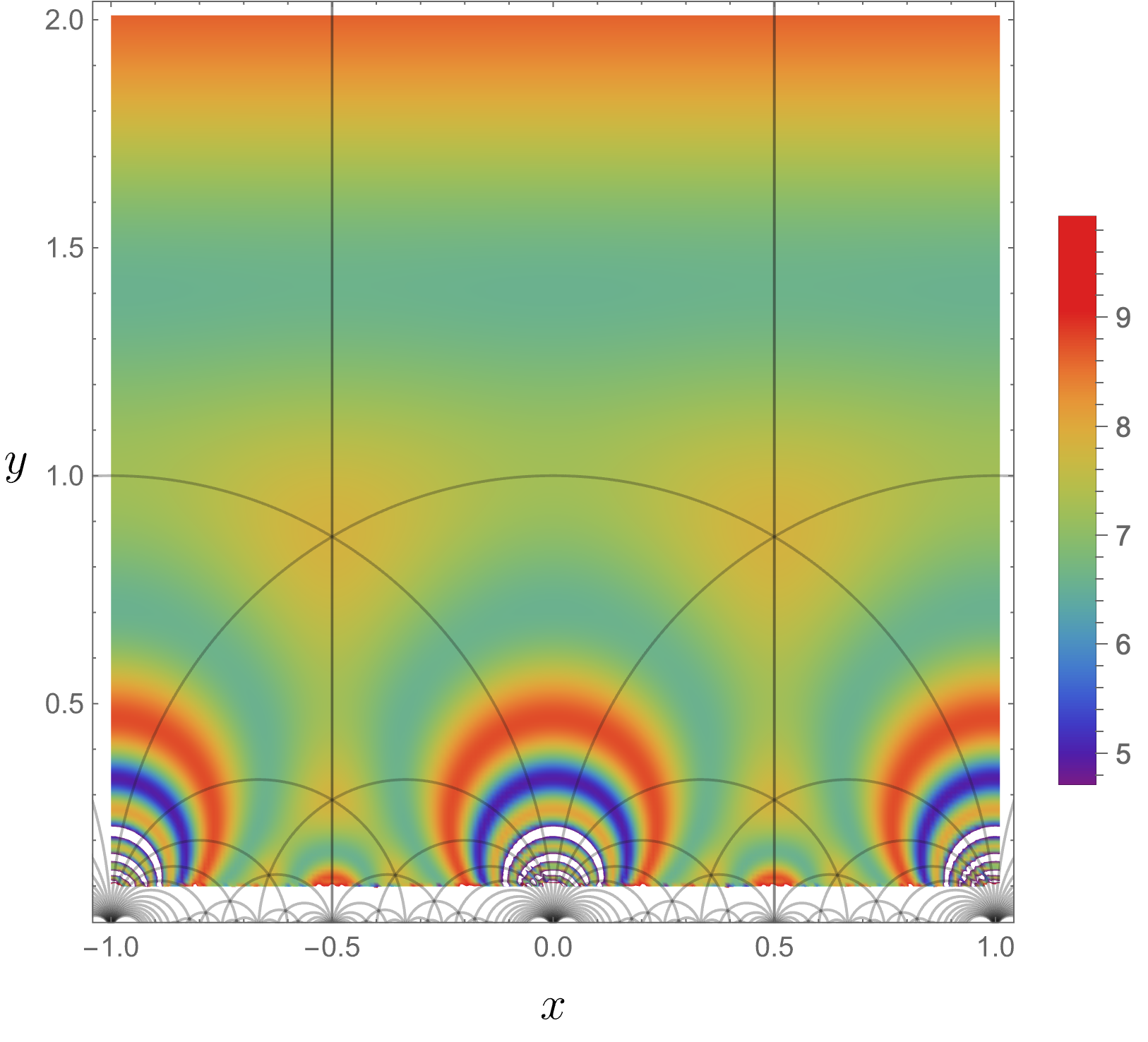}
   
  \caption{The Hartle-Hawking state $|\psi_\r{HH}(\tau)|$ for $\tau=x+iy\in\fh^2$ after including the first 200 terms in \eqref{HHintegerexp2} with $G=0.2$ and 100 Fourier modes.  The convergence is slow and the bottom part of the plot has not converged.}\label{Fig:HHplot}
\end{figure}

    \begin{figure}

    \begin{center}
  \begin{tabular}{cc}
    \subf{\includegraphics[height=6.5cm]{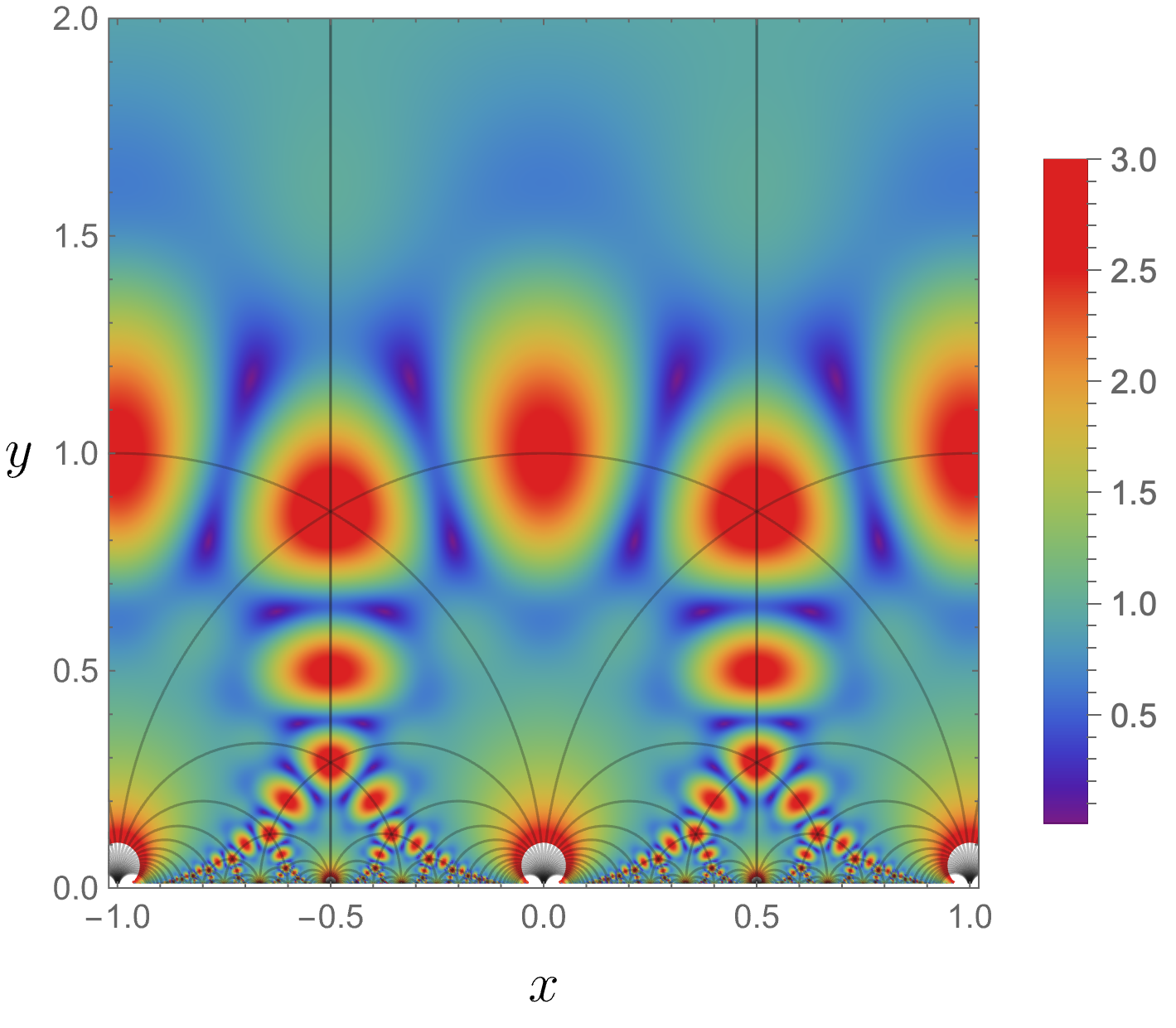}}{$\rho_1 \approx \tfrac12+14.13 i$} &
    \subf{\includegraphics[height=6.5cm]{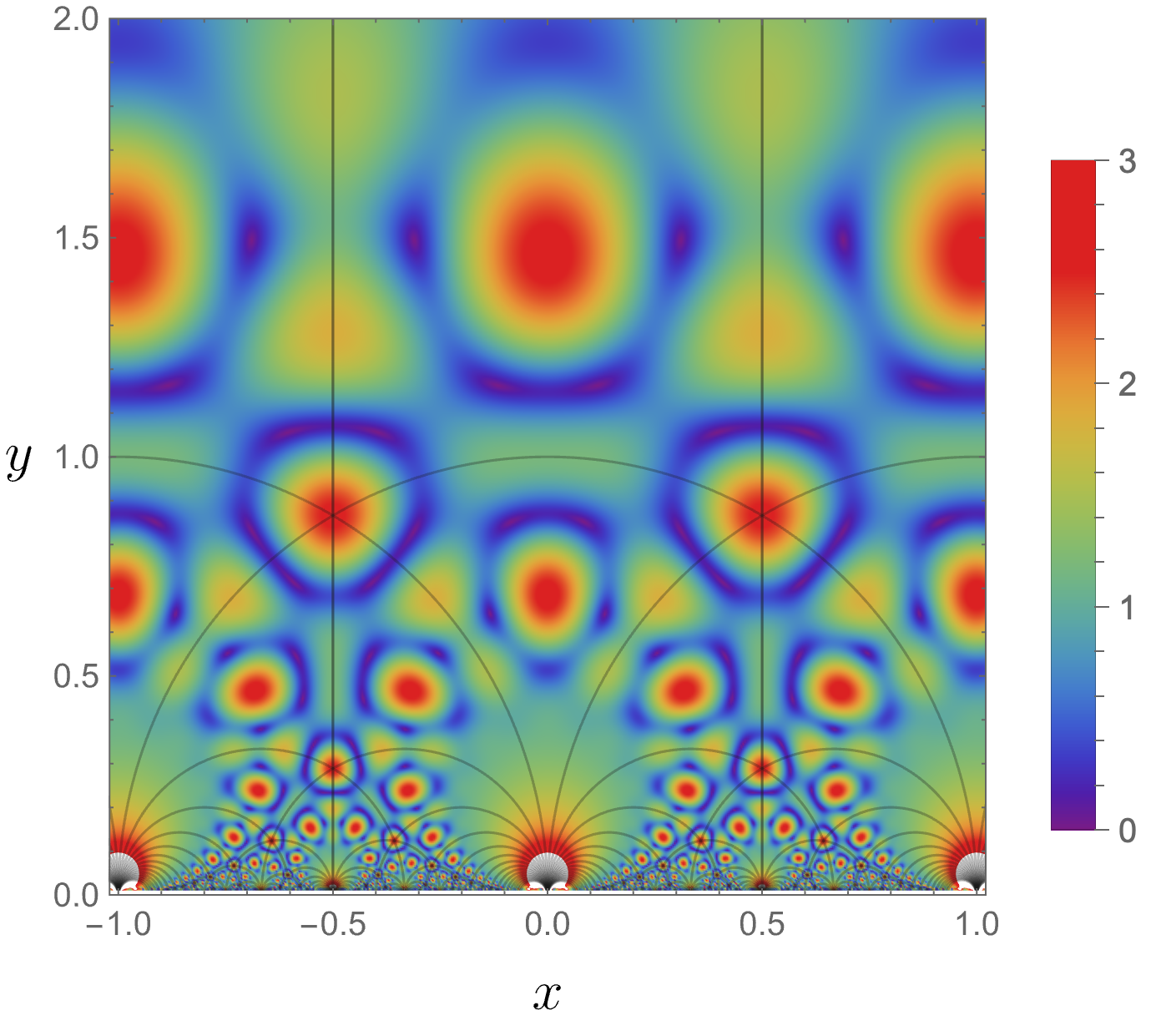}}{$\rho_2 \approx \tfrac12+21.02 i$}\vspace{0.3cm} \\
    \subf{\includegraphics[height=6.5cm]{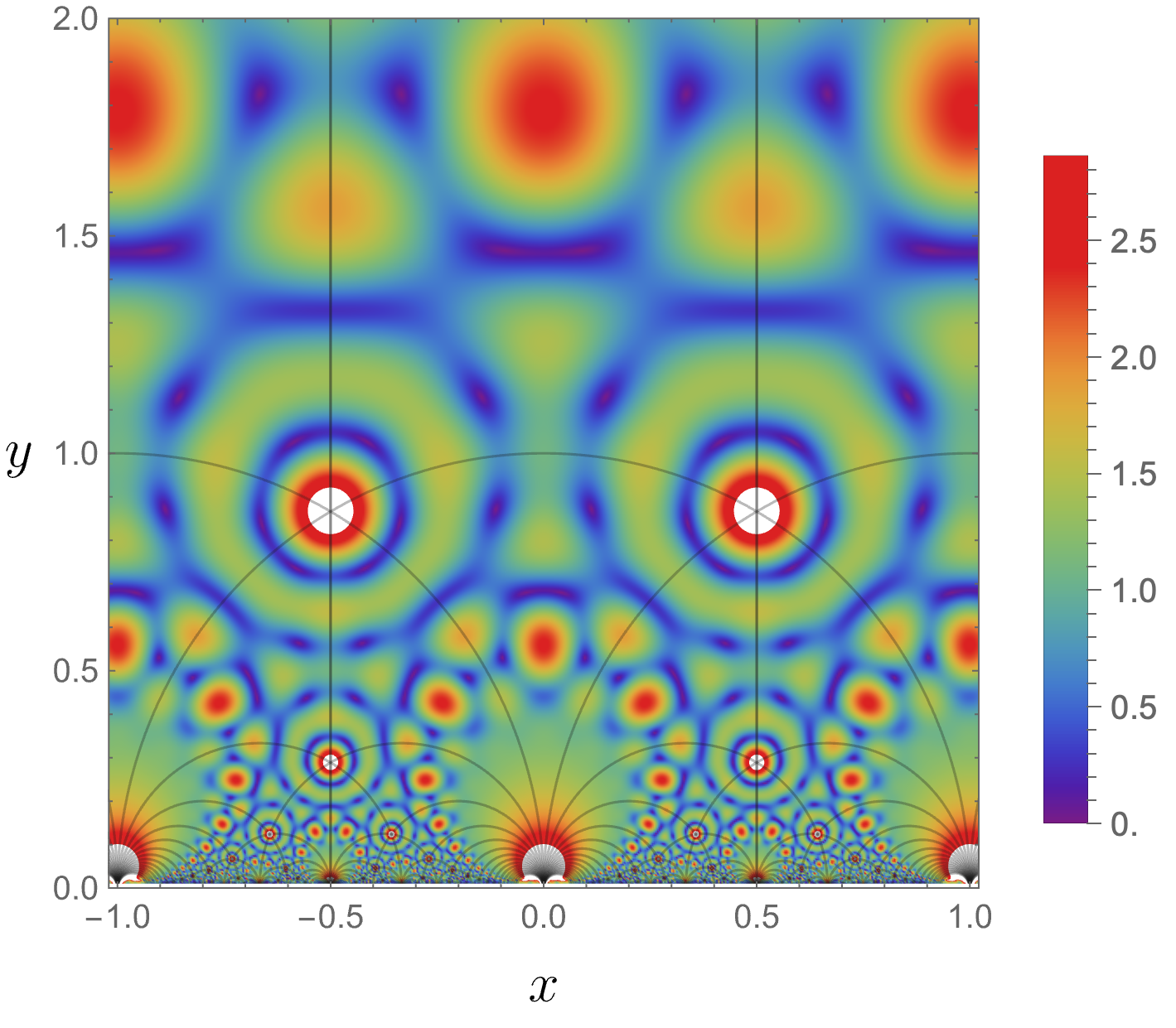}}{$\rho_3\approx \tfrac12+25.01 i $} &
    \subf{\includegraphics[height=6.5cm]{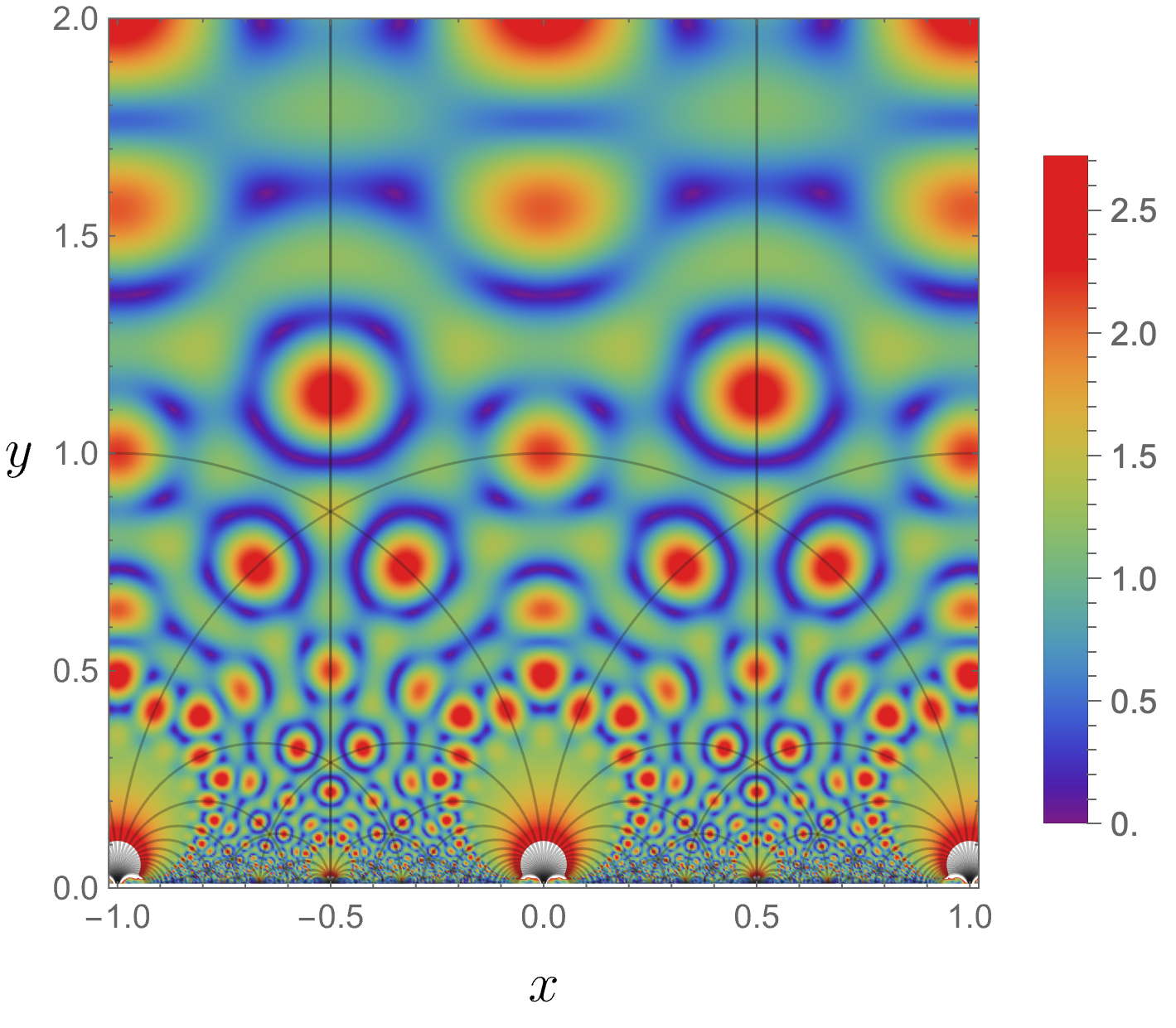}}{$\rho_4\approx \tfrac12+ 30.42i$} 
  \end{tabular}

\end{center}

\caption{We plot the absolute value in $\fh^2$ of the contributions in \eqref{HH2sumzera} of the first non-trivial zeta zeros. They give oscillating corrections of order $G^{1/4}$ on top of the constant term.}\label{Fig:firstfourzeros}
    \end{figure}

We first consider the case of $d=2$. In this case we can use the Riemann functional equation to obtain a simpler expression
\be
\psi_\r{HH}= -{3\, e^{-{i\pi/4}}\/\sqrt{G}}+ {1\/2\pi i}\int ds\,{(C/\pi)^{s-{1\/2} } \/\z(1-2s)}\cE_s(z)~.
\ee
In this representation, the poles are given by the zeros of the zeta function. We have the trivial zeros for   $s={1\/2}+n,n\geq 1$ and the non-trivial zeros $s=(1-\rho)/2$. There is also the pole of $\cE_s$ at $s=1$ which corresponds to the constant term. Note that if $\rho$ is a non-trivial zero, the functional equation implies that $1-\rho$ is also a zero, the complex conjugate of $\rho$ if $\r{Re}(\rho)=\tfrac12$. 

We first sum over the residues on the right half-plane. The pole at $s=1$ cancels the constant term and we obtain the expansion
\be\label{HHintegerexp2}
\psi_\r{HH}(z) = \sum_{n\geq 1}{(4iG)^{-n}\/2\z'(-2n)} \cE_{n+{1\/2}}(z) = \sum_{n\geq 1} {(-C)^n \/n! }E_{n+{1\/2}}(z)
\ee
This gives an expansion of the Hartle-Hawking state as an infinite sum of the Eisenstein series at half-integers, which in fact follows directly from the power series expansion of the exponential. The relation between the two expressions comes from 
\be
\z'(-2n) =\tfrac12 (-1)^n  \pi^{-n}n!\L(n+\tfrac12)~.
\ee
This generalizes  the Maloney-Witten expansion derived in \cite{Maloney:2007ud} to higher Fourier modes, although we haven't included the one-loop corrections here. It gives a convergent expansion, at least in a  region of moduli space, although the convergence is slow for small $G$ with terms that first increase before decreasing.
We plot the magnitude of $\psi_\r{HH}$ on the line $x=0$ in Figure~\ref{Fig:plotHHintline} and in $\fh^2$ in Figure~\ref{Fig:HHplot}.

We obtain a different representation by summing over the poles on the other half plane. This gives an expansion  as a sum over the non-trivial zeta zeros
\be\label{HH2sumzera}
\psi_\r{HH} = -{3\, e^{-{i\pi/4}}\/\sqrt{G}}-\sum_\rho {(4iG)^{\rho/2}\/2\z'(\rho)}\cE_{{1-\rho\/2}}
\ee
where we can also write it equivalently after replacing $\rho\ra 1-\rho$.
Note that the Eisenstein series appearing here is special as its constant term is
\be
\cE_{1-\rho\/2}^{(0)}(y) = \L(\tfrac12(1-\rho)) y^{1-\rho\/2} +\L(\tfrac12(1+\rho))y^{1+\rho\/2}=\L(\tfrac12(1+\rho))y^{1+\rho\/2} ~,
\ee
where the first term vanishes since $\z(\rho)=\z(1-\rho)=0$. Thus this expansion involves the Eisenstein series at Riemann zeta zeros \cite{ZagierZetaEis} and more precisely at poles of the scattering phase $\vphi(s)$, so they are purely ingoing or outgoing waves corresponding to resonances   \cite{EisensteinScattering}.

In $d=2$, this is a highly oscillatory sum. Evaluation of the sum shows that it is oscillatory due to the factor of $i$ in $C=  -{i\pi\/4G} $ which creates exponential enhancement in $\r{Im}(\rho)$ in one direction. We would get a convergent expansion for contours which pick up only half of the zeros, but our contour picks up all of them. So in $d=2$, they must be viewed as a divergent oscillating sum giving small corrections on top of the large constant contribution. We plot the contributions from the first four zeros in $\fh^2$ in Figure~\ref{Fig:firstfourzeros}.

    \begin{figure}
\includegraphics[width=13cm]{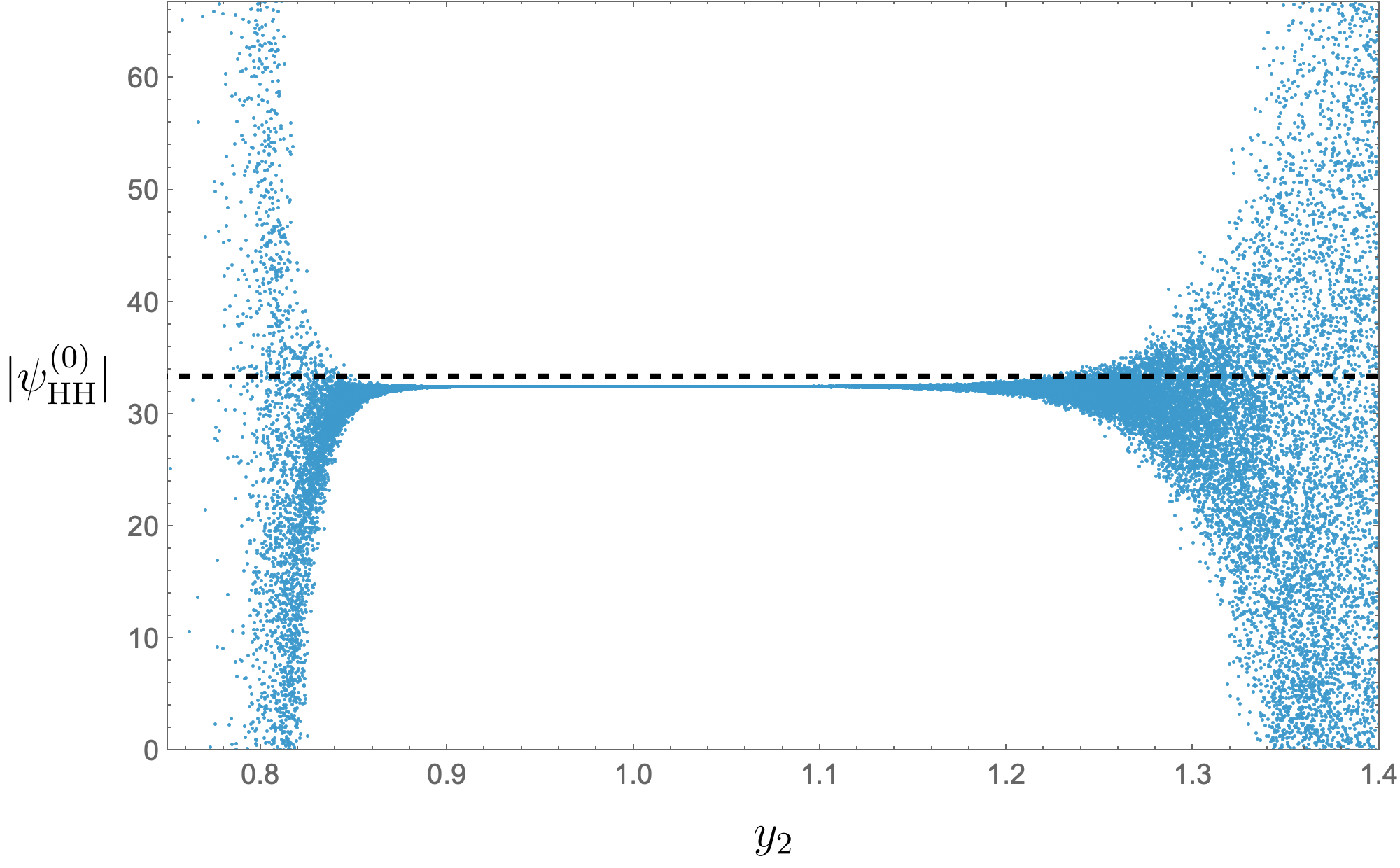}
  \put(0,124){{ $ {2(\frac23\pi)^{5/2}\/\z(3)\sqrt{G}}\approx 33.4$}}
\caption{Magnitude $|\psi_\r{HH}^{(0,0)}|$ of the constant Fourier mode of the Hartle-Hawking state in $d=3$. We sum the first 1000 terms in \eqref{HH3intConstexp} and plot the $y_2$ dependence for $y_1=1$ and $G=0.1$. The sum converges in the displayed region and its value matches with the constant prediction (dashed line)  from the other expansion \eqref{HH3constfluct}.}\label{Fig:HH3intplot}
    \end{figure}

\sss{$d=3$}

In $d=3$, the expansion on the right half-plane gives
\be
\psi_\r{HH}(z) = \sum_{n\geq 1} {(-1)^n C^n\/n!\L({3\/2}(n+{1\/2}))}\cE_{n+{1\/2}}(z)~.
\ee
The pole at $s=1$ has cancelled the constant term and we are left with a sum on the maximal parabolic Eisenstein series at half-integers.

The Hartle-Hawking state can then be defined explicitly from the known Fourier decomposition of the Eisenstein series \cite{friedbergGlobalApproachRankinSelberg1987,buttcaneHigherWeightGL32018a,goldfeldFirstCoefficientLanglands,goldfeldFunctionalEquationsLanglands2023}. For example we can obtain the constant Fourier mode from  that of the maximal parabolic Eisenstein series
\be
\cE_s^{(0,0)} =  \L(\tfrac32s) (y_1^2 y_2)^s +\L(\tfrac32s-\tfrac12)  y_1^{1-s}y_2^s +\L(\tfrac32s-1)(y_1y_2^2)^{1-s}~. 
\ee
There are three infinite sums and we can see that the first one reproduces the leading exponential. Thus we have
\begin{multline}\label{HH3intConstexp}
\psi_\r{HH}^{(0,0)}(y_1,y_2) =  (y_1^2 y_2)^{1/2} \,e^{-C y_1^2 y_2} \\+\sum_{n\geq 1} {(-C)^n\/n! \L(\tfrac32 n+\tfrac34) } \le(  \L(\tfrac32 n-\tfrac14) (y_1 y_2^2)^{{1\/2}-n} + \L(\tfrac32n+\tfrac14) y_1 (y_2/y_1)^{n+{1\/2}} \ri)~.
\end{multline}
We can then compute this sum numerically and we find some regions of convergence, as illustrated in Figure~\ref{Fig:HH3intplot}.

We get the other expansion by summing over the non-trivial zeta zeros at  $s={\rho\/3}$. The critical line is here $\r{Re}(s) = {1\/6}$ so this corresponds to summing over the other side of the contour $\r{Re}(s)={1\/2} $. This gives the representation
\be
\psi_\r{HH} = -{2 (\tfrac23\pi)^{5/2}\/ \z(3) \sqrt{G}} +\sum_\rho C^{ {\rho\/3}-{1\/2}}\psi_\rho(z),\qq 
\psi_\rho(z) =  {\G({1\/2}-{\rho\/3})\/3\,\pi^{-\rho/2}\G({\rho\/2})\z'(\rho)} \cE_{\rho\/3}(z)~.
\ee
  In this case  the constant Fourier mode simplifies as
\be
\cE_{\rho\/3}^{(0,0)} =  \L(\tfrac12\rho-\tfrac12)  y_1 (y_2/y_1)^{\rho/3} +\L(\tfrac12\rho-1)(y_1y_2^2)^{1-\rho/3} ~.
\ee
since one term is cancelled by the fact that $\z(\rho)=0$, which makes these Eisenstein series quite special \cite{EisensteinScattering}.

\begin{figure}
\centering\includegraphics[width=14cm]{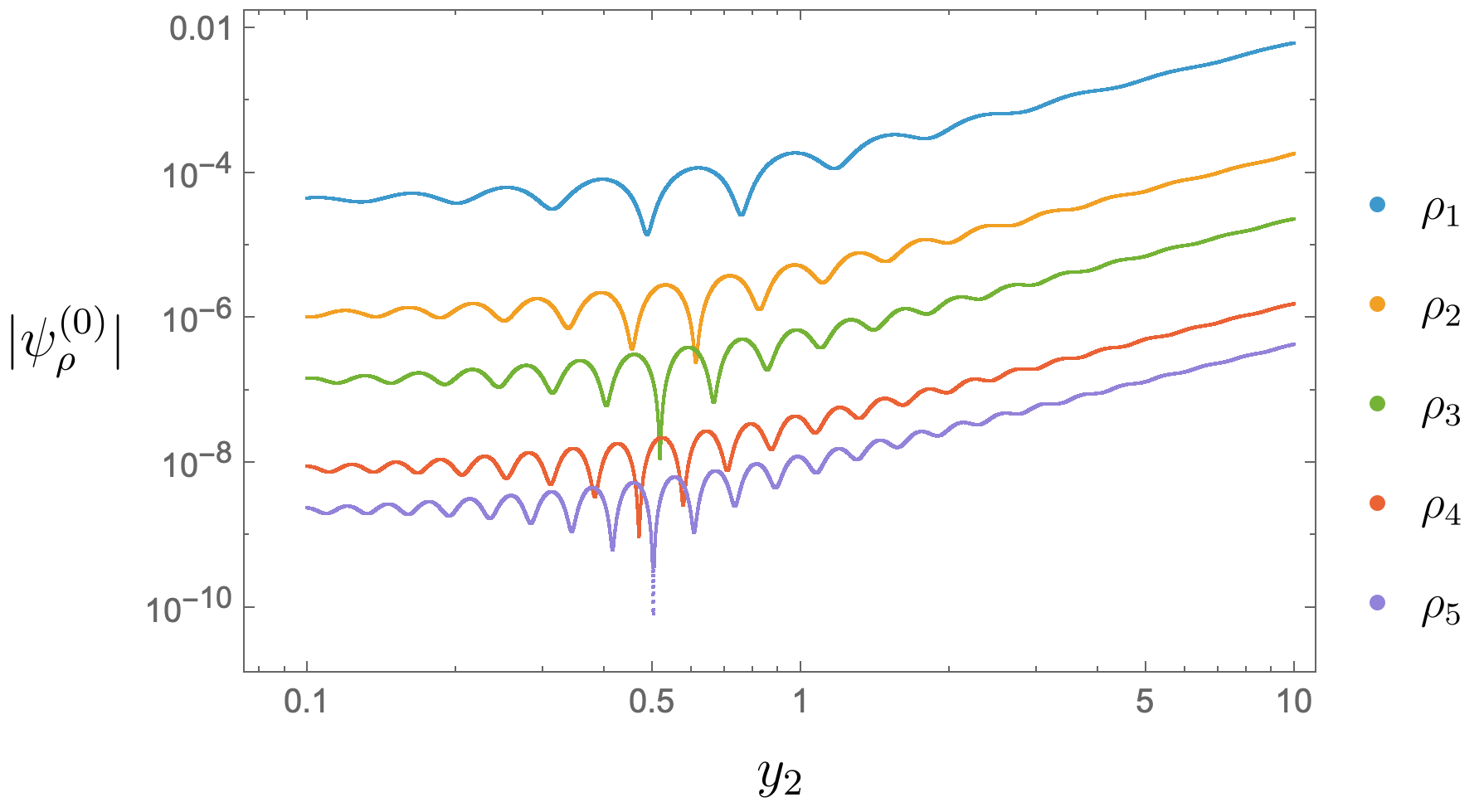}

\caption{Log plot of $|\psi_\rho^{(0)}|$ in $d=3$ for the first non-trivial zeros. We see a clear separation of scales between the zeros so the sum is convergent and well-approximated by the first zero $\rho_1$ and its conjugate.}\label{Fig:logplot3}
    \end{figure}

This gives a convergent  representation for the Hartle-Hawking state as the  non-trivial zeros only contribute small corrections that are rapidly decreasing.  This is illustrated   in Figure~\ref{Fig:logplot3} where we plot the magnitude $|\psi_\rho^{(0)}|$ of the contribution of the first zeros to the constant Fourier mode. As a result, in $d=3$, the Hartle-Hawking state takes the form
\be\label{HH3constfluct}
\psi_\r{HH} =  -{2 (\tfrac23\pi)^{5/2}\/ \z(3) \sqrt{G}}  + \psi_\r{HH}^\text{fluct}
\ee
which is a constant piece with small fluctuations added.  The Riemann hypothesis translates into the statement that these fluctuations are of order $G^{1/3}$. In Figure \ref{Fig:HH3intplot}, we checked that the integer expansion \eqref{HH3intConstexp} does reproduce this constant piece to good accuracy in a region of convergence. The fluctuating piece is the sum over the non-trivial zeta zeros. As the contribution of each zero decreases quickly with its imaginary part, it is dominated by the first zero and we have
\be
\psi_\r{HH}^\text{fluct}(z)\approx C^{-{1\/3}}( C^{i\nu_1/3} \psi_{\rho_1}(z) +  C^{-i\nu_1/3}\psi_{1-\rho_1}(z)) = O(G^{1/3})
\ee
where $\nu_1=\r{Im}(\rho_1)$. This shows that it essentially lies in a two-dimensional Hilbert space generated by $\psi_{\rho_1}$ and $\psi_{1-\rho_1}$.

\sss{General  dimensions}

In general dimensions we have a similar story with a few differences depending on whether the dimension is odd or even due to the phase in the definition  \eqref{Cdefinition} of $C$.

The expansion over the poles at $s=n+{1\/2}$, $n\geq 1$ gives in general
\be\label{HHgenintegerexp}
\psi_\r{HH}(z) = \sum_{n\geq 1} {(-C)^n \/n! }E_{n+{1\/2}}(z)
\ee
where the pole at $s=1$ has cancelled the constant term (as it originated from that pole in the first place). We note that this is just the expansion of the exponential in the Poincaré sum, up to the $E_{1/2}$ ambiguity discussed above. This is expected since these poles correspond to the original poles of the gamma function in the Cahen-Mellin integral.  When viewed as an expansion in $G$,  the expression \eqref{HHgenintegerexp} is an asymptotic expansion as it includes terms $G^{-n}$ with arbitrarily large $n$.

We can also expand on the poles in the other half-plane. There we obtain a sum over the non-trivial zeta zeros
\be\label{HHNT}
\psi_\r{HH}(z) = -{2\sqrt{\pi C}\/d\L({d\/2})} -\sum_\rho  {C^{ {\rho/d}-{1\/2}}\G({1\/2}-{\rho\/d})\/d\,\pi^{-\rho/2}\G({\rho\/2})\z'(\rho)} \cE_{\rho/d}(z)~.
\ee
As commented above, the Eisenstein series that appear are quite special because the fact that $\z(\rho)=0$ will cancel some contributions in the Fourier decomposition \cite{EisensteinScattering}. The convergence of the  sum is better for $C$ real which corresponds to odd $d$. This is because $C\in i\R$ creates a direction in which the contribution of $\rho$ is exponentially enhanced in the imaginary part. Note that the wavefunction is not an observable so a divergent representation may still be useful if it leads to convergent sums in the computation of observables.

  The Riemann hypothesis is then a statement about the order of these fluctuations
\be
\psi_\r{HH}=-{2\sqrt{\pi C}\/d\L({d\/2})}+ O(G^{d-1\/2d})~.
\ee
Thus, the Hartle-Hawking state takes the form of a constant term of order $G^{-{1\/2}}$ with small fluctuations of order $G^{d-1\/2d}$ governed by the non-trivial zeta zeros.

\newpage
\ss{Near-singularity dynamics}

\begin{figure}
\centering\includegraphics[width=12cm]{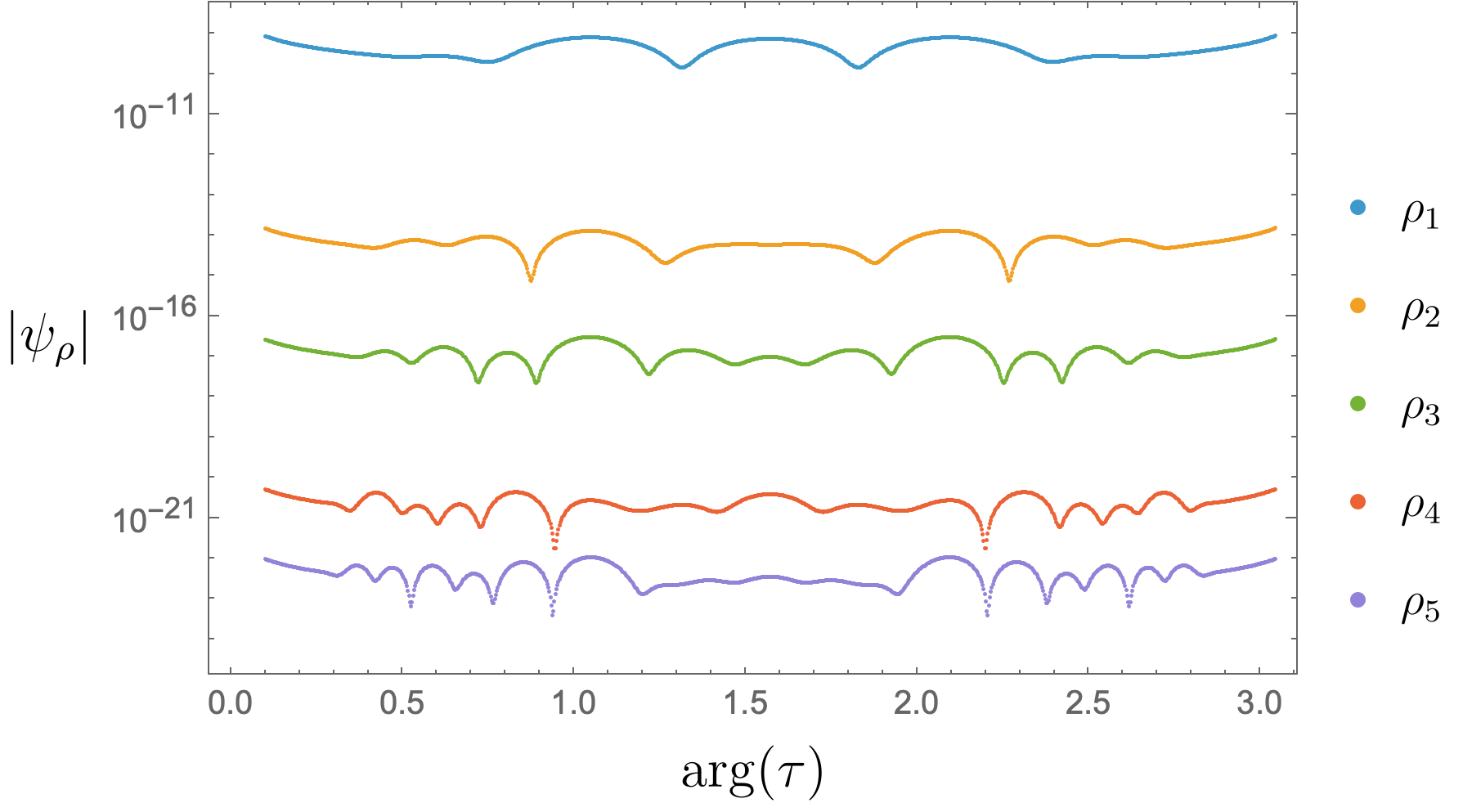}
\caption{The magnitude  of the contribution $\psi_\rho$ to the near-singularity limit of Hartle-Hawking \eqref{psiHHsingexpansion} on the circle $|\tau|=1$. We see a clear scale separation as the wavefunction is   dominated by the first zero $\rho_1$. }\label{Fig:singscalecomp}
    \end{figure}

 For now we have discussed the Hartle-Hawking in the $T\to+\infty$ limit. One can evolve it back in Lorentzian time to obtain the behavior near $T=0$, which is similar to a Big Bang singularity.  The time evolution is governed by the Wheeler-DeWitt equation, the Klein-Gordon equation \eqref{KGauxWDW}, which is solved in terms of Bessel functions. This is how quantum cosmology provides dynamics on the space of automorphic forms  \cite{Godet:2024ich}.

  For simplicity we will restrict the discussion to $d=2$ corresponding to pure three-dimensional cosmology. In this case the dynamics is also equivalent to the $T\bar{T}$ flow with imaginary parameter. The Klein-Gordon equation  takes the form
 \be\label{KGd2}
(\p_T(T^2 \p_T) + M^2 T^2 +\D)\phi_T(T,\tau) =0~,
 \ee
 where $\D$ is the Laplacian \eqref{app:Lapd2} for $\tau\in \fh^2$.
The time-dependent mode $\chi_{s}^\pm(T)$ associated with the Laplace eigenvalue $s(1-s)$ takes the form  
\be
\chi_{{1\/2}+i\mu}^+(T)= e^{{\pi\mu\/2}}\sqrt{\pi \/4 T} H^{(2)}_{i\mu}(MT) ,\qq \chi_{{1\/2}+i\mu}^-(T)=e^{-{\pi\mu\/2}} \sqrt{\pi \/4 T} H^{(1)}_{i\mu}(MT)~.
\ee
which are the normalized modes for the Klein-Gordon inner product \cite{Godet:2024ich}.  They have the asymptotic behavior 
\be
\chi_s^\pm (T) \sim {1\/\sqrt{2M} T} e^{\mp i (MT-{\pi\/4})} ,\qq T\to+\infty
\ee
which shows that $\chi^+_s(T)$ has the right divergent phase for an expanding universe. For the constant mode we get
\be
\chi_0^+(T)={e^{i\pi/4}\/\sqrt{ 2 M}} {e^{-i MT}\/T}~.
\ee
The time-dependent Hartle-Hawking state then takes the form
\be
\phi_\r{HH}(T,\tau)= -{3\, e^{-{i\pi/4}}\/\sqrt{G}} \chi_0^+(T)-\sum_\rho {(4iG)^{\rho/2}\/2\z'(\rho)}\cE_{{1-\rho\/2}}(\tau) \chi_{ 1-\rho\/2}^+(T)~.
\ee
This is  the solution of \eqref{KGd2} with the boundary condition
\be
\lim_{T\to+\infty} {\phi_\r{HH}(T,\tau)\/\chi_0^+(T)}= \psi_\r{HH}(\tau)~,
\ee
where the division by $\chi_0^+(T)$ is the counterterm needed to have a finite $T\to+\infty$ limit. Thus the time-dependent Hartle-Hawking wavefunction should be defined as
\be
\psi^T_\r{HH}(\tau)={\phi_\r{HH}(T,\tau)\/\chi_0^+(T)}~,\qq \lim_{T\to+\infty} \psi^T_\r{HH}(\tau)= \psi_\r{HH}(\tau)~.
\ee
Expanding the Hankel function near $T=0$ gives the near-singularity representation
\be\label{psiHHsingexpansion}
\psi_\r{HH}^T(\tau)=-{3\, e^{-{i\pi/4}}\/\sqrt{G}} + T^{1/4} \sum_{\rho} \Big({4G\/\sqrt{T}}\Big)^{\rho-{1\/2}} \psi_\rho(\tau) + O(T^{3/4})~,
\ee
where we have denoted
\be\label{psirhodefsing}
\psi_\rho(\tau) = -{ \G({\rho\/2})\/2\pi^{\rho/2} \z'(\rho)}\,\cE_{{1-\rho\/2}}(\tau)~.
\ee
This is an Eisenstein series at a pole of the scattering phase $\vphi(s)$ so it can be viewed as a resonance \cite{EisensteinScattering}. Note that the other expansion  does not lead to a well-behaved $T\to0$ limit, as it produces terms $T^{-n}$ for arbitrary $n$. So the expansion on the non-trivial zeta zeros is  the right representation near the singularity.

We see that the constant term has remained so it still provides the leading contribution and in the strict $T=0$ limit the wavefunction becomes this constant. We will be interested in the fluctuations on top so we write
\be
\psi_\r{HH}^T  =-{3\, e^{-{i\pi/4}}\/\sqrt{G}} + T^{1/4} \psi_\text{fluct}+ O(T^{3/4}),\qq T\to0
\ee
and the Riemann hypothesis is the statement that the fluctuations are $\sim T^{1/4}$ so   the purely oscillating fluctuations take the form
\be\label{fluctHHsing}
 \psi_\text{fluct} =\sum_{n\geq 1} \le(  e^{-i \nu_n t}\,\psi_{\rho_n}(\tau) + e^{i \nu_n t}\,\psi_{1-\rho_n}(\tau)  \ri)
\ee
where we have written $\rho_n = \tfrac12\pm i \nu_n$ so that $\{\nu_n\}$ are the set of positive imaginary parts of the non-trivial zeta zeros. The natural time variable that emerges here is 
\be\label{tsing}
t = {-}\log \Big({\sqrt{T}\/4G}\Big)
\ee
written in units where $\l_\r{dS}=1$. In this variable, the singularity $T\to0$ corresponds to $t\to+\infty$. As a result, we see that  the  Hartle-Hawking state oscillates near the singularity with frequencies precisely given by the $\{\nu_n\}$ in the time variable \eqref{tsing}. The Hilbert-Pólya Hamiltonian is defined to  be such that its  eigenvalues are the imaginary parts $\{\nu_n\}$. Then we see that it provides  the time evolution for the  Hartle-Hawking state near the singularity.

\begin{figure}
\hspace{-1cm}
  \begin{tabular}{cc}
    \subf{\includegraphics[height=8cm]{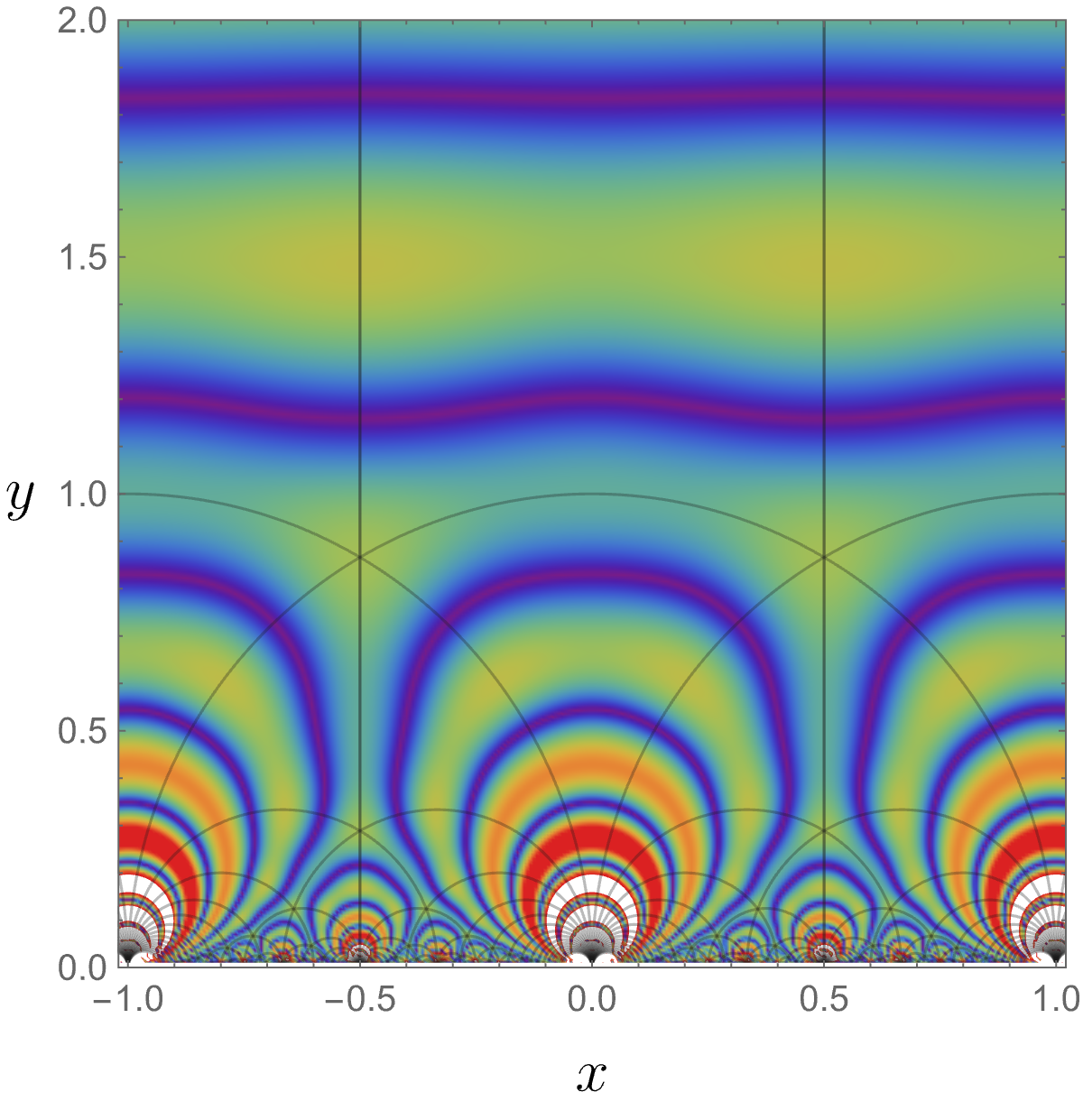}}{\hspace{0.4cm}$\t=0$} &
    \subf{\includegraphics[height=8cm]{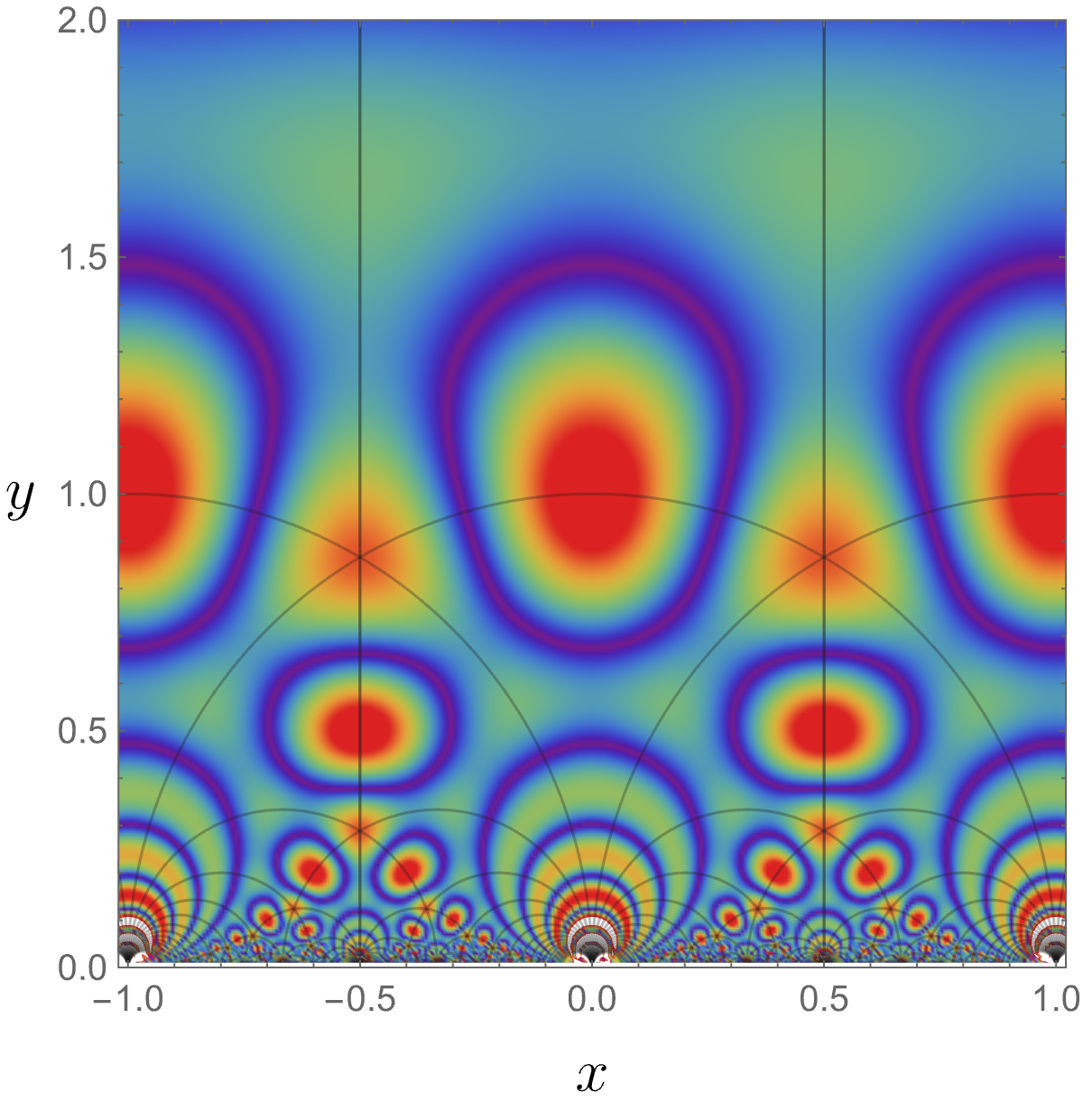}}{\hspace{0.4cm}$\t=2$}\vspace{0.5cm}\\
    \subf{\includegraphics[height=8cm]{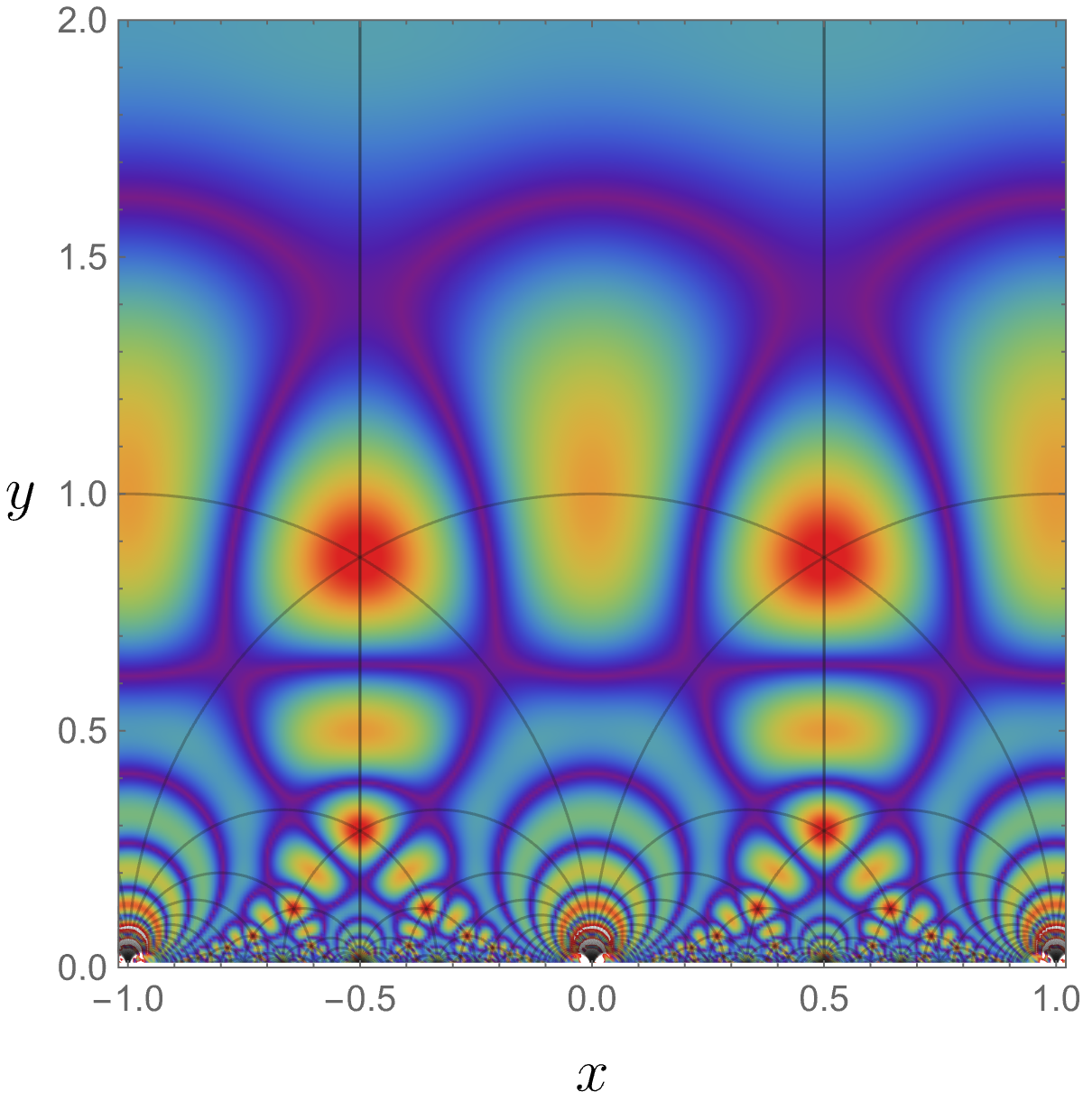}}{\hspace{0.4cm}$\t=4$} &
    \subf{\includegraphics[height=8cm]{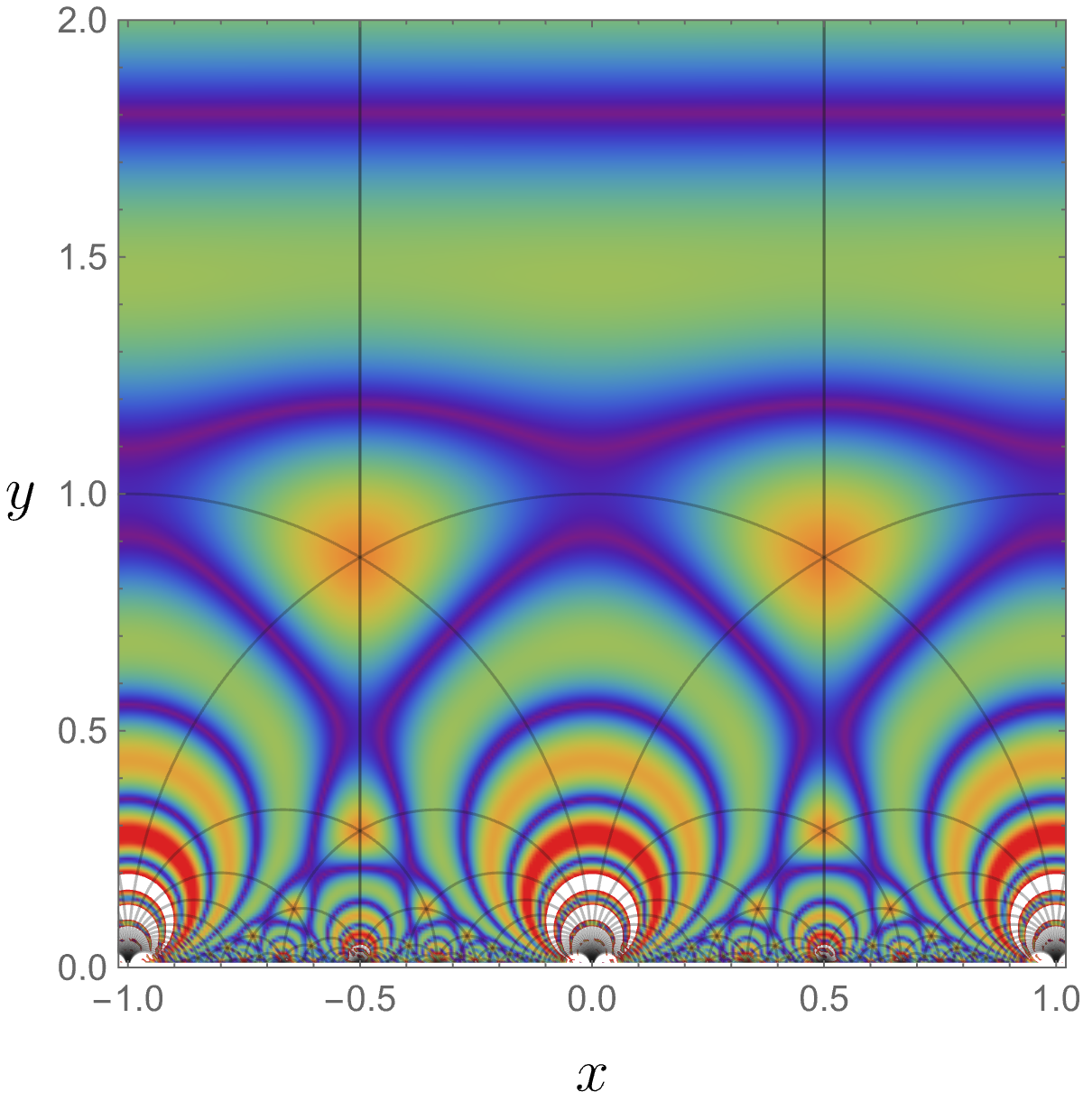}}{\hspace{0.4cm}$\t=6$}
  \end{tabular}

  \caption{The  Hartle-Hawking wavefunction $\psi_\text{fluct}$, where we have removed the leading constant term, in the near-singularity limit $T\to0$. This is dominated by the first non-trivial zero $\rho_1=\tfrac12+i\nu_1$ with $\nu_1\approx 14.13$ so we have $|\psi_\text{fluct}|\approx |\psi_{\rho_1}+e^{i\t}\psi_{1-\rho_1}|$   with an oscillating phase $
\t\sim \nu_1\,\r{log}\,T$. The wavefunction is periodic in the variable $-\tfrac12\r{log}\,\, T$ with frequency $\nu_1$  and we display four snapshots of the magnitude $|\psi_\text{fluct}|$ along the cycle. The other zeros contribute small corrections and the time evolution in the variable $-\tfrac12 \r{log}\,T$ is governed by the Hilbert-Pólya Hamiltonian. }\label{Fig:HHplotsing}
\end{figure}

There is no exponential enhancement here and the sum over $\rho$  appears to converge. 
In fact the contribution of the zeros is suppressed in the imaginary part, see Figure \ref{Fig:singscalecomp}, although not parametrically. Despite being dominated by the constant term, the fluctuations can be approximated by the contribution of the first zero
\be
 \psi_\text{fluct}\approx  \ e^{-i \nu_1 t}\,\psi_{\rho_1}(\tau) + e^{i \nu_1 t}\,\psi_{1-\rho_1}(\tau)  
\ee
where $\rho_1= {1\/2}+ i \nu_1$. The Hartle-Hawking fluctuations then correspond approximately to a two-state system  with oscillations at frequency  $\nu_1\approx 14.13$  in the variable $t$.   We display in Figure~\ref{Fig:HHplotsing} the magnitude $| \psi_\text{fluct}|$ on $\fh^2$ for four snapshots along the oscillation.

The fact that the imaginary parts $\{\nu_n\}$ of the non-trivial zeros should be eigenvalues of a self-adjoint operator has a long history going back to Hilbert and Pólya. They have the same statistics as the eigenvalues of a random Hermitian matrix \cite{Montgomery1973,Odlyzko1987,KatzSarnak1999}, suggesting that a natural realization may be possible, which would in particular recast the Riemann-Weil explicit formula as a  trace formula \cite{hejhalSelbergTraceFormula1976,Connes}. Some attempts to construct physically  the Hilbert-Pólya operator include \cite{BerryKeating1999a,BerryKeating1999b,Sierra:2016rgn, Bender:2016wob}. 

Here, we see that the Hilbert-Pólya Hamiltonian emerges in pure three-dimensional quantum cosmology on the torus, as the  time evolution operator of the Hartle-Hawking state near the singularity. A similar statement can be made in $d=3$ or higher as  the time evolution from \eqref{KGauxWDW} is always solved by Bessel functions, and their near singularity behaviors are $T^{\#}$ with the exponent given by the spectral parameters, proportional to the non-trivial zeta zeros in the expansion \eqref{HHNT}.

\ss{Higher order corrections}\label{sec:higher}

For now we have discussed the leading order Hartle-Hawking state. In this section we briefly explain how to obtain the higher order corrections in $G$.

In $d=2$, the one-loop corrections can be computed exactly \cite{Giombi:2008vd} and give for the leading Hartle-Hawking saddle \cite{Maloney:2007ud}
\be\label{psiHH2oneloop}
\Psi_0(\tau)= {e^{-C y}\/\prod_{n\geq 2} |1-e^{2i\pi n \tau}|^2}~,
\ee
which is expected to be one-loop exact, due to the Virasoro symmetry. As explained above  this must be multiplied by 
\be
\sqrt{Z_\r{bc}} = \sqrt{y}|\eta(\tau)|^2 =\sqrt{y}\,e^{-\pi y/6}\prod_{n\geq 1}|1-e^{2i\pi n\tau}|^2
\ee
so that we get the wavefunction for the leading saddle
\be
\psi_0(\tau)= \sqrt{y}\,e^{-(C+{\pi\/6})y}|1-e^{2i\pi\tau}|^2~.
\ee
In terms of the central charge $c={3i\/2G}+O(1)$, we have $C=-{\pi c\/6}$. Thus the shift $C\ra C+{\pi\/6}$ is the shift   $c\ra c-1$ which corresponds to removing the Virasoro descendants. 

The Hartle-Hawking state is the Poincaré sum $\psi_\r{HH}(\tau)=\sum_{\g\in \G_\infty\bs\r{SL}(2,\Z)} \psi_0(\g\tau)$ and we can derive its spectral representation to be
\bea\nt
\psi_\r{HH} \= {6\,e^{-i\pi/4}\/\sqrt{G}}+{1\/2\pi}\int_\R d\nu\,\pi^{i\nu}\G(-i\nu) \le[ Q^{2i\nu}+\wt{Q}^{2i\nu}  - {2\/\z(2i\nu) }b^{2i\nu} \ri]E_{{1\/2}+i\nu}(\tau)\\\label{HHexactdensity2}
&&\hspace{1cm}+\sum_{j\geq 1}  \rho^\ast_j(1) |\G(i r_j)|^2 (b^{2ir_j}+ b^{-2i r_j}) u_j(\tau)~,
\eea
where $Q,\wt{Q}$ and $b$ are defined in \eqref{QQbParam}. The Eisenstein spectrum was given in \cite{Godet:2024ich} and the Maass spectrum can be derived in a similar way. This is obtained by unfolding the Hartle-Hawking state and using the known Fourier decomposition of the Eisenstein series and Maass forms, reviewed in \ref{app:gl2}. Contributions from  Maass forms arise due to the non-zero Fourier mode in $\psi_0$ and involve their first Fourier coefficient $\rho_j(1)$. We will compute the finite norm of this state in section~\ref{Sec:norm}.

In $d=3$, one should be able to compute the one-loop corrections explicitly using the techniques of  \cite{Giombi:2008vd} as the spacetimes of interest are quotients of $H_4$ (Euclidean AdS$_4$). Here we will suggest a shortcut based on $\r{SL}(3,\Z)$ modular invariance. In $d\geq3$, what replaces the $\eta$ function is the $g$ function and we have from \ref{app:Zbc}:
\be
\sqrt{Z_\r{bc}} =  \r{det}(z)^{1/2}|g(\tz)|^3
\ee
which is an $\r{SL}(3,\Z)$ invariant combination generalizing $\sqrt{y}|\eta(\tau)|^2$ in $d=2$. The $g$ function takes the explicit form
\be\label{defgzhigher}
g(\tz)=\r{exp}\le( - {y_2^{1/2}y_1\/4} \cE_{3/2}^{(2)}(\tau_2) \ri)\prod_{m,n\text{ (mod) }\pm 1} |1-\r{exp}(-2\pi y_1|n\tau_2+m|+2i\pi (mx_1+nx_3))|
\ee
where the product is over $(m,n)\in \Z, (m,n)\neq(0,0)$ and with  $\pm (m,n)$ identified.

We expect that the corrected state takes the form
\be
\Psi(z) = {e^{-C\,\r{det}(z)}\/\prod_{m,n} |1-\r{exp}(\dots)|^3} P(z)
\ee
where the denominator is the product appearing in $|g(\tz)|^3$ and $P(z)$ is a simple function. The multiplication 
by $\sqrt{Z_\r{bc}}$ cancels the infinite product   and we obtain
\be
\psi(z) = \r{det}(z)^{1/2}\,e^{-C\,\r{det}(z) -\d S} P(z)~.
\ee
The exponential factor in the $g$-function gives a correction to the classical action
\be
\d S= {3\/4}\,(y_1^2 y_2)^{1/2} \cE_{3/2}^{(2)}(\tau_2) = {\pi\/8}y_1 + {3\z(3)\/8\pi}y_1y_2^2 + \text{Fourier modes in $x_2$}~,
\ee
which involves the completed Eisenstein series $\cE_{3/2}^{(2)}$  for $\r{SL}(2,\Z)$ in the variable $\tau_2=x_2+iy_2$. This correction should be viewed as the $d=3$ version of the $c\ra c-1$ shift in $d=2$.  Famously, this precise Eisenstein series appears in the type IIB effective action \cite{Green:1997tv}.

This will give corrections to the Eisenstein spectrum as we now have from unfolding
\be
(\psi_\r{HH},E_s)= {3\/2} \int_0^{+\infty}{d\l\/\l^3} \, e^{-C\l^3} \,\Big\ln \r{exp}\Big({-}\tfrac34\l^{3/2}\cE_{3/2}^{(2)}(\tau_2)\Big), E_s^P\Big\rn_{\tau_2}~.
\ee
The Eisenstein series $E_s^{(2)}$ appearing  in \eqref{EismaxP} will contribute. Such inner products involving exponentials of Eisenstein series appear difficult to evaluate but we should be able to obtain approximations at  large $C$. This also gives contributions from Eisenstein series   $E_s^{(j)}$  twisted by $\r{SL}(2,\Z)$ Maass forms $u_j$  as we have
\be
(\psi_\r{HH},E_s^{(j)})= {3\/2} \int_0^{+\infty}{d\l\/\l} \,\l^{3s^\ast-{3\/2}} e^{-C\l^3} \,  \Big\ln \r{exp}\Big({-}\tfrac34\l^{3/2} \cE_{3/2}^{(2)}(\tau_2)\Big), u_j(\tau_2)\Big\rn_{\tau_2}~,
\ee
using that the projection of $E_s^{(j)}$ along $P$ gives $(y_1^2 y_2)^s u_j(\tau_2)$  \cite{Goldfeld_2006}. So we see that the appearance of $E_{3/2}^{(2)}(\tau_2)$ in the classical action leads to non-trivial contributions from the $E_s^{(j)}(z)$ for all $j$.

The minimal parabolic spectrum can be computed from the fact that $(\psi,E_{s_1,s_2})$ is the double Mellin transform in $y_1,y_2$ of the constant term $\psi^{(0,0)}$ \cite{Goldfeld_2006}. At leading order $\psi_0$ depends only on the combination $\l^3= y_1^2y_2=\r{det}(z)$ so this Mellin transform vanishes but higher-order corrections will affect this, and we generically expect contributions from the minimal parabolic Eisenstein series.

Finally, there are also corrections from the function $P(z)$. In $d=2$ it is the factor $|1-e^{2i\pi\tau}|^2$ which arises because  \eqref{psiHH2oneloop} does not include the $n=1$ term in the infinite product. This is because we are computing the vacuum character of a CFT$_2$ and the vacuum has $\r{SL}(2,\R)$ invariance.  In $d=3$, we   should be able to fix the factor $P(z)$ similarly from $\r{SL}(3,\R)$ invariance, as we expect the vacuum character of a CFT$_3$ on $T^3$, and it should correspond to some factors  in the infinite product \eqref{defgzhigher}. We should at least have  
\be
P(z) = |1-e^{2i\pi(x_1+iy_1)}|^2 Q(z)
\ee
where $Q(z)$ contains possible additional  factors.  This factor will project on the first Fourier coefficient in $x_1$ after unfolding, and this should give contributions from $\r{SL}(3,\Z)$ Maass forms, proportional to the corresponding Fourier coefficients, which can be computed  from the  known Fourier decomposition of $\r{SL}(3,\Z)$ Maass forms \cite{Goldfeld_2006}.

Thus we see that higher order corrections will lead to contributions from every automorphic representation appearing in the Langlands spectral decomposition.

\newpage

\section{Möbius  randomness and gravity}\label{sec:Mob}

The Möbius function is defined as  $\mu(n)=(-1)^k$ if $n$ is a product of $k$  distinct primes and zero otherwise. Despite this simple deterministic definition, its statistical properties are similar to a random sequence, and it should be characterized as \emph{pseudo-random}. It is related to the analytic properties of the Riemann zeta function via the  identity 
\be\label{mutozeta}
 \frac{1}{\zeta(s)}=\sum_{n\geq 1} \frac{\mu(n)}{n^s} ~.
\ee
The lack of structure or correlation in the sequence $\mu(n)$ is conjectured to be a  fundamental fact about primes. For instance, the Riemann hypothesis is equivalent to the statement that the Mertens function $M(x) = \sum_{n \leq x} \mu(n)$  exhibits near square-root cancellation, as one would expect  from a sum of random variables. The ``Möbius randomness hypothesis'' suggests that $\mu(n)$ mimics  truly random sequences to a remarkable degree, a fact that underpins many heuristics and conjectures in analytic number theory \cite{sarnakThreeLecturesMobius,GreenTao,Kaisa,TaoBlogPost}. We will see in this section that the Hartle-Hawking state on $T^d$ takes the form of  a Möbius average of CFT$_d$ partition functions. This suggests a more general relationship between Möbius randomness and gravity, to be briefly discussed.

\ss{The compact boson CFT}

We start by discussing  one of the simplest CFT in $d$ dimensions: the compact boson.  This is described by the Euclidean action
\be
S_\text{boson} = {1\/2}\int_{T^d} d^d u \,(g_z^{ij}\p_i\phi \p_j\phi),\qq \phi\sim \phi+R~.
\ee
and the boson is taken to be compact with compactification radius $R$. In $d=2$, this is a standard example of CFT$_2$. In $d=3$, its partition function on $T^3$ was computed in \cite{Chen:2015gma}.

The partition function on $T^d$ can be computed by the path integral. On the torus, the saddle-points correspond to winding solutions
\be
\phi = {R\/2\pi }\,\vn\cdot\vu,\qq \vn\in \Z^d~,
\ee
whose on-shell action is
\be
S_\r{boson} = {1\/2} R^2(2\pi)^{d-2} Q_z(\vn)~.
\ee
where $Q_z$ is the quadratic form on $\Z^d$ associated with the inverse metric
\be
Q_z(\vn) =  \vn\cdot g_z^{-1}\cdot\vn~.
\ee
The leading partition function on $T^d$ is then given by the sum
\be
Z_\r{boson} =\sum_{\vec{n}\in \Z^d} \r{exp} (-D Q_z(\vn)),\qq D = \tfrac12 (2\pi)^{d-2}R^2~,
\ee
a generalized theta function. The Epstein zeta function associated to $Q_z(\vn)$ is defined as
\be
\z_z(s) = \sum_{\vn\in \Z^d,\vn\neq0} Q_z(\vn)^{-s}
\ee
and we can write the partition function as the Mellin integral
\be\label{bosonMellin}
Z_\r{boson}={1\/2\pi i }\int ds\,\G(s) D^{-s} \z_z(s)
\ee
The Epstein zeta function is closely related to the maximal parabolic Eisenstein series, and the relationship is reviewed \ref{app:gfunction}. It takes the form
\be\label{EpEisrelboson}
\cE_{1-s}(z)=\tfrac12 \pi^{-\tfrac{d}{2}s}\G(\tfrac{d}{2}s) \z_z(\tfrac{d}{2}s)~.
\ee
This leads to the spectral representation
\be
Z_\r{boson}(z)={1\/2\pi }(D/\pi)^{-d/4} \int_\R d\mu \,(D/\pi)^{di\mu/2} \cE_{{1\/2}+i\mu}(z)~.
\ee
So the compact boson partition function on $T^d$ is always the Mellin transform of the completed maximal parabolic Eisenstein series, as was found in \cite{Benjamin:2021ygh} for $d=2$.

The functional equation for the Eisenstein series is
\be
\cE_{{1\/2}+i\mu}(z) = \cE_{{1\/2}-i\mu}(\tz)
\ee
which replaces $z$ by its involution $\tz$, defined in \ref{app:involution}. This shows that  if we define
\be
\wh{Z}_\r{boson}(D,z) =(D/\pi)^{d/4} Z_\r{boson}(z)
\ee
then we have the duality
\be\label{bosonReflection}
\wh{Z}_\r{boson}(D,\tz) = \wh{Z}_\r{boson}(D^{-1},z)~.
\ee
This shows that on the compact boson partition function, the involution acts like T-duality: it inverts the compactification radius. In $d=2$, the involution is trivial so T-duality is a symmetry. In higher $d$, it is only when combined with the involution that it is a symmetry. The  one-loop determinant, computed in \ref{app:gfunction}, will break the symmetry \eqref{bosonReflection} as it is not invariant under the involution.

\ss{Hartle-Hawking as an average}

What makes the Hartle-Hawking state different from the compact boson is the relatively prime condition in the sum over $\Z^d$. This comes from the no-boundary condition, as the contracting circle must be primitive to avoid a conical singularity. In the spectral representation this  difference translates in a factor of an inverse zeta function, which is the reason for the existence of an expansion over the zeta zeros.

We obtain a different interpretation if we use instead the Dirichlet expansion \eqref{mutozeta}. From the expression \eqref{HHoverzeta}, we first apply the Riemann functional equation to get
\be\label{HHoverzetaFunc}
\psi_\r{HH}(z)=-{2\sqrt{\pi C}\/d\L(\tfrac{d}{2})}+{1\/2\pi i}\int_{({1\/2})} ds\,{\G(\tfrac12-s) C^{s-{1\/2}} \/\pi^{(ds-1)/2}\G(\tfrac{1}{2}(1-ds))\z(1-ds)} \cE_s(z)~.
\ee
Then using  \eqref{mutozeta},   the Hartle-Hawking state takes the form of a Möbius average
\be
\psi_\r{HH}(z) = \sum_{m\geq 1} \mu(m)m^{{d\/2}-1}Z_\r{CFT}[Cm^d]
\ee
where  what remains is
\be\label{defZCFTMellin}
Z_\r{CFT}[C]={1\/2\pi i}\int ds\,\pi^{(1-d)/2} {\G(\tfrac{1}{2}-s) \G(\tfrac{d}{2}(1-s))\/ \G(\tfrac12(1-ds))}C^{s-{1\/2}}\z_z(\tfrac{d}{2}(1-s))
\ee
where we used the relationship \eqref{EpEisrelboson} to write the Eisenstein series as an  Epstein zeta function. We can evaluate this term-by-term  by summing over the poles of the gamma function, keeping only the leading exponential in the large $C$ limit. 

This takes the form of the partition function of a CFT$_d$ on $T^d$, written explicitly as
\be\label{ZCFTgen}
Z_\r{CFT}[C]=\Big(\frac{d C }{2\pi}\Big)^{d-1\/2} \sum_{\vn\in \Z^d}Q_z(\vn)^{d(d-2)\/4} e^{-C Q_z(\vn)^{d/2}}
\ee
and we will refer to it as the seed CFT.

In $d=2$, we have
\be
Z_\r{CFT}[C] \eqd{2}  \sqrt{C\/\pi}\sum_{(n_1,n_2)\in \Z^2} e^{-C Q_\tau(\vn)}
\ee
so the seed CFT is  the compact boson CFT$_2$. If we define 
\be
Z_\r{boson}[R]\eqd{2}  {R\/\sqrt{2\pi}}\sum_{(n_1,n_2)\in \Z^2}  e^{-{1\/2}R^2 Q_\tau(\vn)}
\ee
where the additional factor of $R$ comes from the zero mode integration, we obtain the Möbius average representation
\be
\psi_\r{HH} = \sum_{m\geq 1}\mu(m)Z_\r{boson}[R = m \sqrt{2C}]~,\qq (d=2)~.
\ee
This gives an expression for  $\psi_\r{HH}$ as a $q$-expansion with integer coefficients, and this interpretation can be generalized to the higher order terms \cite{Godet:2024ich}. In the context of AdS$_3$, this is the Maloney-Witten partition function \cite{Maloney:2007ud}. Here we have $C= -{i\pi\/4G}$ so it is a boson with radius $R \in e^{-i\pi/4}\R$. In AdS$_3$ we would have $C =- {\pi\/4G}$ corresponding to a boson with imaginary radius.

In $d=3$,  the CFT partition function is
\be
Z_\r{CFT}[C]={3C \/2\pi}\sum_{\vec{n}\in \Z^3} Q_z(\vn)^{3/4}e^{-C Q_z(\vn)^{3/2}}~.
\ee
This is not the $T^3$ partition function of the compact boson CFT$_3$ computed above. That said, since it also sums a function  of $Q_z(\vn)$, it can be written as a simple average of the compact boson partition function, using that 
\be
\int_0^{+\infty} dR \,R^2\,\r{exp}\Big( {4\pi^3 R^6\/27 C^2} \Big)\, e^{-\pi R^2 Q_z(\vn)} \approx {3C\/2\pi}  Q_z(\vn)^{3/4} e^{-C Q_z(\vn)^{3/2}}~.
\ee
The saddle-point in the first integral gives the RHS and the one-loop piece can be adjusted to match. This allows us to interpret the Hartle-Hawking state on $T^d$ as an average of the compact boson partition function, involving the Möbius average but also this simpler integral average. In the next section we make a more direct proposal for the seed CFT.

\ss{Liouville CFT as the seed}

The Liouville CFT in $d$ dimensions is defined by the Euclidean action
\be
S_L =\int d^d u\sqrt{g} \le(  {d\/2\Om_d(d-1)!}( \phi\cP_g\phi +2 Q \cQ_g\phi)+\mu \,e^{d b \phi}\ri)~.
\ee
This was introduced in \cite{Levy:2018bdc,Kislev:2022emm} as a generalization of the Liouville CFT$_2$. The action is written using the  conformally covariant GJMS operator \cite{GJMS}  which takes the form 
\be
\cP_g = (-\D)^{d/2}+\text{lower order terms}~,
\ee
and $\cQ_g$ is the Branson $\cQ$-curvature scalar \cite{Branson1985}. We also denote $\Om_d= \r{vol}(S^d) = {2\pi^{d+1\/2}\/\G({d+1\/2})}$  the volume of the sphere.   In odd dimensions the theory is non-local \cite{Kislev:2022emm}. For $d=3$, this theory appeared as part of a proposal for a field theory describing turbulence \cite{Oz:2017ihc,Oz:2018mdq}. In even dimensions, it can be defined rigorously  using probabilistic methods \cite{LiouvilleVargas1,liouvilleDOZZ,Cercle:2019jxx,2021arXiv210513925D}. See also \cite{Diaz:2008hy,Levy:2018xpu,Bugini:2018def, Gaikwad:2023gef}.

The equation of motion is 
\be
\cP_g \phi +Q \cQ_g + \mu \Om_d(d-1)! b e^{bd\phi}=0~.
\ee
In this section we will consider the Liouville CFT$_d$ on the torus $T^d$. The sphere $S^d$ will be discussed in the next section.

The torus partition function of Liouville CFT may be ill-defined or infinite. Here we will define it only in the semi-classical approximation. We take the torus with the flat metric $g_z$ so we have $\cP_g = (-\D)^{d/2}$ and $\cQ_g=0$. The equation of motion reduces to 
\be
\cP_g\phi= -\mu b \,\Om_d (d-1)! \, e^{db\phi}~.
\ee
 We can look for a solution of the form
\be
\phi = \phi_0+\mu \phi_1~,
\ee
where $\D\phi_0=0$ so that $\cP_g\phi_0=0$. Then the equation of motion at leading order is
\be
\cP_g \phi_1= - b\Om_d(d-1)! e^{db\phi_0}~.
\ee
For consistency we see that we must take
\be
\phi_0 = {i \vnu\/db}~,
\ee
where $\vn\in \Z^d$ and the solution takes the form
\be
\phi =  {i \vnu\/db} -\mu b \,\Om_d (d-1)! Q_z(\vn)^{-d/2} e^{i\vnu} + O(\mu^2)~.
\ee
This has compactified $\phi$ according to $\phi\sim \phi+{2i\pi\/db}$ which is the radius consistent with the Liouville interaction. So  we are  really considering  compactified Liouville theory. This compactification can be viewed as a  regularization as the  radius goes to infinity in the semi-classical limit $b\to0$. 

The solution has non-trivial monodromy on the torus which leads to a non-trivial on-shell action. What appears is the integral
\be
\int_0^{2\pi} du\,(i nu)\, e^{inu +\a \mu \,e^{i nu}} = {2\pi \, e^{\a\mu}\/\a\mu},\qq n\in \Z
\ee
which produces a factor of $\mu^{-1}$ necessary to make the action finite in the $\mu\to0$ limit. The result  for the on-shell action is then
\be
S_L= {\pi^{d/2}\/4b^2  \G(1+{d\/2})} Q_z(\vec{n})^{d/2}~.
\ee
Finally, the semi-classical Liouville $T^d$ partition function is the sum over these saddles:
\be
Z_\r{L}[T^d] = \sum_{\vec{n}\in \Z^d} Z_\text{1-loop}\,e^{-C_L Q_z(\vec{n})^{d/2}}
\ee
where
\be
C_L= {\pi^{d/2}\/4b^2\G(1+{d\/2})} = {\Om_{d+1}\/8\pi b^2}~.
\ee
In two dimensions, this is just the partition function of the compact free boson but in higher dimensions the on-shell action has a non-trivial power.

We see that in all dimensions, it matches the partition function of the seed CFT \eqref{ZCFTgen}, at least at the level of the classical action. Thus we can match the parameters using \eqref{Cdefinition} and we see  that the seed CFT is the Liouville CFT$_d$ with the identification
\be
b^2_m = - (-i)^{d-1}  {2 d^d \Om_{d+1}G\/(d-1)(4\pi m )^d}
\ee
which gives for the first dimensions
\be
b_m^2 \eqd{2} {iG\/m^2},\qq b_m^2 \eqd{3}{9 G\/8\pi m^3},\qq b^2_m\eqd{4} -{2iG\/3\pi m^4} ~.
\ee
Thus we can write at leading order
\be
\psi_\r{HH}[T^d] = \sum_{m\geq 1}\mu(m) Z_L^{(b=b_m)}[T^d]~.
\ee
This shows that the Hartle-Hawking state on $T^d$ can be viewed as the Möbius average of the $T^d$ partition function of the  Liouville CFT.

\ss{de Sitter entropy}

Since we can interpret the Hartle-Hawking state on $T^d$ as a Möbius average of the Liouville CFT$_d$, we may expect that such a relationship would go beyond the torus topology.  Here we will see that a similar interpretation is possible for the sphere.

The Hartle-Hawking state on $S^d$ produces the Bunch-Davies vacuum and is related to de Sitter entropy as
\be
\psi_\r{HH}[S^d] = e^{S_\r{dS}/2}
\ee
at leading order. This follows from the fact that the norm
\be
\ln\psi_\r{HH}[S^d] |\psi_\r{HH}[S^d]\rn  = e^{S_\r{dS}}~,
\ee
is computed by the $S^{d+1}$ saddle-point corresponding to Euclidean de Sitter \cite{Gibbons:1976ue}. The  dS$_{d+1}$ entropy is given by
\be
S_\r{dS}= {d \Om_{d+1}\/8\pi G } ={\pi^{d/2}\/2 \G({d\/2})G}~.
\ee
We will show that $\psi_\r{HH}[S^d]$ can also be written in a simple way as a Möbius average of the Liouville CFT$_d$ on the sphere. For the sphere $S^d$ the curvature scalar is $\cQ_g= (d-1)!$ and we can look for constant solutions for which the equation of motion reduces to 
\be
e^{d b\phi} = -{1\/\mu b^2\Om_d}~.
\ee
There are complex constant solutions given by
\be
\phi = {2ik\pi - \log(-\mu b^2\Om_d)\/ db},\qq k\in \Z~.
\ee
The on-shell action takes the form
\be
S_L = s_0 + {2ik\pi\/ b^2}
\ee
where we have isolated the $k$-dependence. We will take a prescription for $\mu$ so that $s_0$ is discarded in the ambiguous normalization.

As a result the semi-classical Liouville sphere partition function is
\be
Z_L[S^d] = -\cN \sum_{k\geq 0} e^{ -{2ik\pi/b^2}}  =\cN {e^{ i\pi /b^2}\/2\,\r{sin}({\pi\/b^2})} =\cN {e^{ 2i\pi /b^2}\/1-e^{2i\pi/b^2}}~.
\ee
For the Liouville CFT$_2$ a similar analysis was performed in \cite{Mahajan:2021nsd}. We expect that the same answer, and perhaps its corrections, can be obtained from the DOZZ formula \cite{Dorn:1994xn,Zamolodchikov:1995aa} and its generalization in higher dimensions \cite{Kislev:2022emm}. This would involve for example integrating three times   $C_\r{DOZZ}(b,b,b)$  and taking a semi-classical limit with an appropriate prescription.

To perform the Möbius average of this quantity, the relevant identity seems to be the Lambert series of the Möbius function
\be
\sum_{m\geq 1}\mu(m) {q^m\/1-q^m} = q~.
\ee
We see that the correct identification is $b=\tilde{b}_m$ with
\be
\tilde{b}_m^2 = {4i\pi\/ m S_\r{dS}} =  {8i   \pi^{1-d/2}\G(\tfrac{d}{2})\/m}G
\ee
which gives
\be
\tilde{b}_m^2 \eqd{2} {8iG\/m},\qq \tilde{b}_m^2 \eqd{3} {4 iG\/m},\qq \tilde{b}_m^2 \eqd{4} {8iG\/\pi m }~.
\ee
This then gives the relation
\be
 \psi_\r{HH}[S^d] = e^{S_\r{dS}/2}= \sum_{m\geq 1}\mu(m) Z_L^{(b=\tilde{b}_m)}[S^d]~.
\ee
This shows that the Hartle-Hawking state on $S^d$ can be written as a Möbius average of the Liouville CFT partition function on $S^d$, in a very similar way as for $T^d$. Note that the precise average performed is different: the values $b_m$ for the torus are different from the values $\tilde{b}_m$ for the sphere. In particular the average is obtained as  $G\ra G/m^d$ for the torus while it is $G\ra G/m$ for the sphere.

The Liouville state in $d=2$, defined as the partition function of the Liouville CFT$_2$ for $c\in 13+i\R$, is related to the complex Liouville string \cite{Collier:2024kmo}. It was suggested in \cite{Collier:2025lux} that  $e^{S_{\r{dS}_3}/2} = N_\r{eff}$ is the effective rank of the matrix model which computes cosmological observables in the Liouville state. This suggests that some cosmological observables in the Hartle-Hawking state could be obtained as Möbius-averaged versions of the complex Liouville string amplitudes.

The Euclidean de Sitter saddle-point has a confusing phase \cite{Polchinski:1988ua,Maldacena:2024spf,Ivo:2025yek,Shi:2025amq} that appears to prevent a direct interpretation in terms of microstate counting. Our interpretation of the de Sitter entropy as a Möbius average is not in tension with a non-trivial phase.

The general Möbius inversion identity \cite{chenMobiusInversionPhysics2010}
\be
G(x)=\sum_{n= 1}^{+\infty}F(nx) \Longleftrightarrow F(x) = \sum_{m= 1}^{+\infty}\mu(m) G(mx)
\ee
suggests that the Möbius average is some kind of gauging. The Hartle-Hawking state for a disconnected topology should be different from the product:
\be\label{NonFactorization}
\psi_\r{HH}[S^d\sqcup S^d]\neq \psi_\r{HH}[S^d]\times \psi_\r{HH}[S^d]
\ee
because of wormhole contributions in
 the path integral. For CFT partition functions we would have equality. This is the factorization paradox \cite{Witten:1999xp,Maldacena:2004rf} which is resolved if we view the gravitational path integral as computing an average \cite{Saad:2019lba,Maloney:2020nni,Chandra:2022bqq,Afkhami-Jeddi:2020ezh,Schlenker:2022dyo,Belin:2023efa,Jafferis:2024jkb}. The natural average to consider here is the Möbius average so this suggests that  the Möbius average should be quenched: we only perform the average at the end, so once in the LHS of \eqref{NonFactorization} and  twice in the RHS. This would give predictions for wormhole contributions similar to those discussed in \cite{Page:1986vw,Cotler:2019nbi,Chen:2020tes,Fumagalli:2024msi}.

\newpage

\section{The Hartle-Hawking norm}\label{Sec:norm}
 
In this section we consider the computation of  the Hartle-Hawking  norm
\be
\norm{\psi_\r{HH}}^2 = (\psi_\r{HH},\psi_\r{HH}) = \int_{\r{SL}(d,\Z)\bs\fh^d} d^\ast z\, |\psi_\r{HH}(z)|^2~.
\ee
In particular we will see that the Hartle-Hawking state has a finite norm.  For $d\geq 3$, we will compute the norm at leading order. For $d=2$, the higher order corrections to the Hartle-Hawking state are known so we will compute the exact norm. We will see that the third-quantized norm takes the suggestive form of an infinite product.

\ss{Leading norm}

The Hartle-Hawking state at leading order takes the form
\be
\psi_\r{HH}(z) = -{2\sqrt{\pi C}\/d\L(\tfrac{d}{2})}+{1\/2\pi i}\int_{({1\/2})} ds\,\G(\tfrac{1}{2}-s) C^{s-\tfrac{1}{2}}E_s(z)~,
\ee
so the norm splits into a constant and Eisenstein contribution
\be
\norm{\psi_\r{HH}}^2= \norm{\psi_\r{HH}^\text{const}}^2+\norm{\psi_\r{HH}^\text{Eis}}^2
\ee
The constant contribution to the norm is then
\be
\norm{\psi_\r{HH}^\text{const}}^2 = V_d  {4\pi |C|\/d^2\L({d\/2})^2}={ \prod_{\l=2}^{d-1} \L(\tfrac{\l}{2})  \/ d\L({d\/2})} 4\pi|C|~.
\ee
where we recall that $V_d=\r{vol}(\r{SL}(d,\Z)\bs \fh^d)$ is given in \eqref{volhdSLd}. 

To compute the Eisenstein contribution, we can use the inner product
\be
(E_{{1\/2}+i\mu},E_{{1\/2}+i\nu}) = {\pi\, V_{d-1}\/2}\d(\mu-\nu),\qq d\geq 3~.
\ee
For $d=2$, the inner product has a reflected piece and this will be discussed in  the next section. We can then compute the Eisenstein piece to the norm
\be\label{normEisIntegral}
\norm{\psi_\r{HH}^\text{Eis}}^2 = {V_{d-1}\/8} \int_\R d\mu\,{(C/C^\ast)^{i\mu}\/\mu\,\r{sinh}(\pi\mu)} ~.
\ee
and we have used $|\G(i\mu)|^2= {\pi\/\mu\,\r{sinh}(\pi\mu)}$. The integral has a divergence at $\mu=0$ but we can shift the contour $\mu\ra\mu\pm i\e$ and the residue at $\mu=0$ vanishes.

Note that $C$ has a periodic phase \eqref{Cdefinition} and we have  $
C/C^\ast=(-1)^{d+1}$. For $d$ odd this is one and the integral evaluates to
\be
\norm{\psi_\r{HH}^\text{Eis}}^2\eqd{3} -{\pi\/12}\log 2
\ee
which is negative after regularization.  In even dimensions, we have $C/C^\ast = -1$ so the integral is logarithmically divergent. In fact, we expect such $G$-independent contributions to be ambiguous and that higher order corrections will correct the norm at this order.  

As a result the constant term always dominates and we have at leading order
\be
\norm{\psi_\r{HH}}^2 \eqd{3} {(\tfrac23 \pi)^5\/\z(3)G},\hspace{1.5cm}\norm{\psi_\r{HH}}^2 \eqd{4} {5(\tfrac34\pi)^2 \z(3)\/2 G}  ~.
\ee
We cannot obtain higher order terms in $d\geq 3$ since we haven't computed the corrections to the Hartle-Hawking state discussed in Section \ref{sec:higher}. In $d=2$, the exact Hartle-Hawking state is known \eqref{HHexactdensity2}, and we will now consider the computation of the exact Hartle-Hawking norm in $d=2$. We expect this to offer a blueprint for a similar computation in higher dimensions.

\ss{Eisenstein contributions}

For $d=2$, we will parametrize the Hartle-Hawking state as
\be\label{QQbParam}
c = 1-6 Q^2 = 25-6\wt{Q}^2 ,\qq Q = b-{1\/b},\qq  \wt{Q} = b+{1\/b}~,
\ee
and we choose $b=e^{i\pi/4}\sqrt{4G}$ to have 
\be
c=13+{3i\/2G}+O(G)i~,
\ee
as required by the Wheeler-DeWitt equation. We  define the elementary Poincaré sums
\bea
\psi[Q]\= \sum_{\g\in \G_\infty\backslash \G}\sqrt{y} |q|^{Q^2/2}|_\g\\
 \chi_1\= \sum_{\g\in \G_\infty\backslash \G} \sqrt{y} |q|^{{1\/2}(b^2+{1\/b^2})}\,2\,\r{cos}(2\pi x)|_\g~.
\eea
where $q=e^{2i\pi \tau}$ and $\tau\in \r{SL}(2,\Z)\bs \fh^2$. The Hartle-Hawking state takes the form
\be
\psi_\r{HH}=\chi_0-\chi_1,\qq \chi_0 = \psi[Q]+\psi[\wt{Q}]~.
\ee
Thus  there are three contributions to the norm
\be
\norm{\psi_\r{HH}}^2 = \norm{\chi_0}^2 - 2\,\r{Re}\,(\chi_0,\chi_1) + \norm{\chi_1}^2~,
\ee
and we will now compute each contribution separately. The spectral decomposition was obtained in \cite{Godet:2024ich} and leads to the expression
\bea
\psi[Q] \= {1\/4\pi} \int_\R d\nu \,\pi^{i\nu}\G(-i\nu)\le[ Q^{2i\nu}+{\z(-2i\nu)\/\z(2i\nu)}Q^{-2i\nu}\ri] E_{{1\/2}+i\nu}~,\-
 \= {1\/2\pi} \int_\R d\nu \,\pi^{i\nu}\G(-i\nu) Q^{2i\nu} E_{{1\/2}+i\nu}~,
 \eea
 where we have used the functional equation to remove the reflected piece. 

 We can compute the Eisenstein inner product by using \eqref{innerprodEis2}. This gives the Eisenstein contributions as integrals which can be evaluated by summing over poles. Alternatively, we can use the unfolding technique to arrive at an equivalent expression. Below we only present the unfolding computation.

 \pg{Computation of $(\chi_0,\chi_0)$.} We compute the inner product by unfolding the second entry:
\be
(\psi[Q_1],\psi[Q_2]) = \int_0^{+\infty} {dy\/y^2}\, \psi[Q_1]^{(0)} \sqrt{y}\, e^{-\pi \ol{Q}_2^2 y}~,
\ee
where $\ol{Q}$ is the complex conjugate of $Q$. This  projects on the constant Fourier mode  
\be
\psi[Q]^{(0)}=\sqrt{y}\Big({-}1+  \,e^{-\pi y Q^2} +\sum_{m,n\geq 1}\mu(m) \,e^{-\pi {n^2\/m^2 Q^2} y}\Big)~.
\ee
This leads to
\be
(\psi[Q_1],\psi[Q_2]) =2 \log \ol{Q}_2-\log ( Q_1^2+\ol{Q}_2^2) -\sum_{m,n\geq 1}\mu(m) \,\r{log}\Big( {n^2\/m^2 Q_1^2} +\ol{Q}_2^2\Big)~,
\ee
where we evaluated the log-divergent integral as
\be\label{logdivintegral}
\int_0^{+\infty} {dy\/y}\,e^{-Cy} = -\log C+\text{ambiguous constant}~.
\ee
This can be derived by taking a derivative with respect to $C$. We should only care about the dependence in the gravitational coupling $G$ and ignore the additive (possibly infinite) ambiguous constants that are independent of $G$.

What appears is the infinite product which we can evaluate as \cite{quineZetaRegularizedProducts1993}
\be
\prod_{n\geq 1} \Big(  {n^2\/a^2}+b^2\Big) = {2\,\r{sinh}(\pi ab)\/b}~.
\ee
Performing the sum over $n$ then gives
\be
(\psi[Q_1],\psi[Q_2]) =-\log ( Q_1^2+\ol{Q}_2^2)-\sum_{m\geq 1}\mu(m) \,\r{log} \,\r{sinh}(\pi m Q_1 \ol{Q}_2)~.
\ee
We can rewrite this in two different ways
\bea
(\psi[Q_1],\psi[Q_2])  \=12\pi Q_1\ol{Q}_2-\sum_{m\geq 1}\mu(m) \,\r{log} (1-e^{-2\pi m Q_1 \ol{Q}_2})-\log ( Q_1^2+\ol{Q}_2^2)~,\\
 \=-12\pi Q_1\ol{Q}_2-\sum_{m\geq 1}\mu(m) \,\r{log} (1-e^{2\pi m Q_1 \ol{Q}_2})-\log ( Q_1^2+\ol{Q}_2^2)~,
\eea
using that ${1\/\z(-1)}=-12$. Among these two expressions, only one will give a convergent answer and this is the one we pick. Note that with our parametrization \eqref{QQbParam} all the $Q_1^2+\ol{Q}_2^2 $ that appear are independent of $G$ so the last logarithmic term never contributes.
 
What appears here is the function
\be
F(\a) = -\sum_{m\geq 1}\mu(m)\log(1-e^{- m \a})~,
\ee
which can be defined by  Mellin transform as
\be\label{Ffromintegral}
F(\a) = {1\/2\pi i }\int_{\e+i\R} ds\,{1\/\z(s)}\a^{-s} \G(s)\z(s+1)~
\ee
with $\e>0$. The equality of these two expressions follows from expanding the two zeta functions in Dirichlet series. The integral  \eqref{Ffromintegral} can be evaluated numerically and is found to be finite.  This function   is closely related  to the Riesz function, \eg \cite{RieszIntegrals}. 

The norm of $\chi_0$ is the sum of the four terms
\be
\norm{\chi_0}^2 =(\psi[Q],\psi[Q]) +(\psi[\tQ],\psi[\tQ])+(\psi[Q],\psi[\tQ])+(\psi[\tQ],\psi[Q]) ~.
\ee
From the above considerations, we obtain
\bea
(\psi[Q],\psi[Q]) =(\psi[\tQ],\psi[\tQ]) \= 12 \pi |Q|^2 + F(2\pi |Q|^2)~,\\
(\psi[Q],\psi[\tQ]) \= 12 \pi i Q^2 + F(2\pi Q^2)~,
\eea 
so this gives the norm of $\chi_0$ as
\be\label{normchi0final}
\norm{\chi_0}^2 = {12\pi\/G} + 2 \,\r{Re}\,F(2\pi Q^2)  + 2 F(2\pi|Q|^2)~.
\ee

 \pg{Computation of $(\chi_0,\chi_1)$.}  For the cross term, the unfolding trick gives
\bea
(\psi[Q],\chi_1) \=\int{dy\/y^2} \psi[Q]^{(1)}\sqrt{y}\,e^{-\pi y(\tb^2+{1\/\tb^2})}
\eea
and projects on the first Fourier mode
\be
\psi[Q]^{(1)} = 2\sqrt{y}\sum_{m\geq1 }\mu(m)\, e^{-\pi y (m^2 Q^2 + {1\/m^2 Q^2}) } ~.
\ee
Thus we obtain
\be
(\psi[Q],\chi_1) = -2 \sum_{m\geq 1}\mu(m)\log P_m[Q],\quad P_m[Q] = m^2 Q^2+{1\/m^2 Q^2} + (b^\ast)^2+{1\/(b^\ast)^2}~,
\ee
using the same integral \eqref{logdivintegral}.  Recalling that $\chi_0=\psi[Q]+\psi[\tQ]$ we get
\be
\r{Re}\,(\chi_0,\chi_1) = - 2 \sum_{m\geq 1}\mu(m) \log |P_m[Q]P_m[\wt{Q}] |~,
\ee
where we have
\be
|P_m[Q]P_m[\wt{Q}] |=  \Big| m^2 Q^2 \Big(1-{1\/m^2(Qb)^2} \Big) \Big(1-{1\/m^2 (Q/b)^2}  \Big)\Big|^2~.
\ee
The Möbius sum can then be computed by defining the function
\bea
T(\a) = {-}\sum_{m\geq 1}\mu(m)\log\Big(1 -{1\/m^2\a^2}\Big)
\eea
and resumming it by expanding the logarithm:
\be
T(\a)= \sum_{k\geq 1} {1\/k \a^{2k}} \sum_{m\geq 1} {\mu(m)\/m^{2k}} = \sum_{k\geq 1} {1\/ k\z(2k)} \a^{-2k}~.
\ee
The last sum converges absolutely for $|\a|>1$. We then obtain
\be\label{Re01final}
\r{Re}\,(\chi_0,\chi_1) =-8 \log|Q| + 4 \,\r{Re}\,T(bQ)+4\,\r{Re}\,  T(Q/b)~.
\ee
We have $|Qb|,|Q/b|>1$ so this gives a finite expression for this contribution. 


\ss{Kuznetsov trace formula}

Finally we need to compute the remaining piece $\norm{\chi_1}^2$ of the Hartle-Hawking norm. The new feature of $\chi_1$ is that it has contributions from Maass cusp forms.

We have the decomposition
\be
\chi_1= \chi_1^\r{Eis}+\chi_1^\r{Maass}~.
\ee
The Eisenstein contribution was computed in \cite{Godet:2024ich} and found to be
\be
\chi_1^\r{Eis}= {1\/\pi} \int_\R d\nu \,\pi^{i\nu}\G(-i\nu) {1\/\z(2i\nu)} b^{2i\nu} E_{{1\/2}+i\nu}~,
\ee
where we used the functional equation to rewrite it as a single term. So the Eisenstein contribution to the norm takes the form
\bea\label{chi1Eispiece}
\norm{\chi_1^\r{Eis}}^2\={2\/\pi}\int_\R dr\,{|\G(ir)|^2 \/|\z(2ir)|^2}\Big( (b b^\ast)^{2ir}+ (b/b^\ast)^{2ir}\Big)~,
\eea
using the inner product \eqref{innerprodEis2}.

The Maass forms for $\r{SL}(2,\Z)$ have the Fourier decomposition \eqref{maassfourierGL2} and we can compute the inner product
\be
(\chi_1,u_j) = \rho^\ast_j(1) |\G(i r_j)|^2 (b^{2ir_j}+ b^{-2i r_j}) 
\ee
using the unfolding trick, which projects on the first Fourier coefficient.  Thus the Maass piece of $\chi_1$ takes the form
\be
\chi_1^\r{Maass}=\sum_j  \rho^\ast_j(1) |\G(i r_j)|^2 (b^{2ir_j}+ b^{-2i r_j}) u_j
\ee
involving the first Fourier coefficients $\rho_j(1)$ of $u_j$, whose numerical values can be found in \cite{lmfdb}. The Maass contribution to the norm is then
\be
\norm{\chi_1^\r{Maass}}^2 =  \sum_{j} |\rho_j(1)|^2  |\G(ir_j)|^4 |b^{2ir_j}+b^{-2i r_j}|^2~.
\ee
Here the sum is over all the Maass forms, \ie the discrete spectrum of the Laplacian on $\r{SL}(2,\Z)\bs\fh^2$. This type of sum  can be computed using trace formulas. Here the relevant trace formula is the Kuznetsov trace formula, which takes the general form  \cite{kuznecovPeterssonsConjectureCusp1981}
\begin{multline}\label{Kuzngen}
\hspace{1cm}\sum_{j} {\rho_j(m)\rho_j^\ast(n)\/\r{cosh}(\pi r_j)} h(r_j) +{1\/\pi}\int_\R dr\,{({m\/n})^{ir}\s_{2ir}(n)\s_{-2ir}(m)\/|\z(1+2ir)|^2} h(r)  \\ 
= {\d_{m,n}\/\pi^2}\int_\R dr\,r\,\r{tanh}(\pi r)\,h(r) +\sum_{c\geq 1} {S(n,m;c)\/c}\vphi\Big({4\pi\sqrt{mn}\/c}\Big)~.
\end{multline}
It is closely related to the Selberg trace formula and gives a way to compute sums over Maass forms  in terms of Kloosterman sums 
\be
S(m,n;c) = \sum_{\substack{1\leq a \leq c \\ (a,c)=1 \\ ad=1 \text{ mod } c} } \r{exp}\Big(2i\pi\Big( a {m\/c}+ d {n\/c}\Big)\Big),\quad \vphi(x) = {2i\/\pi} \int_\R dr\, {r J_{2ir}(x)\/\r{cosh}(\pi r)}h(r)~,
\ee
and the function $\vphi$ is defined as a Bessel integral transform. 

For our purpose we should take $m=n=1$ and   the choice 
\be
h(r)= {|\G(ir)|^4} |b^{2ir}+b^{-2i r}|^2 \,\r{cosh}(\pi r)~,
\ee
Then we see that the LHS of \eqref{Kuzngen} is exactly the norm $\norm{\chi_1}^2$, in particular the second term precisely reproduces the Eisenstein piece \eqref{chi1Eispiece}.  Note that we can write
\be
h(r) = {\pi^2 \/r^2\,\r{sinh}(\pi r) \,\r{tanh}(\pi r)}\Big( (4G)^{2ir}+(4G)^{-2ir}\Big) + {2\pi^2\/r^2 \,\r{tanh}(\pi r)}
\ee
and the second term can be ignored as it is independent of $G$.

The first contribution in the RHS, coming from the constant, can be evaluated to
\be
{1\/\pi^2}\int_\R dr\,r\,\r{tanh}(\pi r)\,h(r) =   -8\log|Q|
\ee
and gives a logarithmic correction. Our final formula takes the form
\be\label{normchi1final}
\norm{\chi_1}^2 = -8\log|Q| +\sum_{c\geq 1} {S(1,1;c)\/c}\vphi\Big({4\pi\/c}\Big)~,
\ee
where the  function $\vphi$ is
\be
\vphi(x) = 2i\pi \int_\R dr\, { J_{2ir}(x)\/r\,\r{sinh}^2 (\pi r)}(  (4G)^{2 ir}+(4G)^{-2ir}) ~.
\ee
This expresses the norm of $\chi_1$ in terms of a Kloosterman sum. It is easily evaluated numerically and found to be finite.

We can now combine \eqref{normchi0final}, \eqref{Re01final} and \eqref{normchi1final} to give the full Hartle-Hawking norm 
\bea
\norm{\psi_\r{HH}}^2 \=  {12\pi\/G}-24 \log|Q| + 2\,\r{Re}\,F(Q^2) +  2F(|Q|^2) \-
&& \hspace{1cm}+  8 \,\r{Re}\,T(Q b)+8\,\r{Re}\,  T(Q/b)+\sum_{k\geq 1} {S(1,1;k)\/k}\vphi\Big({4\pi\/k}\Big)~.
\eea
This gives a finite expression for the loop-corrected Hartle-Hawking norm in $d=2$.

As discussed in Section \ref{sec:third}, the  third-quantized Hartle-Hawking state may be more natural and its norm is obtained by exponentiation $\ln\r{HH}|\r{HH}\rn = e^{\norm{\psi_\r{HH}}^2}$. We obtain
\bea\label{finalHH3norm}
\ln \r{HH} |\r{HH}\rn \=  |Q|^{-24}e^{{12\pi\/G}}  \-
&& \hspace{-2cm}\times \prod_{m\geq 1}e^{{S(1,1,m)\/m} \vphi({4\pi\/m})}\Big|(1-\tfrac{1}{b^2 Q^2 m^2})^4(1-\tfrac{b^2}{ Q^2 m^2})^4 (1-e^{-2\pi m|Q|^2}) (1-e^{-2i\pi m Q^2}) \Big|^{-2\mu(m)} \hspace{-0.5cm} 
\eea
after writing the functions $F$ and $T$ explicitly. It is natural to combine the Kloosterman sum and the Möbius sum since the Möbius function is $\mu(m) =S(1,0;m)$. 
What appears is an  infinite product similar to the definition of the Dedekind eta function but with  Möbius exponents. It suggests a dual interpretation  in terms of a quantum system with bosonic/fermionic states according to the sign $\mu(m)$.

We expect similar corrections in higher dimensions. In $d=3$, the $\r{SL}(3,\Z)$ Maass forms appear at higher orders due to the correction factor $|1-e^{2i\pi(x_1+iy_1)}|^2 $ as discussed in Section \ref{sec:higher}. Their contributions to the norm can be computed in a similar way  using the higher-rank version of the Kuznetsov trace formula \cite{blomerApplicationsKuznetsovFormula2013}.

\newpage

\section{Relation to the Langlands program}\label{sec:Langlands}

We finish with a brief and more speculative discussion about the relationship between quantum cosmology and the Langlands program.  This relationship comes from the fact that the main object of study for both quantum cosmology on $T^d$ and the Langlands program for $\r{GL}(d,\R)$ is the Hilbert space  
\be
\cH= L^2(\r{SL}(d,\Z)\bs\fh^d)
\ee
and the unitary operators acting on it. In fact, the goal of the original Langlands program is to classify the elements of $\cH$  as representations of $\r{GL}(d)$  \cite{langlandsProblemTheoryAutomorphics,muellerGenesisLanglandsProgram2021,borel1979automorphic}.  These automorphic representations are the physical states of quantum cosmology on $T^d$. This relationship should generalize to other topologies and groups. We expect an interesting theory when the moduli space has finite volume so that the Langlands spectral decomposition \cite{langlandsFunctionalEquationsSatisfied1976,Moeglin_Waldspurger_1995} provides the elementary cosmological wavefunctions.


Langlands functoriality is the statement that operations on the representations yield operations on the automorphic forms \cite{jacquet1970automorphic,cogdell2004functoriality,arthur2005introduction}. One example  is the Gelbart-Jacquet lift \cite{ASENS_1978_4_11_4_471_0} which produces an automorphic form for $\r{GL}(3)$ from the symmetric square of an automorphic representation for $\r{GL}(2)$. This gives a procedure to obtain a $T^3$ wavefunction from a $T^2$ wavefunction  and we  expect natural physical realizations of such operations.

The relationship to number theory comes from the fact that some automorphic forms have dual descriptions in terms of Galois representations   \cite{taylorGaloisRepresentations2002}. 
 The  fact that all well-behaved Galois representations should arise from automorphic forms is the Langlands reciprocity conjecture  \cite{Langlands1978ICM}, a vast generalization of the classical reciprocity laws in the non-abelian context. In cosmology, this means that special wavefunctions have a dual description as generalized characters. This is similar to the dS/CFT philosophy \cite{Strominger:2001pn,Maldacena:2002vr,Chakraborty:2023yed} where we view the wavefunction  as a CFT partition function. 
 So one may hope to associate a CFT to a Galois representation in such a way that the CFT partition function on $T^d$ is the corresponding automorphic form.

Among the unitary operators acting on $\cH$, the Hecke operators are of central importance.  They are an infinite set of mutually commuting operators which can be viewed  as discretizations or arithmetic analog of the Laplacian. They act as  an infinite-dimensional global symmetry on the Hilbert space   and encode geometric coverings \cite{Dijkgraaf:1996xw}. They have tangible physical consequences such as the fact that the Laplace spectrum  has Poissonian rather than random matrix  statistics  \cite{PhysRevLett.69.2188}. The natural and modern framework to fully capture the structure and action of these Hecke symmetries is the adelic perspective \cite{tateFourierAnalysisNumber, Langlands1971Euler}. That is, we should replace the base field $\R$ by the adeles $\bA =\R\times \prod_p \bQ_p$ with additional $p$-adic components for each prime \cite{gelbartAutomorphicFormsAdele2016,Moeglin_Waldspurger_1995,Goldfeld_Hundley_2011,Fleig:2015vky}. Thus the wavefunction should really be viewed as an element of the adelic Hilbert space
\be
\psi \in L^2(\r{GL}(d,\bQ)\bs\r{GL}(d,\bA)),
\ee
which suggests that we should replace the variable $z\in \fh^d$ by its adelic generalization. The spatial metric $g_z = z \cdot {}^t z$ then becomes adelic, \ie it acquires $p$-adic components for all primes $p$ in addition to the real component.  Such $\psi$ would be the solution of the Wheeler-DeWitt equation in an adelic version of gravity. 

\acknowledgments

This work benefited from discussions with Scott Collier and Lorenz Eberhardt.

\appendix

\section{Automorphic forms for $\r{GL}(d)$}\label{app:aut}
In this appendix, we review some aspects of the theory of automorphic forms for $\r{GL}(d,\R)$ following  \cite{Goldfeld_2006,bumpAutomorphicFormsGL1984}. Some other useful references include \cite{friedbergGlobalApproachRankinSelberg1987,buttcaneHigherWeightGL32018a,goldfeldFirstCoefficientLanglands,goldfeldFunctionalEquationsLanglands2023,imaiFourierExpansionEisenstein1982,efratGL3AnalogIz1992}.

\ss{The $\r{GL}(2 )$ case}\label{app:gl2}

We first briefly review the $\r{GL}(2,\R)$ case which is the spectral theory of Selberg \cite{Selberg,iwaniecSpectralMethodsAutomorphic2002,terrasHarmonicAnalysisSymmetric1985}. The symmetric space
\be
\fh^2 = \r{GL}(2,\R)/(\r{O}(2)\times \R)\approx \r{SL}(2,\R)/\r{SO}(2)
\ee
is just the Poincaré upper-half plane. We can use the coordinate $\tau=x+iy \in \fh^2$ corresponding to the Iwasawa decomposition of $\r{GL}(2)$
\be \arraycolsep=3pt\def\arraystretch{0.9}
z = \bpm 1 & x \\ 0 & 1\epm\cdot \bpm y & 0 \\ 0 & 1 \epm = \bpm y & x \\ 0 & 1\epm \in \fh^2~.
\ee
The invariant metric on $\fh^2$ is
\be\label{invmeth2}
ds^2(\fh^2) = {dx^2+dy^2\/y^2}~.
\ee
The $\r{SL}(2,\R)$ algebra has a unique Casimir corresponding to the Laplacian
 \be\label{app:Lapd2}
 \D = -y^2(\p_x^2+\p_y^2)~.
 \ee
The element $z$ defines a metric $g_z \propto z\cdot {}^tz$  which is proportional to $|du+\tau dv|^2$. For the $T^2$ topology, the $\r{SL}(2,\Z)$ action on the periodic coordinates correspond to the standard action of $\r{SL}(2,\Z)$ on $\fh^2$. The Hilbert space is then
\be
\cH =L^2(\r{SL}(2,\Z)\bs\fh^2)
\ee
 with the Petersson inner product
 \be
(\psi_1,\psi_2 ) = \int_{\r{SL}(2,\Z)\bs \fh^2} d^\ast\tau\,  \psi_1(\tau)\psi_2^\ast(\tau)~,\qq d^\ast \tau = {dxdy\/y^2}~.
 \ee
The spectral decomposition gives a basis for the Hilbert space in terms of eigenfunctions of the Laplacian. This was obtained by Selberg \cite{Selberg,iwaniecSpectralMethodsAutomorphic2002,terrasHarmonicAnalysisSymmetric1985}. A general element $\psi\in \cH$ is decomposed as
\be
\psi(\tau) = \text{const}+{1\/4\pi i}\int_{({1\/2})} ds\,(\psi,E_s)E_s(\tau) + \sum_{j\geq 1} (\psi,u_j) u_j(\tau)~.
\ee
We will now review the elements appearing in this formula.

 The $\r{GL}(2)$ Eisenstein series are the standard real analytic Eisenstein series defined as the Poincaré sum
\be
E_s(\tau) = \tfrac12\sum_{\g\in P\bs\r{SL}(2,\Z)} \r{Im}(\g\tau)^s = \tfrac12\sum_{\substack{(c,d)\in \Z^2\\ (c,d)=1}} {y^s\/|c\tau+d|^{2s}}~.
\ee
The sum is convergent for $\r{Re}(s)>1$ but the Eisenstein series can be analytically continued as meromorphic function in $s$. This analytic continuation follows from a functional equation. The completed Eisenstein series
\be
\cE_s(\tau)=\L(s)E_s(\tau),\qq \L(s) = \pi^{-s}\G(s)\z(2s)
\ee
is meromorphic in $s$ with only simple poles at $s=0,1$ with residue $-\tfrac12,\tfrac12$ respectively. It satisfies  the functional equation
\be
\cE_{1-s}(\tau)=\cE_s(\tau)~.
\ee
Its Fourier decomposition is
\be
E_s(x,y) = y^{s}+\vphi(s)y^{1-s}+{4 \sqrt{y}\/\L(s)}\sum_{n\geq 1}  c_{s-{1\/2}}(n) K_{s-{1\/2}}(2\pi n  y)\,\r{cos}(2\pi n x)
\ee
which involves 
\be
\vphi(s) = {\L(1-s)\/\L(s)},\qq c_{s-{1\/2}}(n) = n^{s-{1\/2}} \s_{1-2s}(n)  = \sum_{\substack{ a,d\geq 1\\ad = n}} \Big({a\/d}\Big)^{s-{1\/2}}~.
\ee 
We have the delta-function normalization
\be\label{innerprodEis2}
(E_{{1\/2}+i\mu} , E_{{1\/2}+i\nu}) = 2\pi \d(\mu-\nu) + 2\pi \vphi(\tfrac12+i\mu)\d(\mu+\nu)~.
\ee
 The Maass cusp forms $\{u_j\}$ correspond to the discrete spectrum of the Laplacian and satisfy
 \be
\D u_j = (\tfrac14+r_j^2)u_j~.
 \ee
 Their Fourier decomposition takes the form
\be\label{maassfourierGL2}
u_j(x,y) = \sum_{n\neq 0} \rho_j(n) \sqrt{y} K_{ir_j}(2\pi |n| y) e^{2i\pi n x},\qq (u_j,u_j)=1~.
\ee
Although their existence can be proven by the Selberg trace formula \cite{Selberg}, they can only be accessed numerically \cite{lmfdb}.

\ss{Langlands spectral decomposition}

The spectral theory of automorphic forms for $\r{GL}(d)$ is due to Langlands \cite{langlandsFunctionalEquationsSatisfied1976} as a generalization of Selberg's theory for $\r{GL}(2)$.

The generalized upper half-plane is 
\be
\fh^d = \r{GL}(d,\R)/(O(d)\cdot \R^\times) \simeq\ \r{SL}(d,\R)/\r{SO}(d)
\ee
which has dimension
\be
\r{dim}\,\fh^d =  {(d-1)(d+2)\/2}~.
\ee
Coordinates  for $z\in \fh^d$ are given by the Iwasawa decomposition of $\r{GL}(d,\R)$
\be
z = \bpm 1 & x_{1,2} & x_{1,3} & \dots & x_{1,d} \\ & 1 & x_{2,3} & \dots & x_{2,d} \\  & & \ddots & & \vdots \\
& & & 1 & x_{d-1,d} \\
&&&& 1 \epm\cdot \bpm y_1 y_2\dots y_{d-1}  & & & \\ & y_1 y_2\dots y_{d-2} & \\ & & \ddots && \\
& & & y_1 & \\
 & & & & 1\epm\cdot R
\ee
where we can always preserve the upper-triangular form by acting on the right with a rotation matrix $R\in \r{O}(d)$.
The space $\fh^d$ can be viewed as the space of $d\times d$ symmetric matrices which parametrize spatial metrics, defined as 
\be\label{defgzApp}
g_z = {1\/\r{det}(z)^{2/d}}\,\,z \cdot {}^t \,z,\qq \r{det}\,g_z=1~,
\ee
on which the $\r{O}(d)$ action on the right of $z$ has no effect.

If we normalize the variable to $h =z /\r{det}(z)^{1/d}$, such that $\r{det}\,h=1$, we can write the invariant metric on $\fh^d$ as
\be\label{defsymmetrichd}
ds^2(\fh^d) = {1\/2}\,\r{Tr}( h^{-1}\cdot dh \cdot h^{-1}\cdot dh)~.
\ee
Taking $g_z$ to be a metric on the $T^d$ topology, we see that the action of $\r{SL}(d,\Z)$ on the periodic coordinates corresponds to its action on $z\in\fh^d$. This action is defined by acting on the left of $z$ with $\g\in \r{SL}(d,\Z)$ and compensating by an $\r{O}(d)$ transformation on the right to preserve the upper-triangular form \cite{bumpAutomorphicFormsGL1984}.

The volume of $\r{SL}(d,\Z)\bs\fh^d$ is finite and we have \cite{Goldfeld_2006}
\be\label{volhdSLd}
V_d= \r{vol}(\r{SL}(d,\Z)\backslash \fh^d)= d\prod_{\l=2}^d \L(\tfrac{\l}{2})
\ee
where for the first dimensions we record
\be
V_2= {\pi\/3} ,\qq V_3  ={\zeta(3)\/4} , \qq V_4  = {\pi^2\/270}\zeta(3),\qq V_5  = {1\/288}\zeta(3)\z(5)~.
\ee
Quantum cosmology in $3+1$ dimensions corresponds to the $d=3$ case so what appears is the space $\fh^3$ known in this context as DeWitt's mini-superspace \cite{DeWitt:1967yk}. The Iwasawa parametrization is explicitly 
\be\arraycolsep=3pt\def\arraystretch{1.1}
z = \bpm  1& x_{2} & x_{3} \\ 0 & 1 &  x_{1} \\ 0 & 0 & 1 \epm \bpm y_1 y_2 & 0 & 0\\ 0 & y_1 & 0 \\ 0 &0 & 1\epm\cdot R=   \bpm y_1 y_2 & y_1 x_2 & x_3\\ 0 & y_1 & x_1 \\ 0 &0 & 1\epm\cdot R~
\ee
and it is standard to also define $x_4= x_1x_2-x_3$. The invariant metric on $\fh^3$ can be written
\be\label{h3metricApp}
ds^2(\fh^3) = {dx_1^2\/y_1^2} + {dx_2^2\/y_2^2}+ {(dx_3-x_2 dx_1)^2\/y_1^2 y_2^2}+ {4\/3}\Big({dy_1^2\/y_1^2}+{dy_1 dy_2\/y_1 y_2}+{dy_2^2\/y_2^2}\Big)~.
\ee
This is a five-dimensional space with negative Ricci curvature $R=-15/2$. For the $T^3$ topology, we have to quotient by the action of $\r{SL}(3,\Z)$. The Hilbert space is then
\be
\psi\in \cH = L^2(\r{SL}(3,\Z)\bs \fh^3)~.
\ee
There are two generalized Laplacians acting on $\cH$ corresponding to the two Casimir elements of the $\r{SL}(3,\R)$ algebra. They take the form
\be\label{app:Delta1def}
\D_1 = y_1^2 \p_{y_1}^2 + y_2^2\p_{y_2}^2 - y_1 y_2 \p_{y_1}\p_{y_2}+y_1^2(x_2^2 +y_2^2)\p_{x_3}^2 + y_1^2 \p_{x_1}^2+ y_2^2 \p_{x_2}^2+ 2 y_1^2 x_2 \p_{x_1}\p_{x_3}~,
\ee\vspace{-0.8cm}
\begin{multline}\label{GL3Delta2}
\Delta_2 = 
-\,y_1^2 y_2\,\p_{y_1}^2\p_{y_2}
+ y_1 y_2^2\,\p_{y_1}\p_{y_2}^2
- y_1^3 y_2^2\,\p_{x_3}^2\p_{y_1}
+ y_1 y_2^2\,\p_{x_2}^2\p_{y_1}
- 2\,y_1^2 y_2 x_2\,\p_{x_1}\p_{x_3}\p_{y_2}\\
\quad
+(y_2^2-x_2^2 )\,y_1^2 y_2\,\p_{x_3}^2\p_{y_2}
- y_1^2 y_2\,\p_{x_1}^2\p_{y_2}
+2\,y_1^2 y_2^2\,\p_{x_1}\p_{x_2}\p_{x_3}
+2\,y_1^2 y_2^2 x_2\,\p_{x_2}\p_{x_3}^2\\
\quad
+ y_1^2\,\p_{y_1}^2
- y_2^2\,\p_{y_2}^2
+2\,y_1^2 x_2\,\p_{x_1}\p_{x_3}
+ (y_2^2+x_2^2 )\,y_1^2\,\p_{x_3}^2
+ y_1^2\,\p_{x_1}^2
- y_2^2\,\p_{x_2}^2~.
\end{multline}
The invariant inner product is
\be
(\psi_1,\psi_2) = \int_{\r{SL}(3,\Z)\bs\fh^3}{dy_1 dy_2 dx_1 dx_2dx_3\/(y_1y_2)^3}\, \psi_1(z)\psi_2^\ast(z)
\ee
and it makes the Laplace operators Hermitian.

The Langlands spectral decomposition then gives a basis for the Hilbert space
\be
L^2(\r{SL}(3,\Z)\bs\fh^3)
\ee
in which $\D_1$ and $\D_2$ are diagonal. This
takes the form \cite{Goldfeld_2006}
\begin{eqnarray}
\psi(z) &=& \text{const}+{1\/(4\pi i)^2}\int_{({1\/3})}ds_1\int_{({1\/3})}ds_2 \,(\psi, E_{s_1,s_2}) E_{s_1,s_2}(z) \-
&& + {1\/2\pi i }\sum_{j\geq 0} \int_{({1\/2})} ds\, ( \psi, E_s^{(u_j)}) E_s^{(u_j)}(z) + \sum_{k\geq 1} (\psi,v_k)v_k(z)~.
\end{eqnarray}
We will now explain the various elements appearing in this decomposition. The continuous spectrum consists of the minimal parabolic Eisenstein series $E_{s_1,s_2}(z)$ and the Eisenstein series twisted by Maass forms $E_s^{(j)}(z) $. The discrete spectrum consists of  the set $\{v_k\}$ of $\r{SL}(3,\Z)$ Maass cusp forms.

A similar decomposition is available in $d\geq 4$. The Langlands spectral decomposition is known for general reductive groups whenever the quotient space has finite volume \cite{Moeglin_Waldspurger_1995}.

\ss{Eisenstein series}\label{app:Eis}

 Eisenstein series always constitute the continuous part of the spectrum. For $\r{GL}(d,\R)$ this theory is reviewed in \cite{Goldfeld_2006}. We will just give a rough outline.

There is a different type of Eisenstein series for each parabolic subgroup of $\r{SL}(d,\Z)$. These subgroups are generalizations of the subgroup of $\r{SL}(2,\Z)$ generated by $T = \bigl(\begin{smallmatrix}1 & 1 \\ 0 & 1\end{smallmatrix}\bigr)$. They are classified by the partitions of $d$ so for $\r{SL}(3,\Z)$, we have  three parabolic subgroups: the minimal parabolic subgroup $P_{1,1,1}$ and the two maximal ones $P_{1,2}$ and $P_{2,1}$. Explicitly they can be written as \cite{Goldfeld_2006} 
\begin{eqnarray}
P_{1,1,1} & = & \le\{ \bsm \ast & \ast & \ast \\ 0 & \ast &\ast \\ 0 & 0 & \ast \esm\ri\}  = \le\{ \bsm 1 & \ast & \ast \\ 0 & 1 & \ast \\ 0 & 0 & 1\esm \cdot\bsm t_1 & & \\  & t_2 & \\  & & t_3 \esm \cdot \bsm \pm1 & &  \\ & \pm1 & \\ & & \pm1 \esm,   t_1,t_2,t_3>0 \ri\},\-
P_{1,2} & = & \le\{ \bsm \ast & \ast & \ast \\ 0 & \ast &\ast \\ 0 & \ast & \ast \esm\ri\}  = \le\{ \bsm 1 & \ast & \ast \\ 0 & 1 & 0 \\ 0 & 0 & 1\esm \cdot\bsm t_1 & & \\  & t_2 & \\  & & t_2 \esm \cdot  \bsm \pm 1 &0 & 0 \\ 0& a & b\\ 0 & c & d \esm, t_1,t_2>0 ,ad-bc = \pm1 \ri\},\-
P_{2,1} & = & \le\{ \bsm \ast & \ast & \ast \\ \ast & \ast &\ast \\ 0 & 0 & \ast \esm\ri\}  = \le\{ \bsm 1 & 0 & \ast \\ 0 & 1 & \ast \\ 0 & 0 & 1\esm \cdot \bsm t_1 & & \\  & t_1 & \\  & & t_2 \esm\cdot \bsm a & b & 0 \\ c & d & 0 \\ 0 & 0 & \pm 1 \esm,   t_1,t_2>0,ad-bc = \pm1 \ri\}~,
\end{eqnarray}
and we take implicitly the intersection of these sets with $\r{SL}(3,\Z)$. For each parabolic subgroup $P$, there is an invariant $I_P(z,s)$ and the Eisenstein series is defined as
\be
E_s(z,P)= \sum_{\g\in P\backslash\r{SL}(3,\Z)} I_P(\g z,s)~.
\ee
The invariant for each type are given as
\be
I_{P_{1,2}}(z,s)  =   (y_1^{1/2}y_2)^s,\qq I_{P_{2,1}}(z,s)  =   (y_1^{2}y_2)^s,\qq I_{P_{1,1,1}}(z,(s_1,s_2))  =   y_1^{2s_1+s_2} y_2^{2s_2+s_1}~.
\ee
We will now summarize some basic properties of each type of Eisenstein series.

\sss{Maximal parabolic Eisenstein series}\label{sec:maxEis}

The maximal parabolic Eisenstein series is defined as
\be\label{defEmaxapp}
E_s(z) = \tfrac12\sum_{\g\in P\bs \r{SL}(d,\Z)} \r{det}(\g z)^s~.
\ee
This is the central object of this paper so we will review some of its properties.

\pg{Functional equation.}
The series converges absolutely for $\r{Re}(s)>1$ but can be continued analytically. The completed Eisenstein series is defined as
\be
\cE_s(z) = \L(\tfrac{d}{2}s)E_s(z)~,
\ee
and it can be shown that it is meromorphic with only simple poles at $s=0,1$ with residues
\be\label{compEispoles}
\r{Res}_{s=0}\,\cE_s(z) = -\frac1d,\qq 
\r{Res}_{s=1}\,\cE_s(z) = \frac1d~.
\ee
It satisfies the functional equation
\be\label{app:maxparfunc}
\cE_s(z)=\cE_{1-s} (\tz)~
\ee
which can be proven by Poisson resummation \cite{Goldfeld_2006}.
Note that in general the functional equation exchanges $s$ and $1-s$ but also replaces $z$ by its transform $\tz$. This is an effect that is non-trivial only for $d\geq 3$ and we will comment more on this involution below.

The functional equations  for the $\r{GL}(d)$ automorphic forms relate $\L(s)$ to $\L(\tfrac{d}{2}-s)$ and  can be viewed as higher-rank generalization of  the Riemann functional equation $\L(s)=\L(\tfrac12 -s)$ which is really the $\r{GL}(1)$ case \cite{tateFourierAnalysisNumber}.

\pg{Inner product.}  The maximal parabolic Eisenstein appears in the spectral decomposition as the Eisenstein series twisted by the constant Maass form. In general we can define the  Eisenstein series twisted by a Maass form   as  \cite{Goldfeld_2006}
\be
E_s^{(j)}(z) =  \sum_{\g\in P\bs\r{SL}(d,\Z)} \r{det}(\g z)^s u_j(\fm_P(\g z))
\ee
where the sum is twisted by an $\r{SL}(2,\Z)$ Maass form $u_j$ with the projection defined as
\be
\fm_P(z) = \bpm y_1 y_2 & y_1 x_2 & 0 \\0 & y_1 &0 \\ 0&0 & 1\epm
\ee
which corresponds to setting $x_1=x_3=0$.

For the constant Maass form $u_0$ we get 
 \be
E_s^{(u_0)}(z) = {2\/\sqrt{V_{d-1}}} E_s(z)
\ee
is the maximal parabolic Eisenstein series \eqref{defEmaxapp}. Since the Eisenstein series appearing in the spectral decomposition have a normalized inner product 
\be
(E_{{1\/2}+i\mu}^{(u_0)},E_{{1\/2}+i\nu}^{(u_0)}) = 2 \pi\, \d(\mu-\nu)~,
\ee
we obtain the inner product for $E_s$ as
\be
(E_{{1\/2}+i\mu},E_{{1\/2}+i\nu}) = {\pi V_{d-1}\/2} \d(\mu-\nu)~.
\ee
\pg{Explicit representations.}
The maximal parabolic subgroup is
\be
P = \Big\{\bsm * & * & * \\ * & * & * \\ 0 & 0 & 1 \esm \in \r{SL}(3,\Z)\Big\}
\ee
 For any relatively prime triple $(n_1,n_2,n_3)$ there is a unique representative of the coset $P\backslash \SLZ$ whose last row is the triple:
\be
\g = \bsm * & * & * \\ * & * & * \\ n_1 & n_2  & n_3 \esm~,
\ee
and we have
\be
\r{det}(\g z) = {\r{det}(z)\/(y_1^2 |n_1\tau_2+n_2|^2+ (n_1x_3+n_2x_1+n_3)^2)^{3/2}}~.
\ee
This is reviewed for example in \cite{friedbergGlobalApproachRankinSelberg1987} and gives the explicit representation
\bea
E_s(z) \= \tfrac12\sum_{\substack{(n_1,n_2,n_3)\in \Z^3 \\ (n_1,n_2,n_3) = 1}}  {(y_1^2y_2)^s\/(y_1^2 |n_1\tau_2+n_2|^2+ (n_1x_3+n_2x_1+n_3)^2)^{3s/2}}~.
\eea
The quadratic form that appears can be seen to be related to the metric $g_z$ defined in \eqref{defgzApp}. We can see that 
\bea
E_s(z) \= \tfrac12\sum_{\substack{(n_1,n_2,n_3)\in \Z^3 \\ (n_1,n_2,n_3) = 1}} (\vn\cdot g_z \cdot \vn)^{-3s/2}~.
\eea
Note that the completed Eisenstein series can be written
\be\label{compEpsum}
\cE_s(z)=\tfrac12 \pi^{-\tfrac32s}\G(\tfrac32s)\sum_{\substack{\vn \in \Z^3,\vn\neq0 }} (\vn\cdot g_z \cdot \vn)^{-3s/2}
\ee
without the relatively prime condition. This is because the factor $\z(3s)$ is exactly obtained in \eqref{compEpsum} by performing the sum over $\r{gcd}(n_1,n_2,n_3)$.  

\pg{Fourier decomposition.} The Fourier decomposition for the $\r{GL}(3)$ maximal parabolic Eisenstein series is given  in \cite{friedbergGlobalApproachRankinSelberg1987}. The constant term takes the form
\bea
E_s^{(0,0)}(y_1,y_2)\= \int_0^1 dx_1\int_0^1 dx_2\int_0^1 dx_3\, E_s(z) \-
\= (y_1^2 y_2)^s +{\L(\tfrac32s-\tfrac12)\/\L(\tfrac32s)}  y_1^{1-s}y_2^s +   {\L(\tfrac32s-1)\/\L(\tfrac32s)}(y_1y_2^2)^{1-s} ~.
\eea
The Fourier coefficient along the maximal parabolic subgroup is \cite{Goldfeld_2006}
\bea\label{app:EisP21Fourier}
E_s^{P_{2,1}} \= \int_0^1 dx_1\int_0^1 dx_3 \,E_s =\l^{3s} +  \l^{3(1-s)\/2}{\L(\tfrac12(3s-1))\/\L(\tfrac32 s)} E^{(2)}_{3s-1\/2}(\tau_2)
\eea
in terms of $\l^3=y_1^2y_2$ and the $\r{GL}(2)$ Eisenstein series $E_s^{(2)}(\tau_2)$.

\sss{Minimal parabolic Eisenstein series}

The minimal parabolic Eisenstein series is defined as
\be
E_{s_1,s_2}(z) = \sum_{\g\in P_{1,1,1}\backslash \r{SL}(3,\Z)} y_1^{2s_1+s_2}y_2^{s_1+2s_2}|_\g ~.
\ee
From the fact that  $P_{1,1,1}\backslash \r{SL}(3,\Z) $ is in bijection with 
\be
J =\{ (\vec{n},\vec{m})\in \Z^3\times \Z^3 \mid \vec{m}\cdot\vec{n} = 0, (m_1,m_2,m_3)=1, (n_1,n_2,n_3)=1\}~.
\ee
we can write this more explicitly as \cite{bumpAutomorphicFormsGL1984} 
\be
E_{s_1,s_2}(z) ={1\/\z(3s_1/2)\z(3s_2/2)} \sum_{\substack{\vn,\vm\in \Z^3\backslash\{0\} \\ \vn\cdot\vm = 0}} Q_{\tz}(\vn)^{-3s_1/2} Q_z(\vm)^{-3s_2/2}
\ee
where the zeta functions remove the prime condition and we recall that $Q_z(\vn) = \vn \cdot g_z^{-1}\cdot \vn$ and $Q_\tz(\vec{m}) = \vec{m} \cdot g_z\cdot \vec{m}$. 

The completed minimal parabolic Eisenstein series is defined as
\be
\cE_{s_1,s_2}(z)= B(s_1,s_2)E_{s_1,s_2}(z) ~,
\ee
involving a product of three completed zeta functions
\be
B(s_1,s_2)= {1\/4}\L(\tfrac32s_1)\L(\tfrac32 s_2)\L(\tfrac12(3s_1+3s_2-1))~.
\ee
It satisfies the functional equation
\be
\cE_{w(s_1),w(s_2)}(z)=\cE_{s_1,s_2}(z)
\ee
for an arbitrary element of the Weyl group $W=S_3$. This constitutes six elements of $\r{SL}(3,\Z)$ given as
\begin{align}\nt
w_0 & = \bsm 1 & &  \\ & 1 & \\ & & 1 \esm, & w_1 & = \bsm & & -1 \\ & -1 & \\ -1 & & \esm, &w_2 & = \bsm & -1 &  \\  -1 & & \\ & &  -1\esm~, \\\label{weylMat}
 w_3 & = \bsm -1  &   &  \\  &  & -1  \\  & -1  &  \esm, &
w_4 & = \bsm  & 1  &  \\  &  & 1 \\ 1 &  &  \esm, &w_5 & = \bsm &  &  1  \\ 1   &  & \\ & 1 &  \esm~.
\end{align}

Their action on the invariant $I_{P}(z,(s_1,s_2))=y_1^{2s_1+s_2} y_2^{2s_2+s_1}$ defines an action on the spectral parameters $(s_1,s_2)$ from
\be
I_P(w z,(s_1,s_2)) = I_P(z,w(s_1,s_2))
\ee
which is explicitly  \cite{bumpAutomorphicFormsGL1984}
\begin{align}\nt
w_0(s_1,s_2) &= (s_1,s_2), &  w_1 (s_1,s_2)& = (\tfrac23-s_2,\tfrac32-s_1), \\
 w_2(s_1,s_2) &=   (s_1+s_2-\tfrac13,\tfrac23-s_2), &
w_3(s_1,s_2) &= (\tfrac23-s_1,s_1+s_2-\tfrac13),\\ \nt
 w_4(s_1,s_2) &= (1-s_1-s_2,s_1) , & w_5(s_1,s_2) &= (s_2,1-s_1-s_2)~.
\end{align}
The constant Fourier mode 
\be
\cE^{(0,0)}_{s_1,s_2}(y_1,y_2) = \int_0^1 dx_1\int_0^1 dx_2\int_0^1 dx_3\,\cE_{s_1,s_2}(z)
\ee
can be expressed as a sum over the Weyl group
\be
\cE^{(0,0)}_{s_1,s_2}(z) = \sum_{w\in W} B(s_1,s_2)y_1^{2s_1+s_2} y_2^{2s_2+s_1}|_{w}
\ee 
and the rest of the Fourier decomposition can be found in  \cite{bumpAutomorphicFormsGL1984}.

\ss{Determinants and the $g$-function }\label{app:gfunction}

In this section, we review the computation of determinants on $T^d$ and the $g$-function. This makes use of the relationship between the Epstein zeta function and the Eisenstein series, and the Kronecker limit formula. Some references include \cite{efratGL3AnalogIz1992,efratDeterminantsLaplaciansSecond1987,liuKroneckerLimitFormula2015,terrasBesselSeriesExpansions2024,bumpKroneckerLimitFormula,chiuHeightFlatTori1997,2015arXiv150100453A}.

\sss{The involution}\label{app:involution}

We first review the involution on $\fh^d$ defined by the inverse transpose $z\to {}^t z^{-1}$. It can be defined as
\be\label{involutionz}
\tilde{z} = w_1 \cdot {}^t z^{-1} \cdot w_1
\ee
where the long Weyl element $w_1$ given in \eqref{weylMat} is used to preserve the upper-triangular form. This is a non-trivial involution of $\fh^d$ that is trivial in $d=2$, so it is an effect that only exists in $d\geq 3$. For $d=3$, its action on the coordinates is
\be
y_1 \leftrightarrow y_2, \qq    x_1 \leftrightarrow -x_2 , \qq  x_3 \leftrightarrow -x_4~,
\ee
where we recall that $x_4=x_1x_2-x_3$. In terms of the metric $g_z$ defined in \eqref{defgzApp}, the involution corresponds to replacing the metric by its inverse, as we have
\be
g_{\tz} = w_1 \cdot g_z^{-1} \cdot w_1
\ee
and the $w_1$ has no real effect since it is in $\r{SL}(3,\Z)$. Note that this involution is rather different from an  $\r{SL}(d,\Z)$ transformation and automorphic forms are not invariant under it. In fact it leads to the definition of dual forms, defined as $\tilde\psi(z) = \psi(\tz)$ \cite{Goldfeld_2006}.

For the maximal parabolic Eisenstein series, the functional equation actually exchanges $z$ and $\tz$ as given in \eqref{app:maxparfunc}. It can be seen that it exchanges the two maximal parabolic subgroup $P_{2,1}\leftrightarrow P_{1,2}$.  For the compact boson partition function \eqref{bosonReflection}, this   involution acts like T-duality, inverting the compactification radius. For the minimal parabolic Eisenstein series it exchanges the two spectral parameters
\be
E_{s_1,s_2}(z) = E_{s_2,s_1}(\tz)~.
\ee
\sss{Epstein zeta function}
For a real scalar field on the torus $T^d$ with metric $g_z$,  the eigenmodes of the Laplacian $\D_{T^d}$ are just Fourier modes $\phi = \r{exp}(i \vn\cdot\vu)$ and their eigenvalues are given by the quadratic form
\be
Q_z(\vn) =  \vn\cdot g_z^{-1} \cdot \vn~
\ee
The determinant can then be computed from the associated Epstein zeta function 
\be
\z_z(s) = \tfrac12 \sum_{\vn\in \Z^d,\vn\neq0} Q_z(\vn)^{-s}~.
\ee
We use a factor of $\tfrac12$ here since   we use a reality condition which identifies $\vn$ and $-\vn$ in the sum.
The determinant is then computed as a zeta regularized product \cite{quineZetaRegularizedProducts1993}
\be
\r{det}(\D_{T^d}) = \r{exp}\le(- \z_z'(0)\ri)~.
\ee
From the representation \eqref{compEpsum} of the Eisenstein series, we   obtain the relationship
\be
\cE_s(\tz) = \pi^{-\tfrac{d}{2}s}\G(\tfrac{d}{2}s) \z_z(\tfrac{d}{2}s)
\ee
where the Eisenstein series is evaluated at $\tz$ since this replaces the metric with its inverse. This is also discussed in \cite{efratGL3AnalogIz1992,liuKroneckerLimitFormula2015,Goldfeld_2006}.

The functional equation for the Eisenstein series then gives  the functional equation satisfied by the Epstein zeta function \cite{terrasBesselSeriesExpansions2024}. To have a relationship that doesn't involve the involution, we can apply the functional equation \eqref{app:maxparfunc} to obtain
\be
\cE_{1-s}(z)= \pi^{-\tfrac{d}{2}s}\G(\tfrac{d}{2}s) \z_z(\tfrac{d}{2}s)~.
\ee
We see that the $s\to0$ limit of the Epstein zeta function corresponds to the Kronecker limit of $\cE_s$ which is the expansion around $s=1$ \cite{liuKroneckerLimitFormula2015}
\be\label{kronLim}
\cE_s(z) ={ 1/d\/s-1}+\tfrac12(\g -\log(4\pi)) - {1\/d} \log(y_1 y_2^2\dots y_{d-1}^{d-1}) - 2 \log\,g(z) +  O(s-1)
\ee
What appears here is the $g$-function, the higher-rank generalization of the Dedekind eta function, which appears in the Kronecker limit of the $\r{GL}(2)$ Eisenstein series.

This shows that
\be
\r{det}(-\D_{T^3}) = 2\pi (y_1 y_2^2)^{1/3} |g(z)|^2~.
\ee
and in general dimensions we have
\be
\r{det}(-\D_{T^d}) = 2\pi \, \r{det}(\tz)^{1/d} |g(z)|^2~.
\ee
This combination is not invariant under the involution (as is easy to check numerically) so we get another modular invariant combination
\be
\r{det}(-\wt\D_{T^d}) = 2\pi\,  \r{det}(z)^{1/d} |g(\tz)|^2~,
\ee
obtained by exchanging $z\leftrightarrow \tz$. This is the determinant on the dual torus, with the metric inverted, since $g_\tz =g_z^{-1}$. So in $d\geq 3$, there are two different generalizations of the modular invariant $\sqrt{y}|\eta(\tau)|^2$ in $d=2$.

\sss{Faddeev-Popov determinant}\label{app:Zbc}

The gauge-fixing of the inner product \eqref{caninnerGF}  was explained in \cite{Chakraborty:2023los}.  We will use a slightly different gauge-fixing condition
\be
(g_z)_{ij} \p_i h_{jk}-{1\/2} (g_z)_{ij}\p_k h_{ij}=0~,
\ee
a non-covariant version of the  harmonic gauge. The $bc$ ghost action then becomes
\be
S_\r{bc} =\int d^d u\,b^k  (g_z)_{ij}  \p_i \p_j c_k~,
\ee
so the  Faddeev-Popov determinant is
\be
Z_\r{bc} = \r{det}(-\wt\D_{T^d})^d~,
\ee
where $\wt{\D}_{T^d} = (g_z)_{ij}\p_i\p_j$ is the Laplacian of the dual torus. 

In $d=2$, we get 
\be
\sqrt{Z_\r{bc}(\tau)} = y^{1/2}|\eta(\tau)|^2~,
\ee
and $d\geq 3$, this gives
\be\label{Zbctrue}
\sqrt{Z_\r{bc}(z)} = \r{det}(z)^{1/2} |g(\tz)|^d~,
\ee
as computed above.

Let us comment about the involution. A different gauge-fixing could have led to the other combination
\be
\r{det}(\tz)^{1/2} |g(z)|^d~.
\ee
which corresponds to taking the Laplacian on the torus instead of the dual torus. This combination seems to break the $P_{2,1}$ symmetry of the leading Hartle-Hawking saddle. Since $\sqrt{Z_\r{bc}}$ is expected to cancel much of the one-loop corrections, as it does in $d=2$, the right gauge-fixing to use appears to be \eqref{Zbctrue}.

\sss{The $g$-function}\label{app:trueg}

The $g$-function is a $\r{GL}(d)$ analog of the Dedekind eta function $\eta(\tau)$ in $\r{GL}(2)$. It can be defined as the function appearing in the Kronecker limit \eqref{kronLim} of the maximal parabolic Eisenstein series. 

 For $\r{GL}(3)$, it takes the form  \cite{efratGL3AnalogIz1992,terrasBesselSeriesExpansions2024,bumpKroneckerLimitFormula, chiuHeightFlatTori1997} 
\be
g(z) = \r{exp}\le( - {y_1^{1/2}y_2\/4} \cE_{3/2}^{(2)}(\tau_1) \ri)\prod_{m,n\text{ (mod) }\pm 1} |1-\r{exp}(-2\pi y_1|n\tau_1+m|+2i\pi (mx_2+nx_4))|
\ee
where the product is over $(m,n)\in \Z^2,(m,n)\neq 0$ and identifying $(m,n)$ with $-(m,n)$. Its modular properties follow from the fact that the  $\r{SL}(3,\Z)$ invariant combinations are \cite{efratGL3AnalogIz1992}
\be
(y_1y_2^2)^{1/3}|g(z)|^2 ~,\qq (y_1^2y_2)^{1/3}|g(\tz)|^2 ~,
\ee 
which are exchanged by the involution. It is harmonic like the Dedekind eta function:
\be
\D_1 \log g(z)=\D_2\log g(z)=0~.
\ee
and in general $\log g(z)$ is annihilated by the algebra of invariant differential operators.

We now give the definition of $g(z)$ for $\r{GL}(d)$ following \cite{liuKroneckerLimitFormula2015}
\bea
g(z) \= \r{exp}\Big( {-} {1\/4}{(y_1y_2^2\dots y_{d-1}^{d-1})^{1/(d-1)} \cE^{(d-1)}_{d\/d-1}}(z_1)\Big)\-
&& \times \prod_{\substack{\vn\in \Z^{d-1}\\ \vn\neq 0,\, \vn\text{ mod }\pm 1}} \le|1 - \r{exp}\Big(-2\pi w^{1/2} (\vn\cdot  S^{-1}\cdot \vn )^{1/2} +  2i\pi\, \vn\cdot q\Big)\ri|,
\eea
where what appears is the $\r{GL}(d-1)$ Eisenstein series $\cE_s^{(d-1)}(z_1)$. The quantities   are defined from
\be
z \cdot  {}^tz = \bpm m & {}^t r \\ r & S\epm
\ee
so that 
\be
m = (y_1 y_2\dots y_{d-1})^2 + (x_{1,2} y_1y_2\dots y_{d-2})^2 + \dots+x_{1,d}^2
\ee
and
\be
S = z_1 \cdot {}^t z_1,\qq 
z_1 = \bpm   1 & x_{2,3} & \dots & x_{2,d} \\   & \ddots & & \vdots \\
 & & 1 & x_{d-1,d} \\ 
&&& 1 \epm\cdot \bpm   y_1 y_2\dots y_{d-2} & \\  & \ddots && \\
 & & y_1 & \\
  & & & 1\epm~.
\ee
We then define
\be
r = z_1 \bpm x_{1,2} y_1 y_2 \dots y_{d-2} \\ x_{1,3} y_1 y_2\dots y_{d-3} \\ \vdots \\ x_{1,d}\epm,\qq q = S^{-1}\cdot r,\qq w= m -  q\cdot  S \cdot q~.
\ee
The $\r{SL}(d,\Z)$ invariant combinations take the form
\be
\r{det}(\tz)^{1/d}|g(z)|^2,\qq \r{det}(z)^{1/d}|g(\tz)|^2 ~.
\ee 
\bibliographystyle{JHEP}
\bibliography{refs}

\end{document}